\documentclass[oupdraft]{sysbio}
\usepackage{url}
\usepackage{chngcntr}
\counterwithout{equation}{section}  % disables 1.1 style
\usepackage{caption}
\usepackage{subcaption}
\usepackage{multirow}
\usepackage{longtable}
\usepackage{lipsum}  % for generating dummy text
\usepackage{indentfirst}
\usepackage{booktabs}
\usepackage{makecell}
\usepackage{amsmath}
\usepackage{tikz}
\usepackage{tikz-qtree}
\usepackage{float}
\usepackage{natbib}
\usepackage{adjustbox}
\usepackage{algorithm,algpseudocode}% http://ctan.org/pkg/{algorithms,algorithmx}
\usepackage{graphicx}
\usepackage{calc}
\usepackage{placeins} 

\makeatletter
\newcommand\setcurrentname[1]{\def\@currentlabelname{#1}}
\makeatother

\algnewcommand{\Inputs}[1]{%
  \State \textbf{Inputs:}
  \Statex \hspace*{\algorithmicindent}\parbox[t]{.8\linewidth}{\raggedright #1}
}
\algnewcommand{\Initialize}[1]{%
  \State \textbf{Initialize:}
  \Statex \hspace*{\algorithmicindent}\parbox[t]{.8\linewidth}{\raggedright #1}
}

\DeclareMathOperator{\diag}{diag}

\DeclareMathOperator{\pathm}{path}

\DeclareMathOperator{\tr}{tr}

\let\oldalign\align
\let\oldendalign\endalign

\renewenvironment{align}
  {\linenomathNonumbers\oldalign}
  {\oldendalign\endlinenomath}

\begin{document}

\nolinenumbers

% Title of paper
\title{Detection of evolutionary shifts in variance under an Ornstein–Uhlenbeck model}
% Each important word in the title should begin with a capital letter

% List of authors, with corresponding author marked by asterisk
\author{
Wensha Zhang$^{1,\ast}$, Lam Si Tung Ho$^{2*}$, and
Toby Kenney$^{3*}$\\[4pt]
% Author addresses
\textit{$^{1,2,3}$Department of Mathematics and Statistics, Dalhousie University, Nova Scotia, Canada}
\\
\textit{$^{*}$These authors have equal contribution}
\\
\textit{$^{1}$Corresponding Author}
\\[2pt]
% E-mail address for correspondence
E-mails: \{wn209685, Lam.Ho, tb432381\}@dal.ca
}
% Identify the name, address, telephone/fax numbers, and e-mail address for the author who will receive proofs and be designated the "corresponding author" in text.

% Running headers of paper:
\markboth%
% First field is the short list of authors
{W.ZHANG, L.S.T.Ho, T.Kenney} 
% Second field is the short title of the paper
{DETECTION OF SHIFTS IN VARIANCE}
% This should be shortened version of the title and no greater than 50 characters

\maketitle

\begin{abstract}
{Sudden changes in environmental conditions can lead to evolutionary shifts not only in the optimal trait value ($\theta$), but also in the diffusion variance ($\sigma^2$) under the Ornstein-Uhlenbeck (OU) model. While several methods have been developed to detect shifts in optimal values, few explicitly account for concurrent shifts in both evolutionary variance and diffusion variance. We use a multi-optima and multi-variance OU model to describe trait evolution with shifts in both $\theta$ and $\sigma^2$ and analyze how covariance between species is affected when shifts in variance occur along the phylogeny. We propose a new method that simultaneously detects shifts in both variance and optimal values by formulating the problem as a variable selection task using an $\ell_1$-penalized loss function. Our method is implemented in the \texttt{R} package \texttt{ShiVa} (Detection of evolutionary \underline{shi}fts in \underline{va}riance). Through simulations, we compare ShiVa with existing methods that can automatically detect evolutionary shifts under the OU model ($\ell$1ou, PhylogeneticEM, and PCMFit). Our method demonstrates improved predictive ability and significantly reduces false positives in detecting optimal value shifts when variance shifts are present. When only shifts in optimal value occur, our method performs comparably to existing approaches. We apply ShiVa to empirical data on floral diameter in \emph{Rafflesiaceae} (with \emph{Euphorbiaceae} outgroups) and buccal morphology in \emph{Centrarchidae} sunfishes.}

{evolutionary shift detection, Ornstein-Uhlenbeck model, LASSO, trait evolution, phylogenetic comparative methods}
\end{abstract}
\newline

\section{Introduction} 
The process of trait evolution is influenced by environmental conditions. When species experience abrupt environmental changes, their evolutionary processes may also change to adapt to the new environment (\citealp*{losos_2011}; \citealp*{mahler_ingram_revell_losos_2013}). These changes in the evolutionary process leave signatures in the observed traits of present-day species. By detecting such evolutionary shifts based on observed trait data, we gain valuable insights into historical environmental changes and the evolutionary history of species. Such changes can affect not only a lineage's average trait value but also how much that trait varies across the lineage's descendants. An increase in evolutionary variance can reflect niche expansion, adaptive radiation, or competitive release \citep{simpson_1953, schluter_2000}, while a decrease can reflect canalization, constraint, or specialization \citep{waddington_1942, futuyma_2010}. A canonical example is the parasitic plant family \emph{Rafflesiaceae}: the transition to parasitism removed the developmental and photosynthetic constraints on flower size present in autotrophic relatives, and flower size has since diversified into a much wider range than in its sister family \emph{Euphorbiaceae} \citep{barkman_etal_2008, davis2007floral}. Similar patterns appear in other clades: deep-sea teleosts evolve with roughly twice the body-shape disparity of shallow-water relatives \citep{martinez_2021}, body-caudal-fin swimmers exhibit greater variance around body-shape optima than their median-paired-fin counterparts \citep{friedman_2021}, and vertebrate sperm components evolve at markedly different rates depending on fertilisation mode \citep{kahrl_2022}. These examples show that trait variability itself can change over evolutionary time. Some hypothesised shifts have a specific expected location, as in the \emph{Rafflesiaceae} case, while other patterns must be uncovered directly from data.

We typically model evolutionary shifts as changes in the value of the parameters of trait evolution models which assume that traits evolve along a phylogenetic tree according to a continuous time Markov process.
The two most popular trait evolution models for continuous traits are the Brownian Motion (BM) model (\citealp*{felsenstein_1985}) and the Ornstein-Uhlenbeck (OU) model (\citealp*{hansen_1997}).
This paper will focus on detecting evolutionary shifts under the OU model, which describes trait evolution as a stochastic process with a tendency to revert toward a selective optimum.
Although several researchers have studied this topic, much of the existing work focuses only on detecting shifts in the optimal value of the model (\citealp*{butler_king_2004, uyeda_harmon_2014, ho_ane_2014, khabbazian_kriebel_rohe_ane_2016, bastide_mariadassou_robin_2016, zhang_kenney_ho_2024}). Detecting variance shifts has received less attention, despite the biological patterns above. In this paper, the term variance specifically refers to the diffusion variance ($\sigma^2$) of the OU model, which quantifies the background rate of stochastic variation in trait evolution. This variance may increase in unstable environments or decrease under stabilizing conditions. Under the OU model, the stationary variance $\sigma^2/(2\alpha)$, where $\alpha$ is the rate of adaptation toward the optimum $\theta$, determines the equilibrium spread of trait values around the optimum \citep{hansen_1997}. More broadly, differences in evolutionary rate across lineages can arise from variation in population size, generation time, or the nature of selection \citep{omeara_ane_sanderson_wainwright_2006}, and detecting such rate shifts is essential for understanding heterogeneous evolutionary dynamics. Several existing frameworks can accommodate variance shifts when the regime structure is pre-specified. For OU-based analyses, \textit{OUwie} \citep{beaulieu_jhwueng_boettiger_omeara_2012} is the closest supervised analogue for our univariate setting, fitting OU models that allow $\theta$, $\alpha$, and $\sigma^2$ to change between predefined regimes by maximum likelihood. Other frameworks target related but orthogonal problems --- multivariate trait evolution \citep{clavel_escarguel_merceron_2015}, intraspecific variance \citep{Gaboriau2020JIVE}, branch-specific Bayesian estimation \citep{bastide_ho_baele_lemey_suchard_2021}, and L\'{e}vy-process models of instantaneous trait jumps \citep{landis_schraiber_liang_2013, duchen_leuenberger_2017}. Automatic detection of variance shifts, however, is more limited. Within the BM framework, several methods do this automatically \citep{eastman_alfaro_joyce_hipp_harmon_2011, grundler_rabosky_zapata_2022, venditti_2011}. Under the OU model, \textit{PCMFit} \citep{Mitov_Bartoszek_Stadler_2019} is, to our knowledge, the only method that automatically detects variance shifts without predefined regimes; however, PCMFit assumes that all model parameters shift together across the same regimes, so shifts in $\alpha$, $\theta$, and $\sigma^2$ cannot occur at independent locations on the tree.

In this paper, we propose a novel LASSO-based method that detects both types of shifts within a unified variable selection framework. Our method uses a multi-optima and multi-variance OU process model, similar to the models used in OUwie (\citealp*{beaulieu_jhwueng_boettiger_omeara_2012}), but innovates by formulating the detection of both shifts in optima and shifts in variance as a single variable selection problem. This approach allows flexible and efficient detection of shifts across the entire tree, without requiring \textit{a priori} specification of regime boundaries. We implement this approach in a new \texttt{R} package, ShiVa (Detection of Evolutionary \underline{Shi}fts in \underline{Va}riance), which detects both shifts in optimal values and variance under the Ornstein–Uhlenbeck model for ultrametric trees and univariate traits. Through simulation studies, we compare ShiVa to existing methods, including $\ell$1ou, PhylogeneticEM and PCMFit evaluating performance in terms of both shift detection accuracy and mean integrated squared error in estimating the optimal value and diffusion variance.

The remainder of the paper is organized as follows. The \emph{Trait evolution with shifts in both optimal value and variance} section introduces the trait evolution model incorporating shifts in both optimal value and variance, and formally defines the shift detection problem. The \emph{Methods} section describes the LASSO-based shift detection algorithm. The \emph{Simulations} section presents simulation results, including comparisons with existing methods. The \emph{Case study} section illustrates the application of ShiVa to two empirical datasets: floral diameter in the parasitic plant family \emph{Rafflesiaceae} (with \emph{Euphorbiaceae} outgroups), using the \texttt{flowerTree} phylogeny \citep{davis2007floral}, and buccal morphology in \emph{Centrarchidae} sunfishes \citep{Revell_Collar_2009}.

\bigskip

\section{Trait evolution with shifts in both optimal value and variance}\setcurrentname{Trait evolution with shifts in both optimal value and variance} \label{sec2}
\subsection{Trait evolution models}
The phylogenetic tree is reconstructed from DNA sequences and is assumed to be known in this paper. The phylogenetic tree reveals the correlation structure between trait values of different species. The trait values are correlated based on the shared evolutionary history of species. The trait values of internal nodes are hidden and only trait values of tip nodes can be observed. We assume the phylogenetic tree is ultrametric.

Trait evolution models are used to model how the trait values change over time. Brownian Motion and Ornstein–Uhlenbeck are two commonly used models to model the evolution of continuous traits. For both models, we let $Y$ denote the vector of observed trait values at the tips, and $Y_i$ denote the trait value of taxon $i$. For a single branch, we let $Y_i(t)$ denote the trait value at time $t$. These two models assume that conditioning on the trait value of a parent, the evolutionary processes of sister species are independent. We therefore only need to specify the model on one branch. For specifying the model on a single branch, we let $Y(t)$ denote the trait value at time $t$ on a fixed branch.

\subsubsection*{Brownian Motion Model}%\textbf{Brownian Motion Model. }
Brownian motion (BM) was first applied to model the evolution of continuous traits over time by \citet{felsenstein_1985}. Under this model, the trait value evolves following a BM process. The process can be written as a stochastic differential equation:
\[
dY(t) = \sigma(t)\, dB(t),
\]
where $\sigma^2(t)$ is the instantaneous diffusion variance at time $t$, and $B(t)$ is a standard Brownian motion. The evolutionary processes of two species are independent, given the trait value of their most recent common ancestor. Therefore, the correlation between the trait values of two species depends only on the shared branch length between them. Based on the properties of the BM process, the observed trait values $\mathbf{Y}$ follow a multivariate Gaussian distribution. For an ultrametric tree of height $1$, each $Y_i$ has mean $\mu_0$ and variance at time $t$, $\sigma^2(t)$, so the covariance between $Y_i$ and $Y_j$ is $\int_0^{t_{ij}} \sigma^2(t) dt$, where $t_{ij}$ is the shared branch length between species $i$ and $j$.

\subsubsection*{Ornstein–Uhlenbeck Model} 
\citet{hansen_1997} uses an OU process to model the evolutionary process.
A selection force that pulls the trait value toward a selective optimum $\theta$ is included in OU models. An OU process $Y(t)$ is defined by the following stochastic differential equation

$$dY(t)=\alpha[\theta(t)-Y(t)]dt+\sigma(t) dB(t)$$

where $ dY(t)$ is the infinitesimal change in trait value; $B(t)$ is a standard BM; $\sigma^2(t)$ measures the intensity of random fluctuation at time $t$; $\theta(t)$ is the optimal value of the trait at time $t$; and $ \alpha \geq 0$ is the rate of adaptation toward the optimum. We assume that $\alpha$ is fixed throughout the tree; when $\alpha$ is unknown, a reasonable estimate can be obtained by fitting a null (no-shift) model using a package such as \texttt{phylolm} \citep{tung2014linear}. For the BM model, the variance of the trait among lineages $\int_0^T \sigma^2(t) dt$ is unbounded when $T$ increases. On the other hand, for the OU model, the variance of the trait among lineages $\int_0^T \sigma^2(t) e^{-2\alpha(T-t)}dt$ is bounded (\citealp{hansen_1997}).  Here, $T$ represents the present time of the species being observed.

\subsection{Evolutionary shifts in optimal value and variance}
\citet{butler_king_2004} formulate the multi-optima OU model for adaptive evolution. This model assumes that the optimal value $\theta(t)$ is constant along a branch and may be different between branches. An abrupt change in $\theta(t)$ on a branch is considered an evolutionary shift in optimal value. However, in most previous work, it is assumed that the variance $\sigma^2(t)$ is constant throughout the tree. In reality, when a change in the environment happens, not only the optimal values, but also the variances are likely to change. In this paper, we use a multi-optima multi-variance model (\citealp*{beaulieu_jhwueng_boettiger_omeara_2012}) which allows the variances $\sigma(t)$ to change over different branches. We first derive the formulas assuming a fixed root, and later extend them to account for a random root.

Solving the OU process equation, the trait value at time $t$ is given by

$$Y(t) = y_0e^{-\alpha t}+\alpha\int_{0}^{t}e^{-\alpha(t-s)}\theta(s)ds+\int_{0}^{t}\sigma(s)e^{-\alpha(t-s)}dB(s)
$$
Solving this, the trait value of species $i$ follows a normal distribution with expectation given by:

$$E(Y_i) = y_0e^{-\alpha T}+\alpha\int_{0}^{T}e^{-\alpha(T-s)}\theta(s)ds$$

By Itô isometry, the element $\mathbf{\Sigma}_{ij}$ (covariance of $Y_i$ and $Y_j$) of covariance matrix $\mathbf{\Sigma}$ is given by

$$\mathbf{\Sigma}_{ij} = e^{-2\alpha (T-t_a)} \int_{0}^{t_{a}}\sigma^2(s)e^{-2\alpha(t_{a}-s)}ds$$

Where $a$ is the separating point of species $i$ and $j$. For ultrametric trees, $t_a = t_{ij}$ is the time from the root to the separating point (i.e., the shared branch length introduced above). The distance between i and j is $2t_{ai} = 2t_{aj} = 2(T-t_{a})$. For shifts in optimal value, \citet{ho_ane_2014} showed that the exact location and number of shifts on the same branch are unidentifiable. We therefore also assume that the optimal value is constant along a branch. We let $\theta_b$ denote the optimal value on branch $b$, $\Delta\theta_b$ denote the change in optimal value on branch $b$ and $\pathm(\textrm{root},i)$ denote the branches on the path of the phylogenetic tree from the root to the tip $i$. The expectation can be writen as the following equation (\citealp*{khabbazian_kriebel_rohe_ane_2016}).

$$E(Y_i) = y_0e^{-\alpha T} + (1-e^{-\alpha T})\theta_0 + \sum_{b \in \pathm(\textrm{root},i)}(1-e^{-\alpha (T-t_{b})})\Delta\theta_{b}$$
%\todo{Add the definition of $\pathm(\textrm{root},i)$}

Let $\beta_0 = y_0e^{-\alpha T} + (1-e^{-\alpha T})\theta_0$ and $\beta_b = (1-e^{-\alpha (T-t_{b})})\Delta\theta_{b}$. Then,
\begin{equation}
\label{eqn:mean}
    E(Y_i) = \beta_0+\sum_b\beta_b \mathbf{(X_b)}_{i}
\end{equation}
where $\mathbf{X_b}$ is a vector defined by $\mathbf{(X_b)}_{i}=0$ if taxon $i$ is not under branch $b$, and $\mathbf{(X_b)}_{i}=1$ if the taxon $i$ is under branch $b$, and $t_b$ denotes the time from the root to the beginning of branch $b$.

While shifts in the optimal value ($\theta$) affect only the expectation of each trait $Y_i$, shifts in variance ($\sigma^2$) affect both the variance of $Y_i$ and the covariance between traits $Y_i$ and $Y_j$. This makes the location and magnitude of variance shifts potentially more identifiable from the data than shifts in optima. Suppose species $i$ and $j$ diverged at time $t_a$, and along their shared ancestry there are variance shifts at times $t_1, \ldots, t_m$ with magnitudes $\Delta\sigma_1^2, \ldots, \Delta\sigma_m^2$. Then their covariance $\mathbf{\Sigma}_{ij}$ is given by:
\begin{align}
\mathbf{\Sigma}_{ij} = \frac{e^{-2\alpha (T - t_a)}}{2\alpha} \left( (1 - e^{-2\alpha t_a}) \sigma_0^2 + \sum_{l = 1}^{m} \left(1 - e^{-2\alpha (t_a - t_l)}\right) \Delta\sigma_l^2 \right) \nonumber
\end{align}

This expression shows that each variance shift contributes to the total covariance based on its position along the evolutionary path. Unlike shifts in optima, whose location along a branch is not identifiable~\citep{ho_ane_2014}, the position of a single variance shift along a branch \emph{is} identifiable when several descendant species are available, because their pairwise covariances respond differently depending on when the shift occurred. However, when two variance shifts occur on the same branch their joint contribution to the covariance reduces to a sum that can be reproduced by a single equivalent shift, so two shifts on the same branch are not identifiable. The full derivations for both cases are given in the Supplementary Materials section \emph{Identifiability of variance shifts along a branch}.

For model and computational simplicity, we assume that $\sigma^2$ is constant over a branch like the assumption for $\theta$. That is, we assume that any shift in variance on a branch occurs at the beginning of that branch. For shifts in mean, this assumption did not limit the space of possible models due to the unidentifiability. However, for shifts in variance, this does restrict the space of models for the trait values at the leaves, and could therefore adversely affect shift detection. We conducted simulation studies to assess the impact of these assumption violations and found that they generally do not affect detection performance, except in the case of two opposing shifts on the same branch—an edge case where the assumptions matter more. Overall, these assumptions are not a major concern. The corresponding simulation results are presented in the Supplementary Materials.

Under our assumptions, we let $\sigma^2_b$ denote the variance on the branch $b$, $\Delta\sigma^2_b$ denote the magnitude of shift in variance on the branch $b$: if no shift occurs on the branch $b$, $\Delta\sigma^2_b=0$. The covariance between species $i$ and $j$ is given by:

\begin{align}
\label{eqn:cov}
    \mathbf{\Sigma}_{ij}  %&= e^{-2\alpha (T-t_a)}\sum_{b \in \pathm(\textrm{root},a)} \int_{t_b}^{t_{\nodeend{(b)}}} \sigma^2_b e^{-2\alpha(t_a-s)}ds \\
 %   &=e^{-\alpha (T-t_a)}\sum_{b \in \pathm(\textrm{root},i)}\frac{\sigma^2_b e^{-2\alpha t_a}}{2\alpha}(e^{2\alpha t_{\nodeend{(b)}}}-e^{2\alpha t_{b}}) \\
   % &=\frac{e^{-2\alpha (T-t_a)}}{2\alpha}\left(\sum_{b \in \pathm(\textrm{root},a)}\sigma_0^2(e^{-2\alpha \left(t_a-t_{\nodeend{(b)}}\right)}-e^{-2\alpha \left(t_a-t_{b}\right)})\\&+\sum_{b \in \pathm(\textrm{root},a)} \sum_{b \in \pathm(\textrm{root},b)}\left(e^{-2\alpha \left(t_a -t_{\nodeend(b)}\right)}-e^{-2\alpha \left(t_a-t_b\right)}\right)\Delta\sigma^2_{b}\right) \\
    %&= \frac{e^{-2\alpha (T-t_a)}}{2\alpha}\left((1-e^{-2\alpha t_a})\sigma_0^2+ \sum_{b \in \pathm(\textrm{root},a)} \sum_{b \in \pathm(b,a)}\left(e^{-2\alpha \left(t_a-t_{\nodeend(b)}\right)}-e^{-2\alpha \left(t_a-t_b\right)}\right)\Delta\sigma_{b}^2\right) \\
    &= \frac{e^{-2\alpha (T-t_a)}}{2\alpha}\left((1-e^{-2\alpha t_a})\sigma_0^2+ \sum_{b \in \pathm(\textrm{root},a)}(1-e^{-2\alpha \left(t_a-t_{b}\right)})\Delta\sigma^2_{b}\right) \nonumber \\
    &= \frac{e^{-2\alpha (T-t_a)}}{2\alpha}\left((1-e^{-2\alpha t_a})\sigma_0^2+ \sum_{b \in \pathm(\textrm{root},a)}\left(1-e^{-2\alpha t_a}+e^{-2\alpha t_a}-e^{-2\alpha \left(t_a-t_{b}\right)}\right)\Delta\sigma^2_{b}\right) \nonumber \\
    &=\frac{e^{-2\alpha (T-t_a)}}{2\alpha}\left((1-e^{-2\alpha t_a})\sigma_0^2+ \sum_{b \in \pathm(\textrm{root},a)}\left((1-e^{-2\alpha t_a})+e^{-2\alpha t_a}(1-e^{2\alpha t_{b}})\right)\Delta\sigma^2_{b}\right) \nonumber \\
    &= \frac{e^{-2\alpha (T-t_a)}}{2\alpha}\left(1-e^{-2\alpha t_a}\right)\sigma_0^2+\sum_{b \in \pathm(\textrm{root},a)}\frac{e^{-2\alpha (T-t_a)}}{2\alpha}(1-e^{-2\alpha t_a})\Delta\sigma^2_{b} \nonumber \\
    &\qquad+\sum_{b \in \pathm(\textrm{root},a)}\frac{e^{-2\alpha T}}{2\alpha}\left(1-e^{2\alpha t_{b}}\right)\Delta\sigma^2_{b}.
\end{align}

For the ancestral state at the root node, two different assumptions are commonly used, fixed value or stationary distribution. We assume that the ancestral state at the root is a fixed value for the above process. Equation~\eqref{eqn:cov} is the covariance of node $i$ and $j$ for the OU model with fixed root. %\todo{explain the terms fixed root and random root}
For the OU model, the ancestral state at the root is also often assumed to have the stationary distribution.
%\textcolor{red}{(this sentence is not complete)}. 
In this case, the variance of the root node is $\sigma_0^2/(2\alpha)$ (\citealp*{ho_ane_2013}). In this case, the covariance with shifts can be written as:
%\begin{equation}
    \begin{align} 
    \mathbf{\Sigma}_{ij}  &= \frac{e^{-2\alpha (T-t_a)}}{2\alpha}\left(1-e^{-2\alpha t_a}\right)\sigma_0^2+\sum_{b \in \pathm(\textrm{root},a)}\frac{e^{-2\alpha (T-t_a)}}{2\alpha}(1-e^{-2\alpha t_a})\Delta\sigma^2_{b} \nonumber \\ 
    &\qquad +\sum_{b \in \pathm(\textrm{root},a)}\frac{e^{-2\alpha T}}{2\alpha}\left(1-e^{2\alpha t_{b}}\right)\Delta\sigma^2_{b} + \frac{e^{-2\alpha T}}{2\alpha}\sigma_0^2 \nonumber \\
    & = \frac{e^{-2\alpha (T-t_a)}}{2\alpha}\sigma_0^2+\sum_{b \in \pathm(\textrm{root},a)}\frac{e^{-2\alpha (T-t_a)}}{2\alpha}(1-e^{-2\alpha t_a})\Delta\sigma^2_{b}\nonumber\\
    &\qquad +\sum_{b \in \pathm(\textrm{root},a)}\frac{e^{-2\alpha T}}{2\alpha}\left(1-e^{2\alpha t_{b}}\right)\Delta\sigma^2_{b} \nonumber
    \end{align} 
%\end{equation}

Let ${\gamma}_b =  \Delta\sigma_{b}^2$, $\mathbf{V}_{ij}=\frac{e^{-2\alpha (T-t_a)}}{2\alpha}(1-e^{-2\alpha t_a})$, $\mathbf{U}_{ij}=\frac{e^{-2\alpha (T-t_a)}}{2\alpha}$, $q_b = \frac{e^{-2\alpha T}}{2\alpha}\left(1-e^{2\alpha t_{b}}\right) $; ${\gamma}_b$ is the shift in variance on branch $b$, $\mathbf{V}$ is the phylogenetic covariance matrix when $\sigma^2 = 1$ (no shift in variance) with fixed root, $\mathbf{U}$ is the phylogenetic covariance matrix when $\sigma^2 = 1$ (no shift in variance) with stationary distributed root. $\mathbf{U}$ and $\mathbf{V}$ only depend on the phylogeny and $\alpha$. The covariance between species $i$ and $j$ can be expressed as the following equation.
\begin{align}     \mathbf{\Sigma}_{ij}  =  \begin{cases}
     \mathbf{V}_{ij}\sigma_0^2 + \sum_b {\gamma}_b \mathbf{(X_b)}_{i} \mathbf{(X_b)}_{j}\mathbf{V}_{ij} + \sum_b {\gamma}_bq_b \mathbf{(X_b)}_{i} \mathbf{(X_b)}_{j} & \text{OU model with fixed root} \\
      \mathbf{U}_{ij}\sigma_0^2 + \sum_b {\gamma}_b \mathbf{(X_b)}_{i} \mathbf{(X_b)}_{j}\mathbf{V}_{ij} + \sum_b {\gamma}_bq_b \mathbf{(X_b)}_{i} \mathbf{(X_b)}_{j} & \text{OU model with random root} \nonumber
     \end{cases}
     \end{align}

It consists of 3 terms. The first term is the original covariance without any shift. The last 2 terms show the influence of shifts in variance. In addition to the shift sizes $\gamma_b$, the second term is influenced by the phylogenetic structure ($\mathbf{V}_{ij}$): species with higher original covariance experience a larger change in covariance for the same shift size. The third term is influenced by the start time of branch $b$ ($q_b$): earlier shifts lead to a larger change in covariance between two species.

For simplicity, we let $\mathbf{R}$ denote the phylogenetic covariance matrix when there is no shift in variance and $\sigma_0^2 = 1$. 
%\todo{$\mathbf{R}$ is the covariance matrix when there is no shift in variance?}
So $\mathbf{R = V}$ when the root is fixed and $\mathbf{R = U}$ when the trait value at the root follows the stationary distribution.

Using matrix notation, the covariance matrix can be expressed as:
\begin{align}
\label{eqn:covariance}
    \mathbf{\Sigma} &= \sigma_0^2\mathbf{R} + \left(\sum_{b=1}^p{\gamma_b}\mathbf{X_bX_b^T}\right)\odot \mathbf{V} + \sum_{b=1}^p{\gamma_b}q_b\mathbf{X_bX_b^T}\nonumber \\
    &= \sigma_0^2\mathbf{R} + \left(\mathbf{X} \diag(\boldsymbol{\gamma}) \mathbf{X^T}\right)\odot \mathbf{V}+ \mathbf{X} \diag(\boldsymbol{\gamma} \odot \mathbf{q}) \mathbf{X^T}
\end{align}

where $\odot$ denotes elementwise multiplication, and $\diag(\cdot)$ denotes a diagonal matrix with the entries of the vector inside the parentheses on its diagonal. The trait values at tips can be written as:
\[   \mathbf{Y} =\beta_0\mathbf{1}+\sum_b\beta_b \mathbf{X_b}+\mathbf{\epsilon} \]

Where $\epsilon$ follows a normal distribution with mean 0 and covariance
matrix $\Sigma$, given by Equation~\eqref{eqn:covariance}.
The main task is to select the branches that have $\beta_b \neq 0$ and $\gamma_b \neq 0$; and estimate the values of $\beta_b$ and $\gamma_b$. 

\bigskip
\section{Methods} \setcurrentname{Methods}\label{sec3}
%There are some existing approaches to detect the shifts in optimal values. \citet{uyeda_harmon_2014} propose a Bayesian framework to detect shifts in the selective optimum of OU models. However, results of Bayesian approaches are deeply influenced by prior distributions and the computation cost is relatively high.  \citet{ho_ane_2014} illustrate the limitation of traditional model selection criteria (AIC, BIC) in the shift detection task and propose to use forward-backward selection with modified BIC (\citealp*{zhang_siegmund_2006}). \citet{khabbazian_kriebel_rohe_ane_2016} formulate the shift detection problem into a variable selection problem and combine the OU model with LASSO to detect the shift points, which they implement in the $\ell$1ou \texttt{R} package.  \citet{bastide_mariadassou_robin_2016} develop a maximum likelihood estimation procedure based on the EM algorithm (implemented in the \texttt{R} package PhylogeneticEM). 

In this section, we propose a new method to simultaneously detect the shifts in both variance and optimal values based on minimizing the loss function with L1 penalty. When $Y$ follows a multivariate normal distribution with mean and variance given by Equations~\eqref{eqn:mean} and~\eqref{eqn:covariance}, the log likelihood is
\begin{align}
    l(\beta,\gamma) &=  -\frac{1}{2}(\mathbf{Y}-\beta_0-\mathbf{X}\mathbf{\beta})^T\mathbf{\Sigma}^{-1}(\mathbf{Y}-\beta_0-\mathbf{X}\mathbf{\beta})-\frac{1}{2}\log{\det{(\mathbf{\Sigma})}} \nonumber \\
    &= -\frac{1}{2}(\mathbf{Y}-\beta_0-\mathbf{X}\mathbf{\beta})^T\left(\sigma_0^2\mathbf{R} + \left(\mathbf{X} \diag(\boldsymbol{\gamma}) \mathbf{X^T}\right)\odot \mathbf{V} + \mathbf{X} \diag(\boldsymbol{\gamma} \odot \mathbf{q}) \mathbf{X^T}\right)^{-1}(\mathbf{Y}-\beta_0-\mathbf{X}\mathbf{\beta}) \nonumber \\ &\qquad-\frac{1}{2}\log{\det{\left(\sigma_0^2\mathbf{R} + \left(\mathbf{X} \diag(\boldsymbol{\gamma}) \mathbf{X^T}\right)\odot \mathbf{V} + \mathbf{X} \diag(\boldsymbol{\gamma}\odot \mathbf{q}) \mathbf{X^T}\right)}} \nonumber  
\end{align}

To select the shifts in optimal values ($\beta_i\neq 0$) and the shifts in variance ($\gamma_i\neq 0 $), we use an L1 penalty in the loss function to conduct the feature selection as in LASSO. Therefore, the loss function to be optimized is given by:
\begin{align}
\label{eqn:loss}
    L(\beta_1,...,\beta_p,\gamma_1,...,\gamma_p)     &= l(\beta_1,...,\beta_p,\gamma_1,...,\gamma_p;\textbf{Y}) + \text{Penalty} \nonumber \\
    &= \frac{1}{2}(\mathbf{Y}-\beta_0-\mathbf{X}\mathbf{\beta})^T\left(\sigma_0^2\mathbf{R} + \left(\sum_{i=1}^p\gamma_i\mathbf{X_iX_i^T}\right)\odot \mathbf{V} + \sum_{i=1}^p\gamma_iq_i\mathbf{X_iX_i^T}\right)^{-1} \nonumber \\&\qquad(\mathbf{Y}-\beta_0-\mathbf{X}\mathbf{\beta})+\frac{1}{2}\log{\det{\left(\sigma_0^2\mathbf{R} + \left(\sum_{i=1}^p\gamma_i\mathbf{X_iX_i^T}\right)\odot \mathbf{V} + \sum_{i=1}^p\gamma_iq_i\mathbf{X_iX_i^T}\right)}} \nonumber \\&\qquad+\lambda_1\|\boldsymbol{\beta}\|_1+\lambda_2\|\boldsymbol{\gamma}\|_1
\end{align}
%\todo{link this loss function with the log likelihood function, aka, L = loglik + penalty}

\subsection{Optimization}
In this paper, we do not estimate $\alpha$ jointly with the shift parameters, but instead fix it at an ad-hoc estimate obtained from the null model fitted by the \texttt{phylolm} R package. This is because $\alpha$ enters the covariance matrix nonlinearly and interacts with all shift coefficients. Joint estimation with the LASSO problem is therefore ill-conditioned in high-shift regimes. In our simulations, we demonstrate that the estimation of $\alpha$ does not significantly affect the detection results. The primary objective here is to find the parameters $\boldsymbol{\beta}$, $\boldsymbol{\gamma}$, and $\sigma_0^2$ that minimize the loss function described in Equation~\eqref{eqn:loss}. When $\boldsymbol{\gamma}$ is fixed, the problem reduces to a standard LASSO formulation, which can be efficiently solved.

For optimizing $\boldsymbol{\gamma}$, we employ coordinate-wise proximal gradient descent, as described by \cite{parikh_boyd_2014}. The proximal gradient algorithm is a powerful tool for handling non-differentiable  optimization problems. It decomposes the objective function into two components: a differentiable part and a non-differentiable part, allowing for iterative updates that combine gradient descent with the proximal operator. The proximal operator serves to handle the non-differentiable component, promoting sparsity and effectively reducing the influence of irregularities in the model. By applying the proximal gradient algorithm in our optimization process, we can efficiently minimize the objective function even when the problem involves non-smooth terms, such as LASSO regularization. 

The loss function for $\gamma_k$ can be written as:
$$L(\gamma_k) = g(\gamma_k) + \lambda_2\|\gamma_k\|_1.$$

The derivative of $g(\gamma_k)$ is given by:
\begin{align}
\nabla g(\gamma_k) &= -\frac{1}{2}\mathbf{r}^T\mathbf{\Sigma^{-1}}\left(\mathbf{\left(X_kX_k^T\right)} \odot \mathbf{V}\right)\mathbf{\Sigma^{-1}}\mathbf{r}-\frac{1}{2} q_k\|\mathbf{X_k^{T}\Sigma^{-1}r}\|^2+\frac{1}{2}\tr\left(\left(\mathbf{\left(X_kX_k^T\right)} \odot \mathbf{V}\right)\mathbf{\Sigma}^{-1}\right) \nonumber \\ &\qquad+\frac{1}{2}q_k\mathbf{X_k^T\Sigma^{-1} X_k}, \nonumber
\end{align}
where $\mathbf{r} = \mathbf{Y}-\beta_0-\mathbf{X^T}\mathbf{\beta}$. The proximal operator is given by:
\[
\text{prox}_{\lambda_2 h}(z) = S_{\lambda_2}(z) = \begin{cases}z-\lambda_2 & z>\lambda_2, \\0 & -\lambda_2 \leq z \leq\lambda_2, \\z+\lambda_2 & z<-\lambda_2. \end{cases}
\]
The proximal algorithm here steps in the direction $\nabla g(\gamma_k)$ but setting any values that are close to zero equal to zero.
For $\sigma_0^2$, there is no penalty on the parameter. Therefore, we use gradient descent to update $\sigma_0^2$. To avoid negative values of $\sigma_0^2$, we use $\tau_0 = \log(\sigma_0^2)$ for optimization. The gradient of $\tau_0$ is given by:
$$\begin{aligned}
    \frac{\partial L}{\partial \tau} &= \left(-\frac{1}{2}\mathbf{r^T}\mathbf{\Sigma}^{-1}\mathbf{R}\mathbf{\Sigma}^{-1}\mathbf{r}+\frac{1}{2}\tr\left(\mathbf{R}\mathbf{\Sigma}^{-1}\right)\right)e^{\tau_0}.
    \end{aligned}$$

We initialize $\boldsymbol{\beta}$ and $\boldsymbol{\gamma}$ as $\mathbf{0}$. In each iteration, we first update the covariance matrix $\mathbf{\Sigma}$ and transform $\mathbf{X}$ and $\mathbf{Y}$ with the current $\mathbf{\Sigma}^{-1/2}$. Then we update $\boldsymbol{\beta}$ by applying LASSO on the transformed data. Then we update $\boldsymbol{\gamma}$ using the proximal algorithm in an elementwise manner. After that, we update $\tau_0^2$ using gradient descent. The algorithm is summarized in Algorithm~\ref{alg:opti}.

\begin{algorithm}[htbp]
  \caption{Optimization with $\boldsymbol{\beta}$, $\boldsymbol{\gamma}$ and $\tau_0$}
  \begin{algorithmic}[1]
   %\scriptsize
    \Inputs{$M$: maximum number of steps; $t$: step size; $\epsilon$:
      error tolerance; $\mathbf{V}$; $\mathbf{R}$; $\mathbf{Y}$; $\mathbf{X}$; $\lambda_2$}
    \Initialize{\strut$\beta_i^0 \gets 0$, $i=1,\ldots,p$ \\ \strut$\gamma_i^0 \gets 0$, $i=1,\ldots,p$ \\
    \strut$\tau_0 \gets 0$ \\
    \strut$L \gets \infty$}
    \For{s = 1 to M}
      \State $\mathbf{\Sigma} \gets e^{(\tau_0)}\mathbf{R} + \left(\sum_{i=1}^p\gamma_i\mathbf{X_iX_i^T}\right)\odot \mathbf{V} + \sum_{i=1}^p\gamma_iq_i\mathbf{X_iX_i^T}$
      \State $\mathbf{Y'} = \mathbf{\Sigma^{-1/2}Y}$; $\mathbf{X'} = \mathbf{\Sigma^{-1/2}X}$ 
      \State $\beta \gets \mathrm{LASSO}(\mathbf{Y'},\mathbf{X'})$
      \For{k = 1 to p}
        \State calculate $\nabla g(\gamma_k)$
        \State update $\gamma_k \gets S_{\lambda_2t}(\gamma_k-t\nabla  g(\gamma_k))$
        \State update $\mathbf{\Sigma}$
          \EndFor
       \State calculate the gradient for $\tau_0$: $\frac{\partial L}{\partial \tau_0}$
       \State update $\tau_0 \gets \tau_0-t\frac{\partial L}{\partial \tau_0}$
        \State update the loss function $L$ with Equation~\eqref{eqn:loss}
        \If{update for the loss function $L$ in the iteration $< \epsilon$}
        \State break
        \EndIf
    \EndFor
  \end{algorithmic}
\label{alg:opti}
\end{algorithm}

\subsection{Model selection}
For the model selection process, we use a strategy to fine-tune the parameters $\lambda_1$ and $\lambda_2$ to strike an optimal balance between model simplicity and performance.

\begin{itemize}
    \item \textrm{Fixing $\lambda_2$}: We begin by setting $\lambda_2$ to a fixed value, which controls the degree of regularization for variance shifts. With $\lambda_2$ fixed, we focus on adjusting $\lambda_1$, which regulates the penalization of shifts in the optimal values.
    \item \textrm{Cross-validation for $\lambda_1$}: For each fixed value of $\lambda_2$, we use cross-validation to find the best $\lambda_1$. We first de-correlate the data with the estimated covariance matrix. We then use the \texttt{cv.glmnet} function from the R package \texttt{glmnet} to efficiently perform cross-validation here. We use cross-validation because \texttt{cv.glmnet} allows for quick and easy computation of cross-validation results, making it more convenient and efficient for selecting the optimal parameters. 
    \item \textrm{Iterating over different $\lambda_2$ values}: After optimizing $\lambda_1$ for one fixed $\lambda_2$, we repeat this process for multiple values of $\lambda_2$, producing a set of candidate models, each associated with different pairs of $\lambda_1$ and $\lambda_2$.
    \item \textrm{Selecting the best model using BIC}: To identify the optimal model, we compare all candidate models across the range of $\lambda_2$ values by using the Bayesian Information Criterion (BIC). The model with the lowest BIC is selected as the final model, as it achieves the best trade-off between goodness of fit and model complexity, minimizing the risk of overfitting.
\end{itemize} 

We choose BIC over other criteria based on empirical findings in \citet{zhang_kenney_ho_2024}, where pBIC was found to be a more conservative criterion. While a conservative selection method can be beneficial in some cases, detecting shifts in variance is inherently more challenging than detecting shifts in optimal values. A less conservative criterion like BIC is preferable here, as it increases sensitivity to shifts in variance while still penalizing overly complex models.

\bigskip
\section{Simulations}
\subsection{Comparison of performance}\label{sec:performance}
We generated simulation data under an Ornstein–Uhlenbeck (OU) model with a fixed root at a value $y_0 = \theta_0 = 0$, using the Anolis lizard phylogeny from \citet*{mahler_ingram_revell_losos_2013}. This is an ultrametric tree and we rescaled the tree height to 1. The adaptation rate parameter was set to $\alpha = 1$, which corresponds to a phylogenetic half-life of approximately $t_{1/2} = \log(2)/\alpha \approx 0.693$. The diffusion variance was set to $\sigma_0^2 = 2$. 
Each parameter setting was repeated across 50 simulation replicates. To assess the robustness of our conclusions, we also performed additional simulations with different values of $\alpha$, which are presented in the supplementary materials and yield results consistent with the main findings. In the following simulations, $\alpha$ is assumed to be unknown. For ShiVa, we estimate $\alpha$ by fitting a null model (without shifts) using the \texttt{phylolm} package.

We considered three simulation scenarios:
(1) a single shift in the optimal value, placed on branch~71 of the lizard phylogeny,
(2) a single shift in the variance, placed on branch~195, and
(3) one shift in the optimal value (branch~71) combined with one shift in the variance (branch~195).
Branch indices follow the edge order of \texttt{lizard.tree} in the \texttt{l1ou} package.

For the optimal value shift, we varied the shift size across the values $-5$, $-3$, $-1$, $1$, $3$, and $5$.  
For the variance shift scenarios, we varied the shift magnitude $\gamma_b$ across the values $-1.5$, $-1$, $1$, $1.5$, $2$, $3$, $5$, and $7$. 
In the combined scenario, we first fixed the variance shift at $\gamma_b = 5$ and varied the optimal value shift as in Scenario (1), and then fixed the optimal value shift at $5$ and varied the variance shift as in Scenario (2). For the setting of $\gamma_b$, the largest magnitude considered was 7, which corresponds to a variance that is 3.5 times greater than the original value, given that the baseline $\sigma^2$ is 2. In the case study section, we observe examples where the estimated variance shift reaches up to 7 times the original value. Therefore, we believe that the simulated shift magnitudes fall within a realistic and meaningful range. Traitgrams of simulated data across the shift sizes used here are provided in Supplementary Figure~\ref{fig:traitgram-supp} for visual inspection.

\begin{figure}[htbp]

    \centering
    % First subplot: 1 shift in optimal value
    \begin{subfigure}[b]{0.9\textwidth}
        \centering
        \begin{tikzpicture}
            \node[anchor=south west, inner sep=0] (img) {\includegraphics[width=\textwidth,height=0.21\textheight,keepaspectratio]{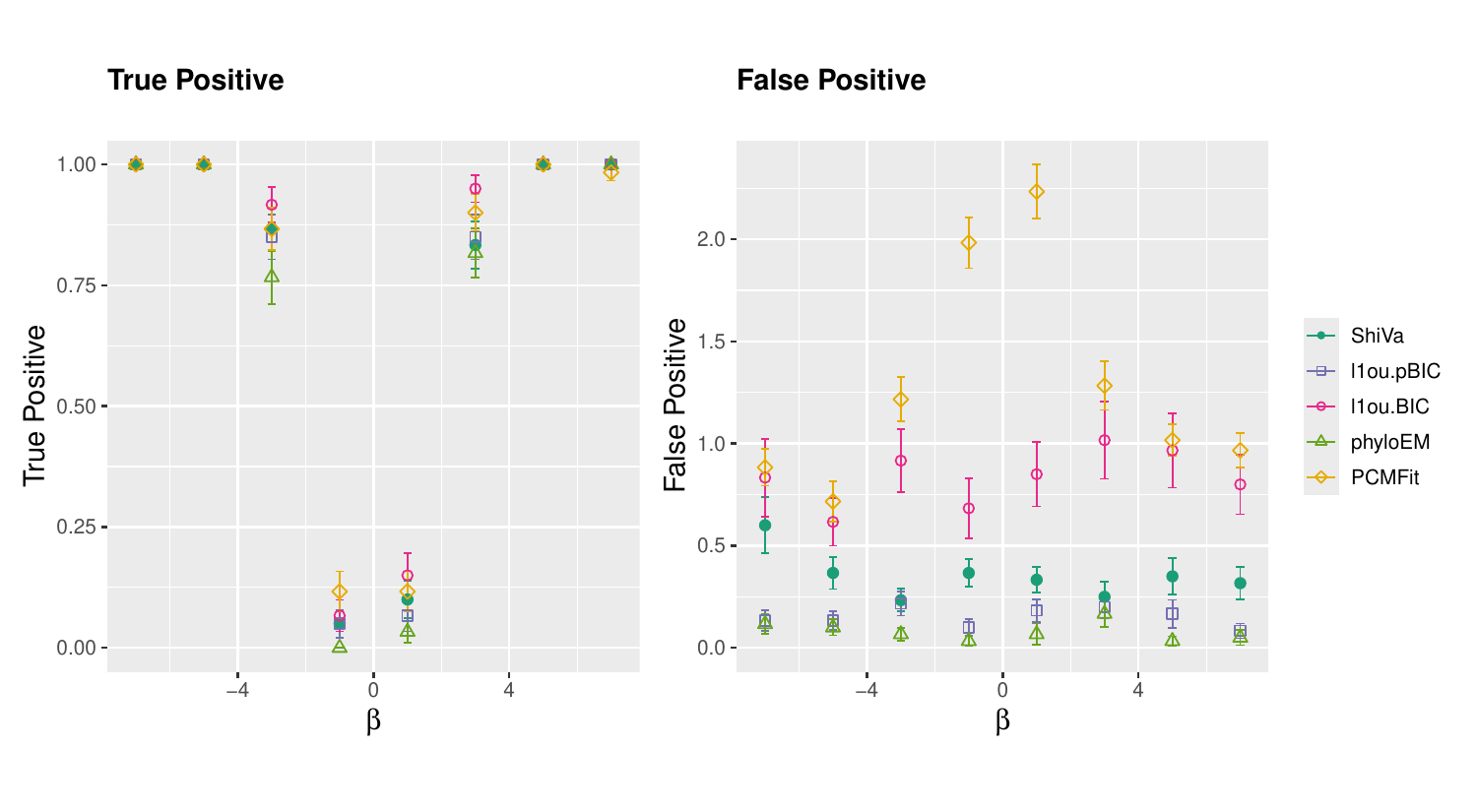}};
            \node[anchor=north west, inner sep=1pt, xshift=-10pt, font=\footnotesize] at (img.north west) {a)};
        \end{tikzpicture}
        \label{fig:2-shift-optimal-value}
    \end{subfigure}

    % Second subplot: 1 shift in variance
    \begin{subfigure}[b]{0.9\textwidth}
        \centering
        \begin{tikzpicture}
            \node[anchor=south west, inner sep=0] (img) {\includegraphics[width=\textwidth,height=0.21\textheight,keepaspectratio]{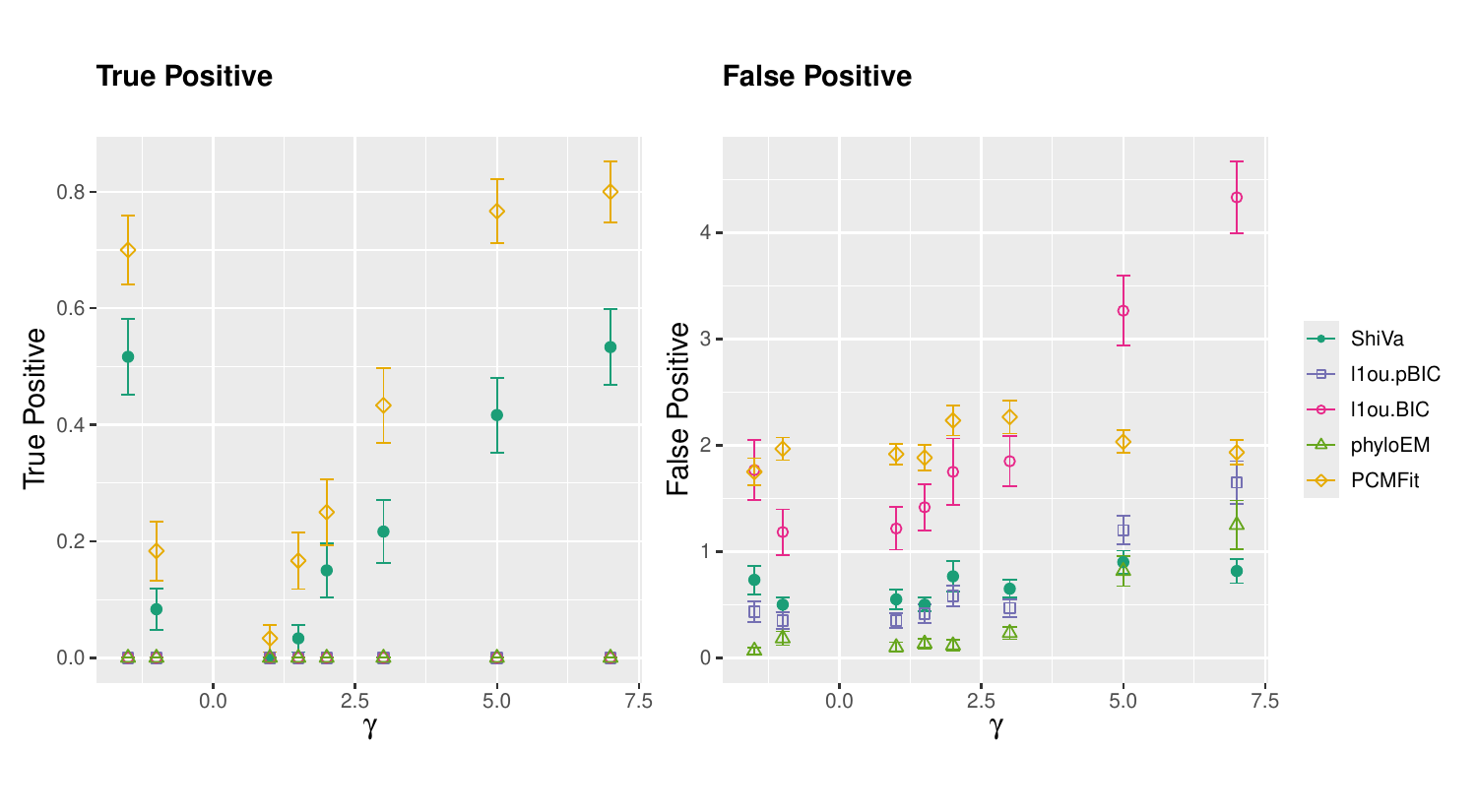}};
            \node[anchor=north west, inner sep=1pt, xshift=-10pt, font=\footnotesize] at (img.north west) {b)};
        \end{tikzpicture}
        \label{fig:2-shift-variance}
    \end{subfigure}

    % Third subplot: 1 shift in optimal value and 1 shift in variance
    \begin{subfigure}[b]{0.9\textwidth}
        \centering
        \begin{tikzpicture}
            \node[anchor=south west, inner sep=0] (img) {\includegraphics[width=\textwidth,height=0.21\textheight,keepaspectratio]{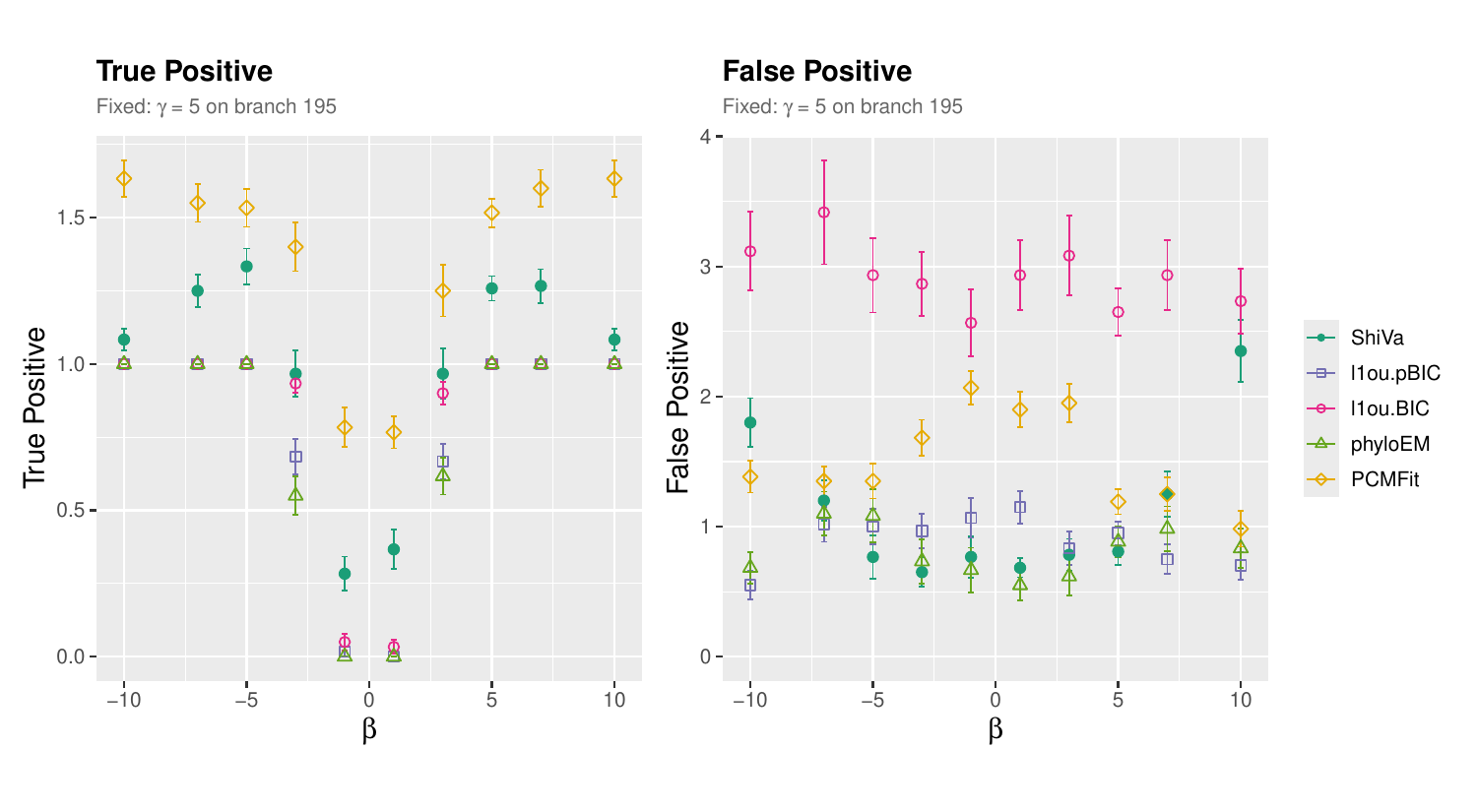}};
            \node[anchor=north west, inner sep=1pt, xshift=-10pt, font=\footnotesize] at (img.north west) {c)};
        \end{tikzpicture}
        \label{fig:2-shift-both-beta}
    \end{subfigure}

    \begin{subfigure}[b]{0.9\textwidth}
        \centering
        \begin{tikzpicture}
            \node[anchor=south west, inner sep=0] (img) {\includegraphics[width=\textwidth,height=0.21\textheight,keepaspectratio]{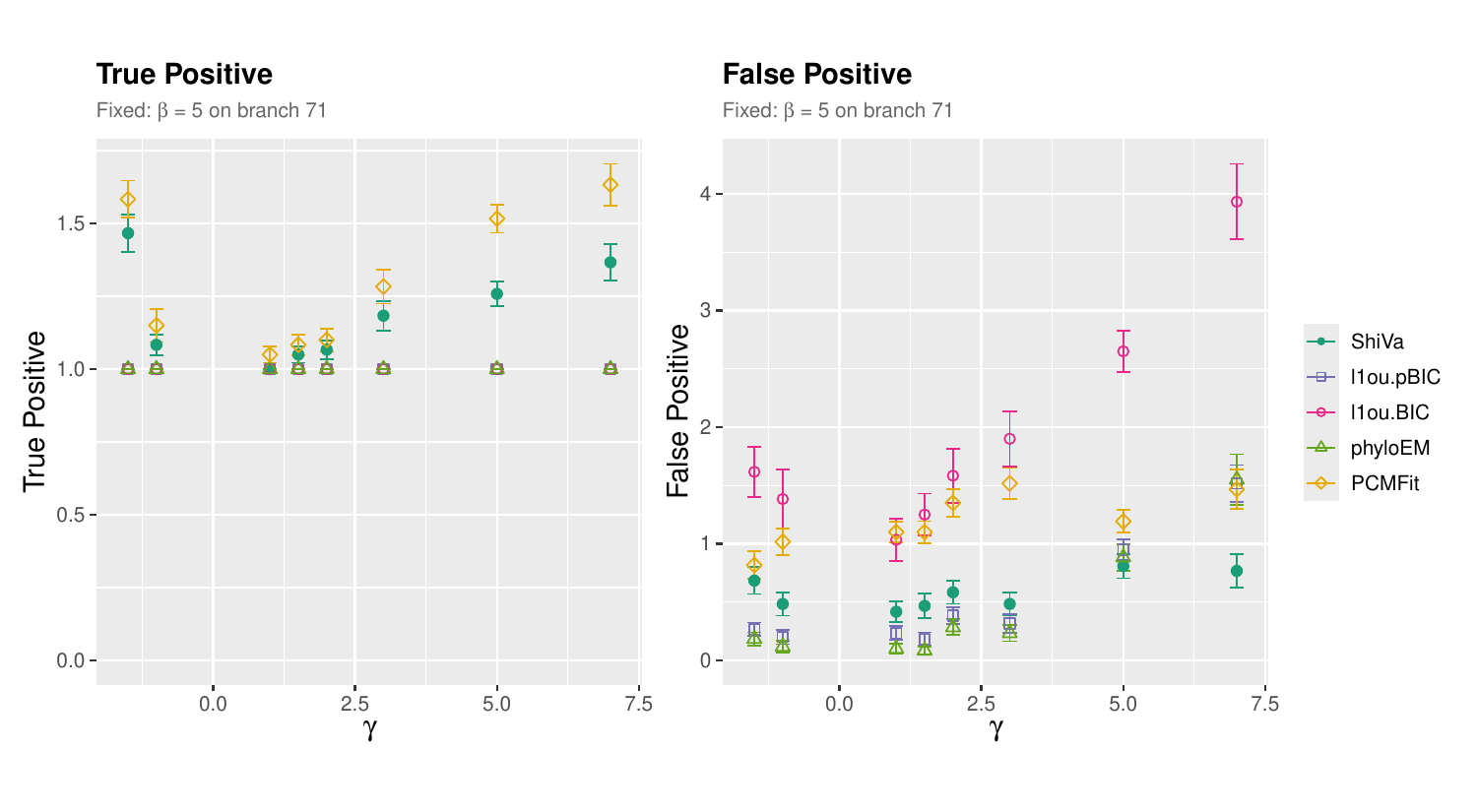}};
            \node[anchor=north west, inner sep=1pt, xshift=-10pt, font=\footnotesize] at (img.north west) {d)};
        \end{tikzpicture}
        \label{fig:2-shift-both-gamma}
    \end{subfigure}
    \caption{Comparison of True positives and False positives across five methods under four scenarios: a) 1 shift in optimal value; b) 1 shift in variance; c) 1 shift in optimal value + 1 shift in variance (fixed $\gamma$); d) 1 shift in optimal value + 1 shift in variance (fixed $\beta$). \textbf{Note:} In panel a) (left), when $\beta = -5$ or $\beta = 5$, all methods correctly detect the shift (True Positives = 1), resulting in overlapping points.}
    \label{fig:2-TP-FP}
\end{figure}

We compared the performance of different methods using five key metrics: True Positives (TP), False Positives (FP), mean integrated squared error (MISE) of $\sigma^2$ and $\mu$, predictive log-likelihood, and computational time. We count the number of correctly detected true shifts as True Positives (TP), and the number of detected shifts that do not correspond to any true shift as False Positives (FP). TP and FP are distinguished by shift type: a detection on the correct branch but with the wrong shift type (e.g., a variance shift detected where the truth is an optimal-value shift) counts as a false positive, not a true positive. The positions of shifts are not always identifiable: for example, under a 3-branch tree, any two shifts can lead to an identical model. In such cases, TP and FP may not be reliable metrics, as they would count the true model as including one false positive and missing one true positive, even though the models are effectively equivalent. However for the simulation studies presented here, the true shifts are identifiable.
Figure~\ref{fig:2-TP-FP} illustrates the TP and FP results across the various methods under different scenarios. When there is only a shift in the optimal value, ShiVa achieves a well-balanced trade-off between True Positives and False Positives. It effectively detects shifts in the optimal value without generating excessive False Positives. The performance of ShiVa is comparable to other methods that only consider shifts in the optimal value even if only a shift in optimal value is present. In contrast, PCMFit produces a much higher number of False Positives, particularly when the signal size is small. 

In the case of shifts in variance, detection is inherently harder than for shifts in optimal value: even at the largest shift magnitude considered ($\gamma_b = 7$), ShiVa's True Positive rate plateaus near $0.5$. Within this regime, however, ShiVa is the only method that detects a substantial fraction of the true variance shifts while maintaining tight control over False Positives. Although PCMFit shows a higher True Positive rate compared to ShiVa, it comes at the cost of generating significantly more False Positives. Methods like $\ell$1ou and phyloEM, which are designed to detect shifts in optimal value, face challenges when a shift in variance is present, since the variance signal can be misinterpreted as multiple shifts in the optimal value. This effect is most pronounced for $\ell$1ou, whose False Positives in the optimal value increase substantially as the variance shift grows; phyloEM behaves more conservatively, so its False Positives remain low or comparable to ShiVa's even at large variance shifts. In contrast, ShiVa successfully distinguishes between variance and mean shifts, effectively controlling the False Positive rate for optimal value shifts, even as the variance shift signal strengthens. We have more simulations with varying $\alpha$ values in the Supplementary Material and the conclusion remains similar. 

\begin{figure}[htbp]
  \centering
    \includegraphics[width=\textwidth]{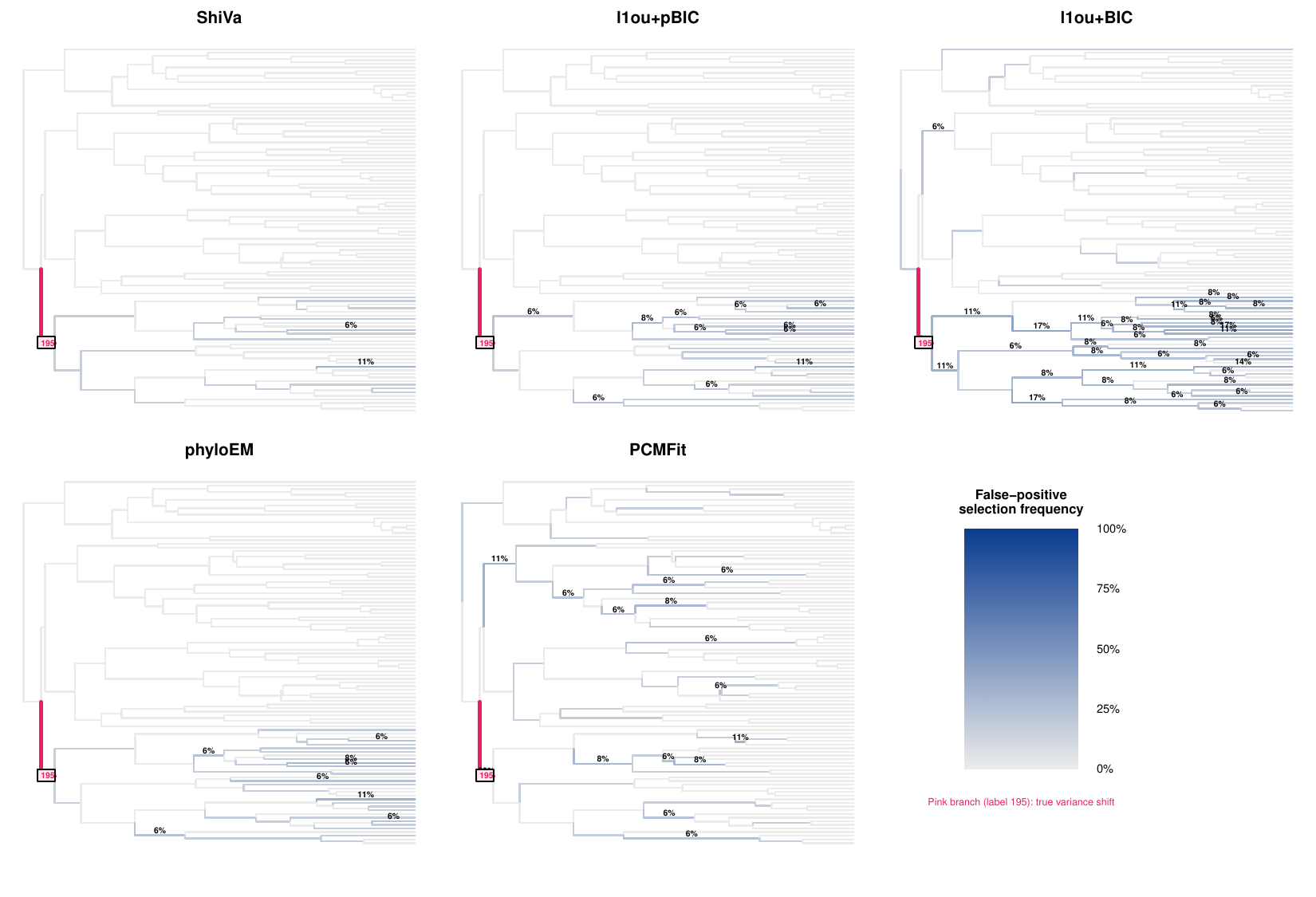}
     \caption{\label{FP-analysis} Detection frequencies of false positive shifts in the optimal value under a true variance shift. The pink box highlights the true variance shift on branch 195 (with shift magnitude $\gamma_b = 7$). Branch 195 refers to the edge label number of the shifted branch.}
\end{figure}

Figure~\ref{FP-analysis} shows the per-branch frequency at which each method falsely detects an optimal-value shift, when the only true shift is a variance shift on branch~195 with magnitude $\gamma_b = 7$. Branches whose false-positive frequency exceeds $5\%$ are labelled in the figure. ShiVa labels only $1.0\%$ of the $197$ non-target branches above the $5\%$ threshold, the smallest among all methods; phyloEM and $\ell$1ou+pBIC sit at $4.1\%$ and $5.6\%$ respectively, while $\ell$1ou+BIC labels $18.3\%$ of branches with non-trivial false-positive frequencies (peaks up to $17\%$). For these three methods, the false-positive branches are concentrated on descendants of branch~195, reflecting the way optimal-value-only methods spread the variance signal as multiple spurious mean shifts. PCMFit labels $7.1\%$ of branches above the $5\%$ threshold, but its false positives are scattered across the tree rather than concentrated near branch~195. Overall, ShiVa is the method that produces the fewest spurious optimal-value detections in the presence of a true variance shift.

When both shifts in optimal value and variance are present, ShiVa continues to perform competitively. Although it detects slightly fewer True Positives compared to PCMFit, its False Positive rate is much lower, indicating a better balance between sensitivity and specificity. Compared to $\ell$1ou and phyloEM, ShiVa carries a somewhat higher optimal-value False Positive rate at small to moderate $|\gamma|$; this is the cost of also modelling variance shifts, which $\ell$1ou and phyloEM are structurally unable to detect. PCMFit, while capable of identifying more shifts overall, tends to produce excessive False Positives. This leads to reduced reliability in shift detection, especially in noisy data.

While TP and FP indicate whether the estimated and true shift sets agree at the branch level, they do not reflect how close the estimated per-branch parameters are to the truth. Moreover, the OU model is identifiable when no shifts are present, but previous studies have shown that the exact number and locations of shifts along an edge cannot always be fully determined due to identifiability issues---for example, any two of the three branches connected to a node can yield equivalent models. Even if the selected shifts do not match the true model exactly, choosing a close surrogate shift may be preferable to missing the shift entirely. In such cases, the true positive versus false positive framework may misrepresent performance: a method selecting a surrogate shift is penalized with a false positive, even though it provides a more reasonable approximation than failing to detect the shift. We therefore also report the mean integrated squared error (MISE) of the estimated diffusion variance and optimal trait value along the tree. MISE is well-defined under these identifiability ambiguities, since models with different shift locations but the same effective per-branch parameter functions yield the same MISE. For a tree with branches $b = 1, \ldots, B$ and branch lengths $\ell_b$, let $\sigma^2_b$ denote the true effective diffusion variance on branch~$b$ (the piecewise-constant value obtained after applying the variance shifts to the ancestral $\sigma^2_0$) and $\hat{\sigma}^2_b$ its estimate; similarly let $\theta_b$ denote the true optimal trait value on branch~$b$ (the root value $\theta_0$ plus the sum of all upstream mean shifts) and $\hat{\theta}_b$ its estimate. We define
\[
\mathrm{MISE}(\sigma^2) = \sum_{b=1}^{B} \ell_b \,(\sigma^2_b - \hat{\sigma}^2_b)^{2}, \qquad
\mathrm{MISE}(\mu) = \sum_{b=1}^{B} \ell_b \,(\theta_b - \hat{\theta}_b)^{2}.
\]
These are branch-length-weighted $L^2$ distances between the true and estimated branch-wise parameter functions, so that errors on longer branches are penalised proportionally.

\begin{figure}[htbp]
    \centering
    \begin{subfigure}[b]{0.9\textwidth}
        \centering
        \begin{tikzpicture}
            \node[anchor=south west, inner sep=0] (img) {\includegraphics[width=\textwidth,height=0.21\textheight,keepaspectratio]{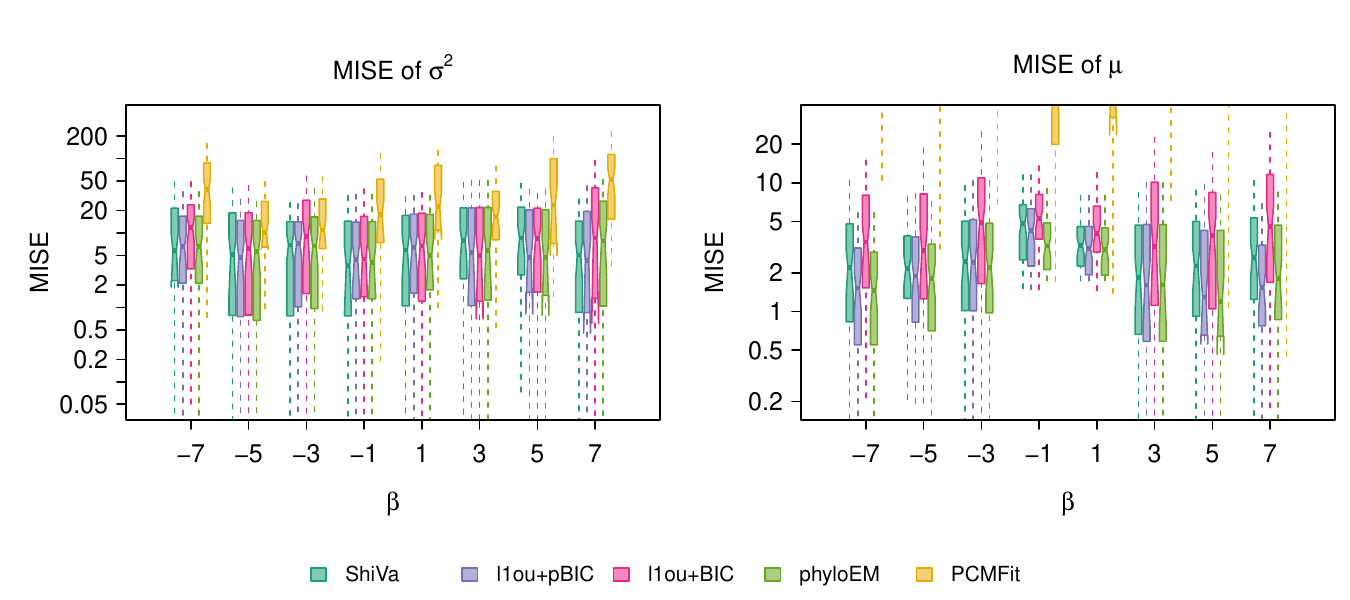}};
            \node[anchor=north west, inner sep=1pt, xshift=-10pt, font=\footnotesize] at (img.north west) {a)};
        \end{tikzpicture}
        \label{fig:mise-main-a}
    \end{subfigure}

    \begin{subfigure}[b]{0.9\textwidth}
        \centering
        \begin{tikzpicture}
            \node[anchor=south west, inner sep=0] (img) {\includegraphics[width=\textwidth,height=0.21\textheight,keepaspectratio]{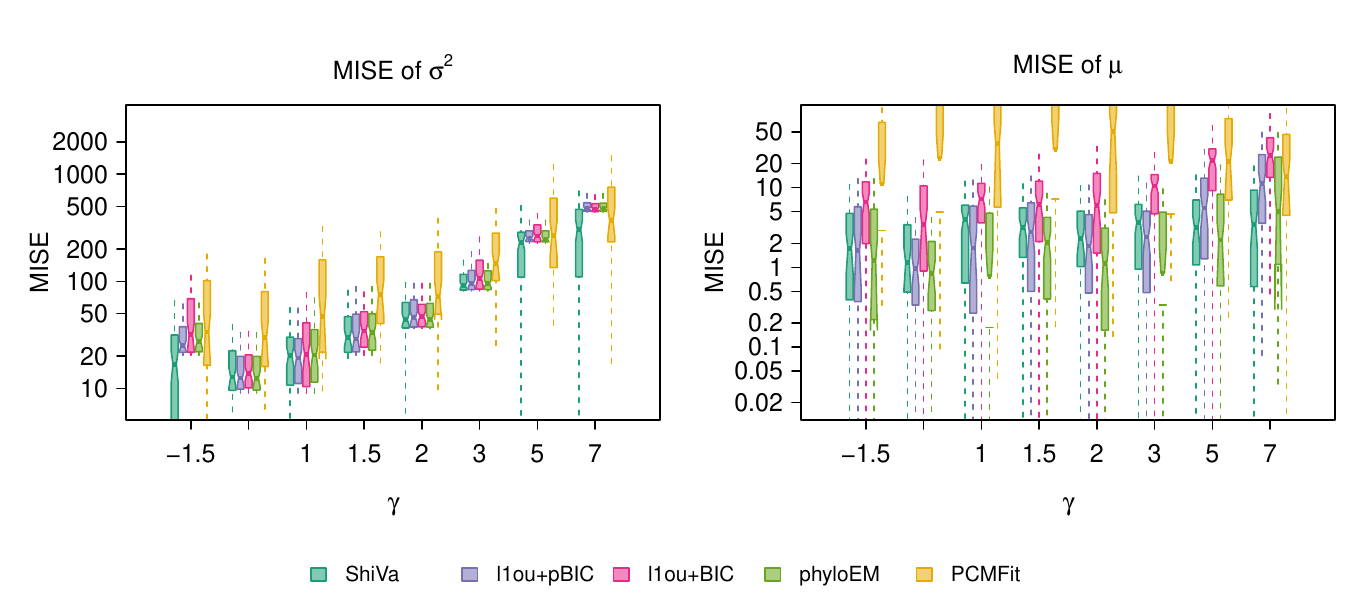}};
            \node[anchor=north west, inner sep=1pt, xshift=-10pt, font=\footnotesize] at (img.north west) {b)};
        \end{tikzpicture}
        \label{fig:mise-main-b}
    \end{subfigure}

    \begin{subfigure}[b]{0.9\textwidth}
        \centering
        \begin{tikzpicture}
            \node[anchor=south west, inner sep=0] (img) {\includegraphics[width=\textwidth,height=0.21\textheight,keepaspectratio]{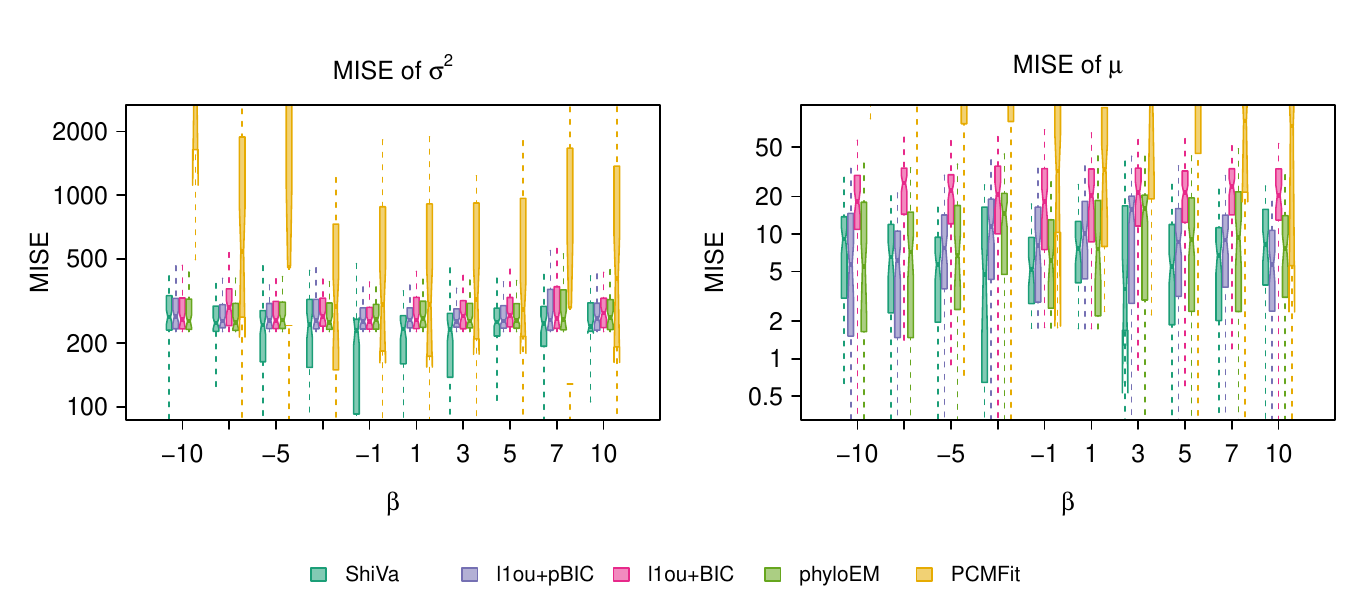}};
            \node[anchor=north west, inner sep=1pt, xshift=-10pt, font=\footnotesize] at (img.north west) {c)};
        \end{tikzpicture}
        \label{fig:mise-main-c}
    \end{subfigure}

    \begin{subfigure}[b]{0.9\textwidth}
        \centering
        \begin{tikzpicture}
            \node[anchor=south west, inner sep=0] (img) {\includegraphics[width=\textwidth,height=0.21\textheight,keepaspectratio]{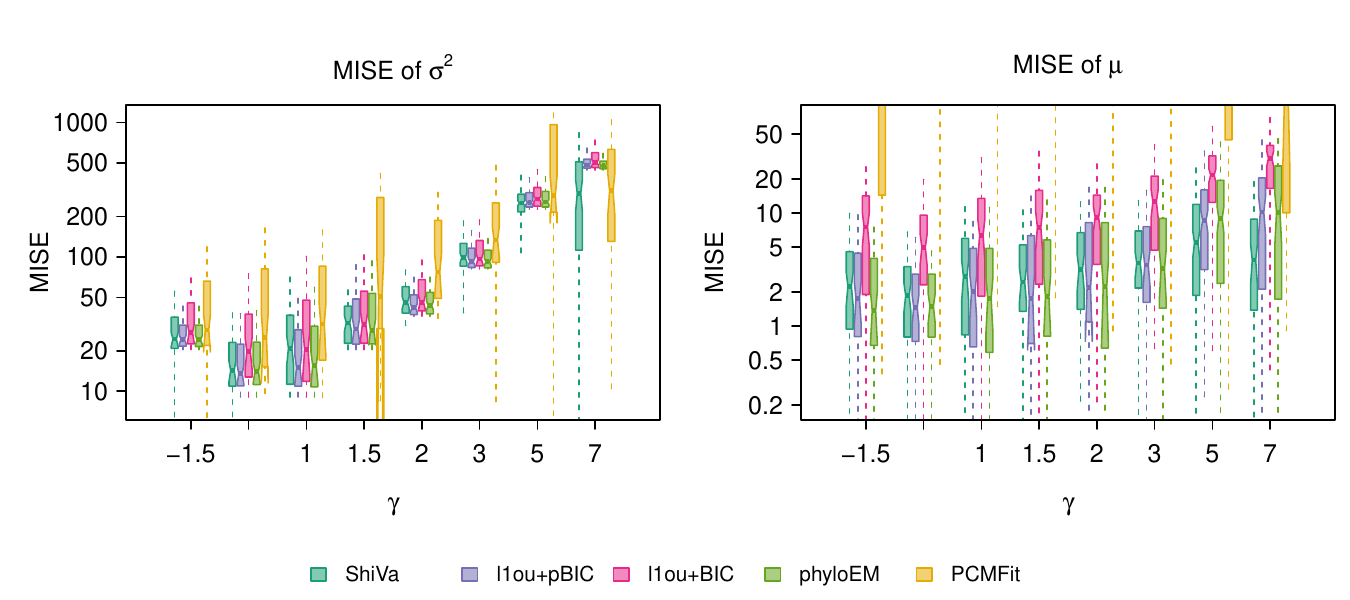}};
            \node[anchor=north west, inner sep=1pt, xshift=-10pt, font=\footnotesize] at (img.north west) {d)};
        \end{tikzpicture}
        \label{fig:mise-main-d}
    \end{subfigure}
    \caption{Per-replicate distributions of MISE of $\sigma^2$ (left in each panel) and MISE of $\mu$ (right in each panel) across the five methods, shown for: a) only mean shifts on branch~71, varying $\beta$; b) only variance shifts on branch~195, varying $\gamma$; c) both shifts, $\gamma=5$ fixed, varying $\beta$; d) both shifts, $\beta=5$ fixed, varying $\gamma$. Notched boxplots over the replicate datasets, with the $y$-axis on a log scale. The $y$-limits are chosen from the four shift-detection methods so that their boxplots remain readable; PCMFit values that exceed this range are clipped from view. Median values, including PCMFit, are tabulated in Supplementary Table~\ref{tab:mise-all}.}
    \label{fig:mise-main}
\end{figure}

Figure~\ref{fig:mise-main} summarises the per-replicate MISE distributions for the four configurations of Figure~\ref{fig:2-TP-FP}; each panel pairs MISE$(\sigma^2)$ on the left with MISE$(\mu)$ on the right, with the $y$-axis on a log scale. Median values in tabular form are given in Supplementary Table~\ref{tab:mise-all}. For MISE of $\sigma^2$, ShiVa attains the lowest median in panels~c)~and~d) when the variance-shift signal is moderate or large ($|\gamma|\ge 2$), and its advantage is largest in panel~c), where ShiVa is the only method that can model the variance shift directly while the others must absorb it into the global $\sigma^2$. In panel~a), where the true model contains only mean shifts, ShiVa, $\ell$1ou+pBIC, $\ell$1ou+BIC and phyloEM produce comparable values, since none of them face a variance-shift signal and all four methods estimate a single global~$\sigma^2$.

For MISE of $\mu$, phyloEM and $\ell$1ou+pBIC achieve the lowest values in most rows of panels~a),~b)~and~d), with ShiVa typically within a factor of two. ShiVa overtakes the others in panel~c), where the strong variance-shift signal causes $\ell$1ou+pBIC and phyloEM to allocate spurious optimal-value shifts and inflates their MISE$(\mu)$. $\ell$1ou+BIC has a consistently larger MISE$(\mu)$ than $\ell$1ou+pBIC because its looser penalty admits further spurious shifts in the optimal value.

PCMFit's MISE values are one to three orders of magnitude larger than those of the other methods in many configurations, so its boxes occasionally extend beyond the plotted $y$-range; the $y$-limits are set from the four shift-detection methods to keep their boxplots readable, and the exact PCMFit values are reported in Supplementary Table~\ref{tab:mise-all}. PCMFit's mixed-model framework allows different OU/BM models on different clades, and we found that with strong mean-shift signals it frequently selects regimes whose optima are far from the truth on the wrong subset of branches. The resulting per-branch error in~$\theta$ accumulates over most of the tree, producing the very large MISE$(\mu)$ values reported in the supplementary table.

Taken together, ShiVa offers the best variance-recovery accuracy when variance shifts are actually present. Importantly, ShiVa's joint modelling also benefits researchers whose primary interest is in shifts of the optimal value: when variance heterogeneity is present in the data, methods that ignore it ($\ell$1ou, phyloEM) tend to mis-attribute the variance signal to spurious optimal-value shifts, producing inflated MISE$(\mu)$ --- as seen in panel~c). By absorbing the variance signal into $\gamma$ rather than into extra mean shifts, ShiVa yields cleaner $\mu$ estimates whenever variance is non-uniform. In scenarios with only mean shifts or only mild variance heterogeneity, phyloEM and $\ell$1ou+pBIC are equally accurate or slightly better on the optimal-value estimates. Because the underlying shift structure is unknown a priori, the joint perspective afforded by ShiVa is the safer choice in practice.

\begin{figure}[htbp]

    \centering
    % First subplot: 1 shift in optimal value
    \begin{subfigure}[b]{0.9\textwidth}
        \centering
        \begin{tikzpicture}
            \node[anchor=south west, inner sep=0] (img) {\includegraphics[width=\textwidth,height=0.21\textheight,keepaspectratio]{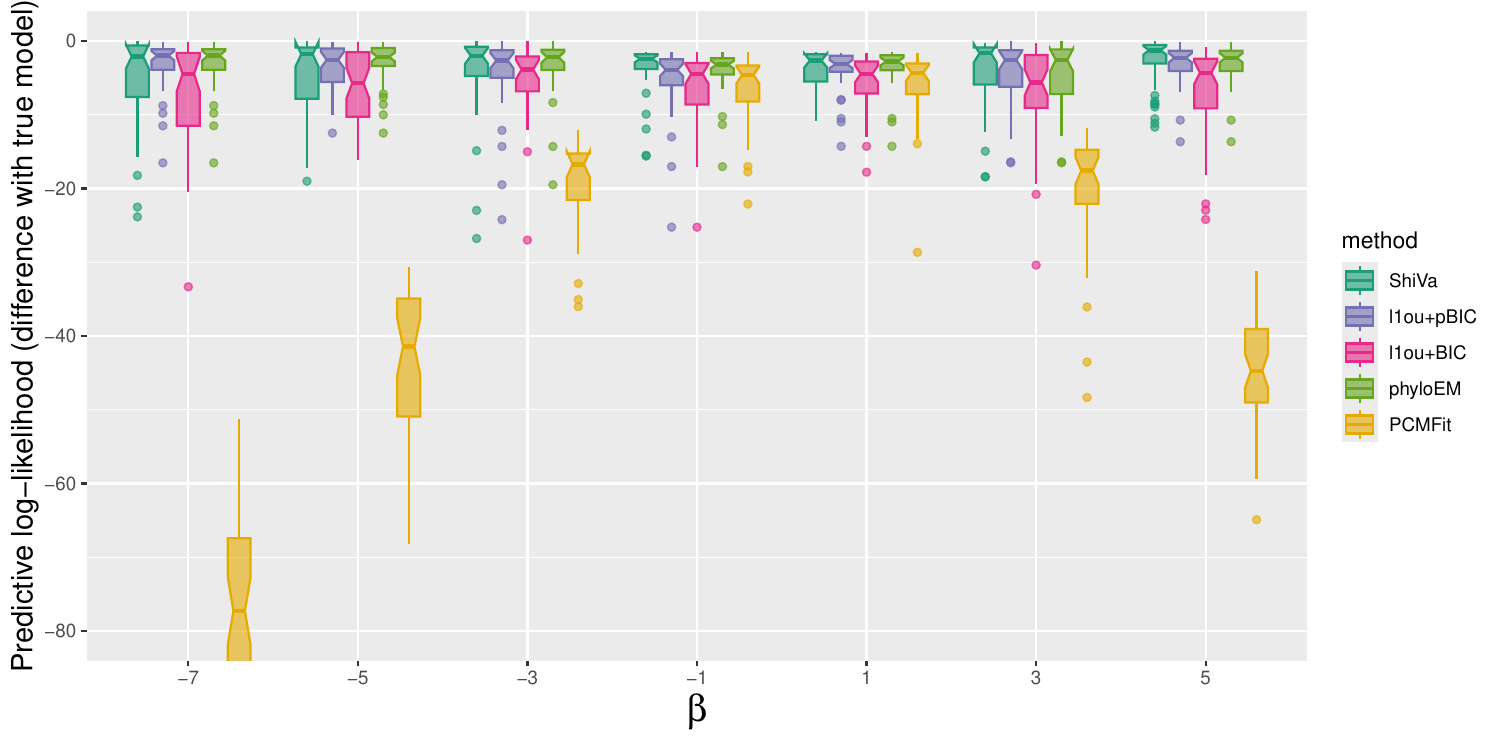}};
            \node[anchor=south west, inner sep=1pt, xshift=-10pt, yshift=2pt, font=\footnotesize] at (img.north west) {a)};
        \end{tikzpicture}
        \label{fig:2-loglik-mean}
    \end{subfigure}

    % Second subplot: 1 shift in variance
    \begin{subfigure}[b]{0.9\textwidth}
        \centering
        \begin{tikzpicture}
            \node[anchor=south west, inner sep=0] (img) {\includegraphics[width=\textwidth,height=0.21\textheight,keepaspectratio]{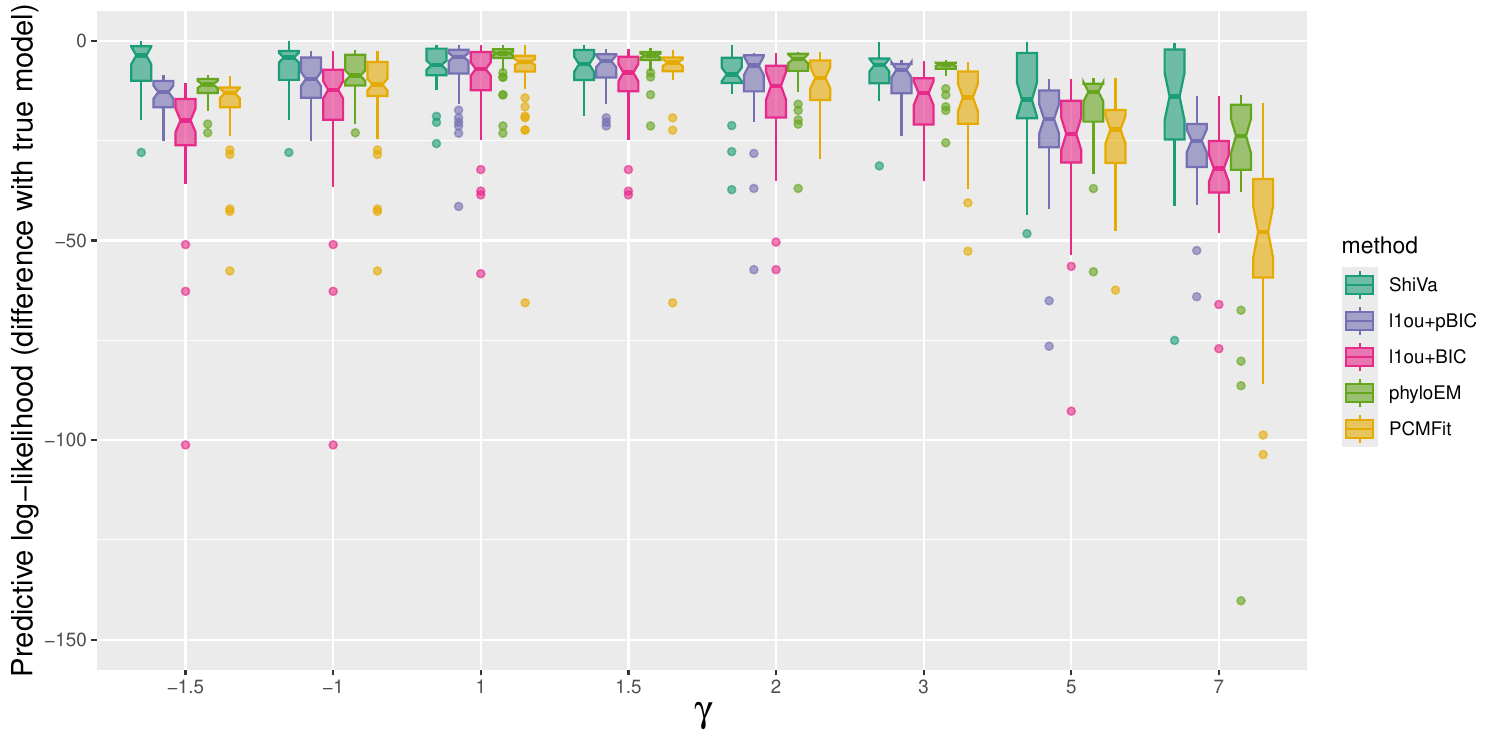}};
            \node[anchor=south west, inner sep=1pt, xshift=-10pt, yshift=2pt, font=\footnotesize] at (img.north west) {b)};
        \end{tikzpicture}
        \label{fig:2-log-var}
    \end{subfigure}

    % Third subplot: 1 shift in optimal value and 1 shift in variance
    \begin{subfigure}[b]{0.9\textwidth}
        \centering
        \begin{tikzpicture}
            \node[anchor=south west, inner sep=0] (img) {\includegraphics[width=\textwidth,height=0.21\textheight,keepaspectratio]{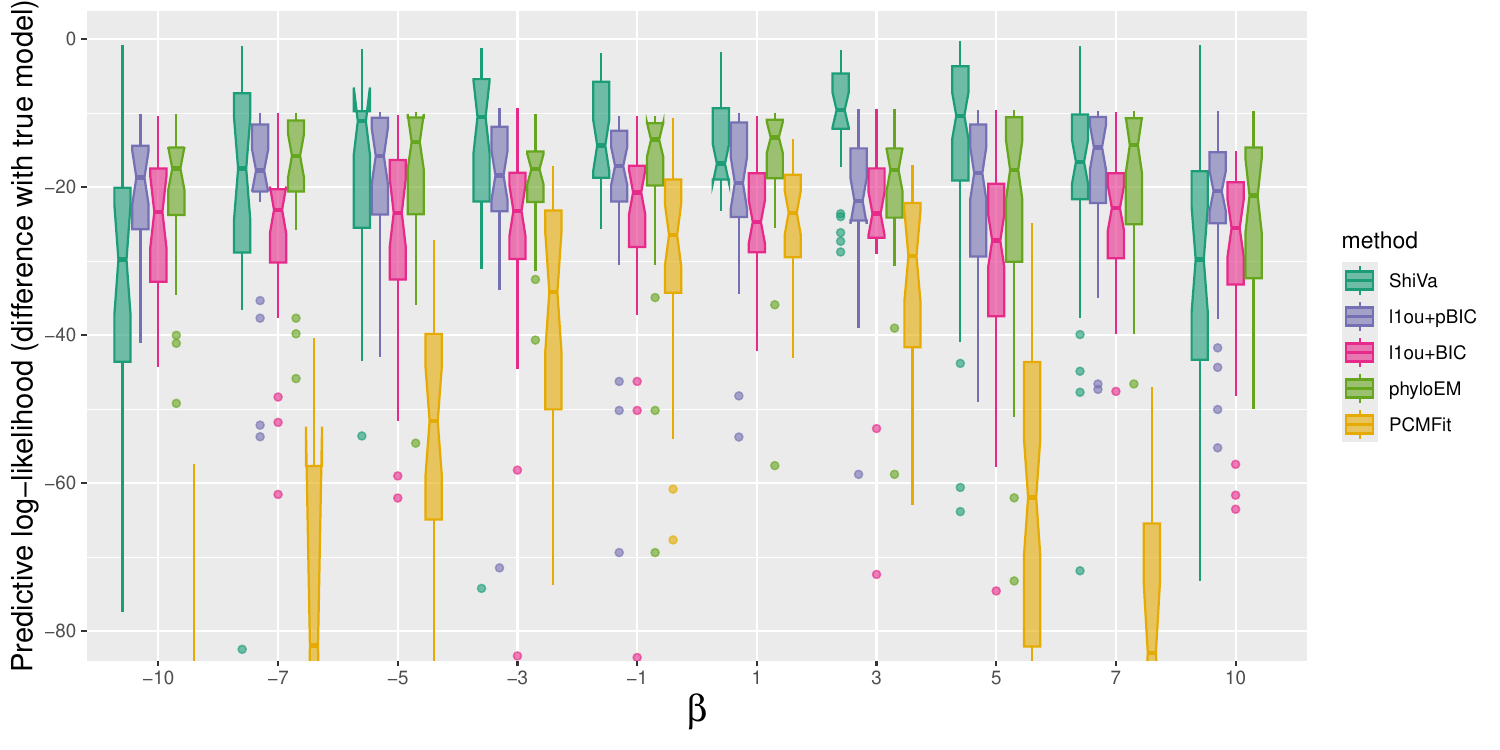}};
            \node[anchor=south west, inner sep=1pt, xshift=-10pt, yshift=2pt, font=\footnotesize] at (img.north west) {c)};
        \end{tikzpicture}
        \label{fig:2-loglik-both-beta}
    \end{subfigure}

    \begin{subfigure}[b]{0.9\textwidth}
        \centering
        \begin{tikzpicture}
            \node[anchor=south west, inner sep=0] (img) {\includegraphics[width=\textwidth,height=0.21\textheight,keepaspectratio]{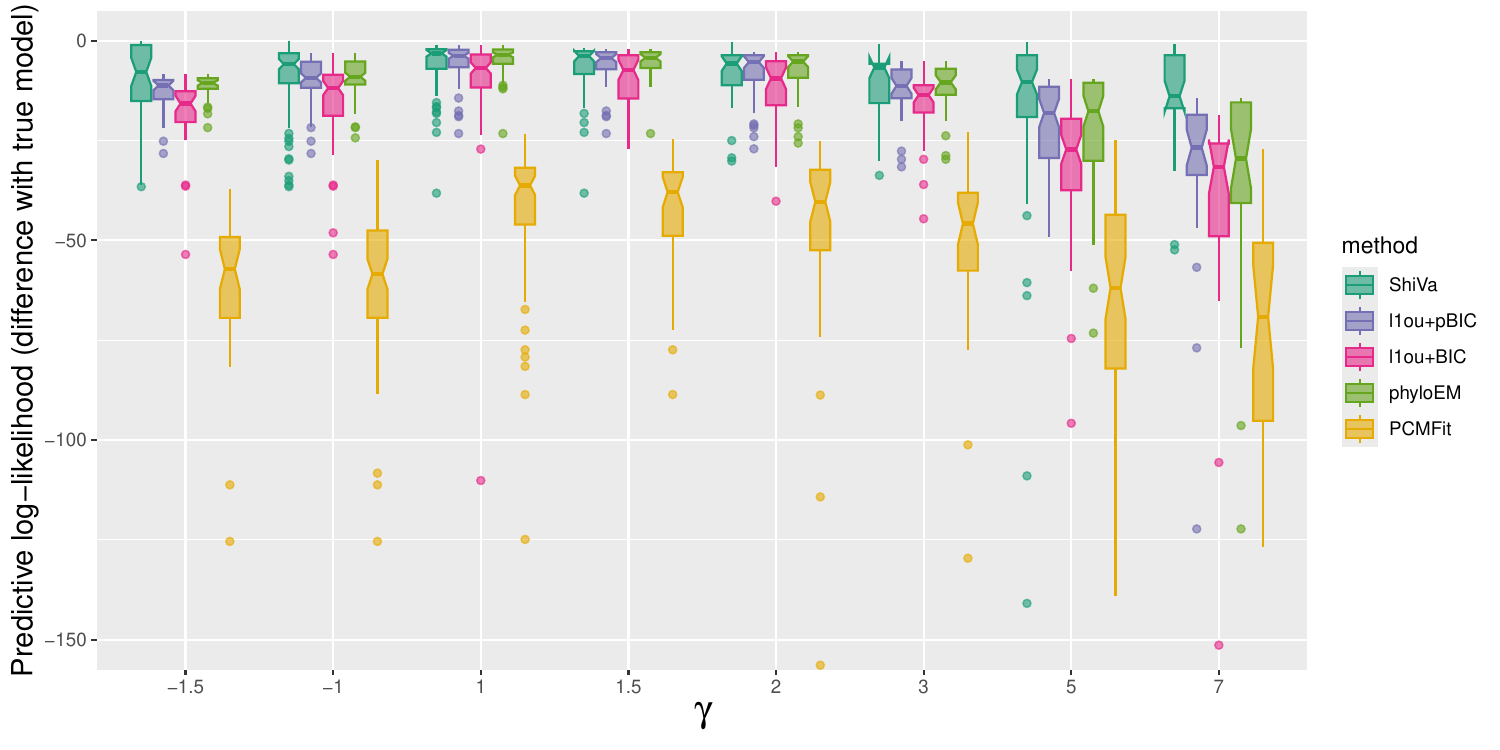}};
            \node[anchor=south west, inner sep=1pt, xshift=-10pt, yshift=2pt, font=\footnotesize] at (img.north west) {d)};
        \end{tikzpicture}
        \label{fig:2-loglik-both-gamma}
    \end{subfigure}
    \caption{Comparison of predictive log-likelihood across the five methods for four scenarios: a) 1 shift in optimal value; b) 1 shift in variance; c) 1 shift in optimal value + 1 shift in variance (fixed $\gamma$); d) 1 shift in optimal value + 1 shift in variance (fixed $\beta$).}
    \label{fig:2-loglik}
\end{figure}

In addition to TP, FP and MISE, we also evaluated predictive log-likelihood to assess the prediction accuracy of the models estimated by different methods and the similarity between the estimated models and the generating model. Specifically, for each of the 50 training datasets, we estimate model parameters (including the locations and magnitudes of shifts) and use them to compute the expected mean vector $\boldsymbol{\mu}$ and covariance matrix $\boldsymbol{\Sigma}$ under the fitted model; $(\boldsymbol{\mu}, \boldsymbol{\Sigma})$ define the multivariate normal density that will be used to score new datasets. We then generate 1,000 new test datasets independently using the same phylogeny and true evolutionary process as the training data. For each of these test datasets $\mathbf{Y}^{(k)}$, we evaluate the log-likelihood $\log \mathcal{N}(\mathbf{Y}^{(k)} \mid \boldsymbol{\mu}, \boldsymbol{\Sigma})$ under the estimated model from the corresponding training set, and take the average to obtain a predictive log-likelihood for that training replicate. This process results in 50 predictive log-likelihood values, from which we report summary statistics (e.g., mean and median). This metric assesses how similar the inferred model is to the generating evolutionary process, providing insight into the biological relevance of the estimated shifts. Since the log-likelihoods of the true models vary, we compute the predictive log-likelihood difference as the predictive log-likelihood of each estimated model minus the log-likelihood of the corresponding true model. This normalization allows for fair comparison across methods, where a higher score indicates better performance. The normalized score is a Monte Carlo estimate of the Kullback--Leibler divergence $-\mathrm{KL}(p \parallel \hat{p})$, where $p$ is the true model and $\hat{p}$ is the fitted model, and so provides an information-theoretic comparison between fitted and true models \citep{burnham_anderson_2002}. Figure~\ref{fig:2-loglik} illustrates the differences in predictive log-likelihood between the estimated models and the true model. When shifts occur only in optimal value, ShiVa performs comparably to $\ell$1ou and phyloEM, demonstrating similar predictive precision. However, when shifts in variance are introduced, ShiVa's median predictive log-likelihood is higher than those of $\ell$1ou, phyloEM, and PCMFit. Although the full boxplot distributions partly overlap, the notched $95\%$ confidence intervals for the medians are non-overlapping at larger $|\gamma|$, indicating a consistent statistical improvement once the variance-shift signal is strong enough. In contrast, PCMFit consistently exhibits the lowest predictive log-likelihood across all scenarios, likely due to overfitting.

Finally, we compare the average computational time across different methods. Methods that only account for shifts in optimal values, such as $\ell$1ou, are the fastest, with $\ell$1ou being particularly efficient due to its use of the LASSO path method. ShiVa performs similarly to phyloEM but is slightly slower in comparison. On the other hand, PCMFit requires significantly more time due to its handling of complex computations, especially when both shifts in optimal values and variance are present. The current PCMFit implementation allows parallel computing to partially compensate for the disadvantage of the method in terms of computational complexity. For the comparisons in Table~\ref{tab:2-computation}, we ran all methods single-threaded to provide a fairer comparison of computation required by all methods.

\begin{table}[htbp]
\centering
\small
\caption{Mean (standard deviation) computation time in seconds across parameter settings within each scenario.}
\label{tab:2-computation}
\begin{tabular}{lccccc}
\toprule
\textbf{Scenario} & \textbf{ShiVa} & \textbf{l1ou-pBIC} & \textbf{l1ou-BIC} & \textbf{phyloEM} & \textbf{PCMFit} \\
\midrule
Only in optimal value ($\beta$) & 89.3 (27.9) & 8.3 (0.2) & 8.5 (0.2) & 69.2 (1.0) & 13{,}062 (1{,}708) \\
Only in variance ($\gamma$)     & 89.0 (11.5) & 8.1 (0.3) & 8.3 (0.3) & 68.4 (1.8) & 16{,}956 (2{,}132) \\
Both shifts                     & 85.2 (16.0) & 8.2 (0.2) & 8.6 (0.2) & 69.5 (0.8) & 19{,}892 (4{,}088) \\
\bottomrule
\end{tabular}
\end{table}

\subsection{Relationship between shift position and detection difficulty}
\begin{figure}[htbp]
    \centering
    \includegraphics[width=\linewidth]{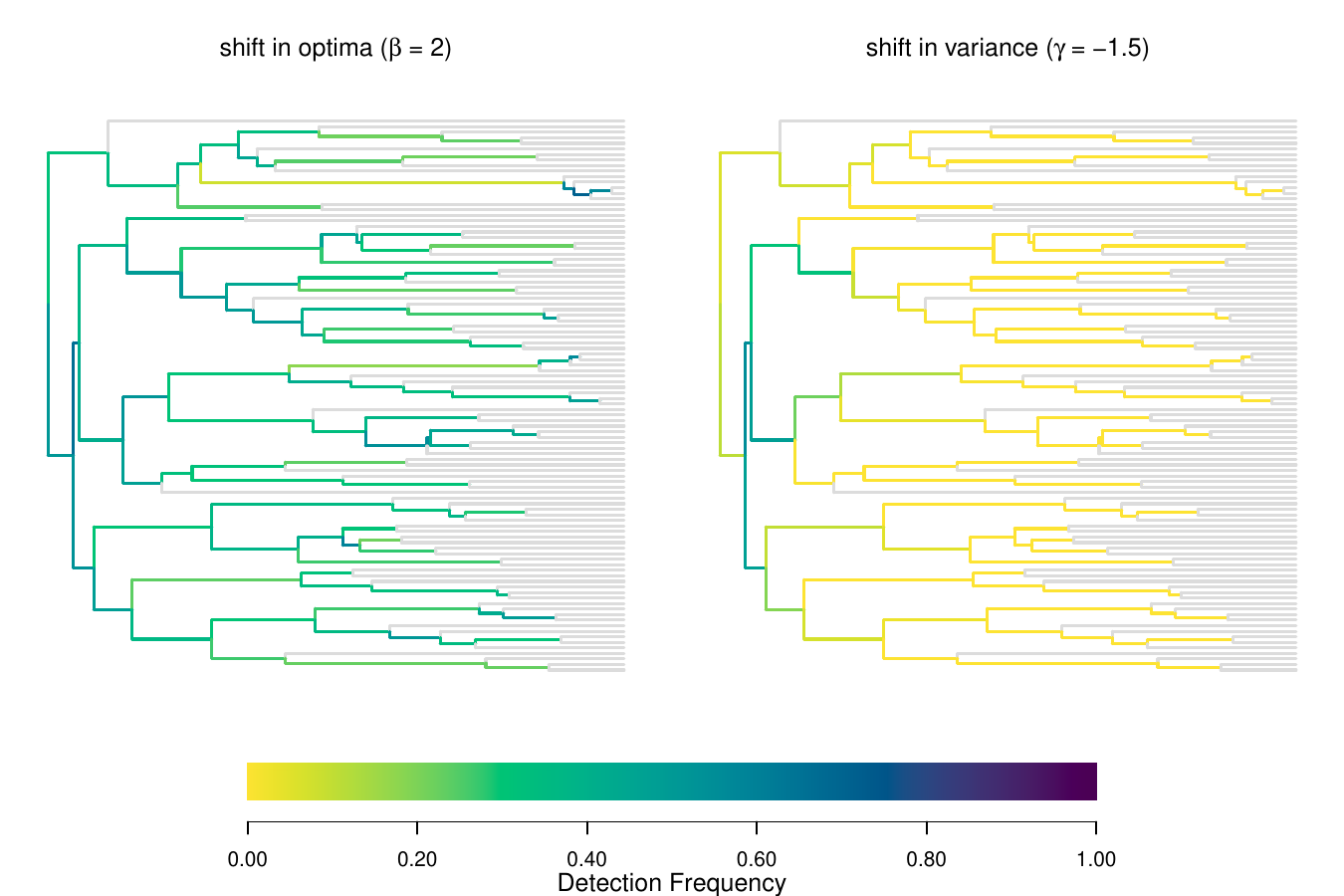}
    \caption{The detection probability of a shift on different branches using ShiVa.}
    \label{fig: 2-detection-distribution}
\end{figure}

To investigate the difficulty of detecting shifts at specific positions on a phylogenetic tree, we conducted simulations for every internal branch of the tree. For each branch, we performed 50 simulations where a shift in the optimal value occurred and 50 simulations where a shift in variance occurred. We then calculated the detection probabilities for each shift using ShiVa, considering different positions on the tree. Figure~\ref{fig: 2-detection-distribution} presents the detection probabilities for various branches, highlighting the different detection patterns for shifts in optima versus shifts in variance.

The results indicate that shifts in optima and shifts in variance exhibit distinct detection patterns. Shifts in the optimal value tend to be more detectable when they occur near the root or near the tips, while those occurring in the middle regions of the tree are generally harder to identify. In contrast, shifts in variance are more challenging to detect overall. Our findings show that shifts in variance are more likely to be detected when they occur on branches close to the root and with a greater number of descendant tips, both of which amplify the signal of the variance change. In our data the effect of absolute branch length is small and, if anything, weakly negative after controlling for branch position. Notably, the two branches immediately below the root node, branches~197 and~198, show lower per-branch detection probabilities than their own descendants. For a variance shift of $\gamma = -1.5$, ShiVa detects the true shift on branch~197 or branch~198 only $16.7\%$ and $3.0\%$ of the time, respectively; by contrast, the immediate descendants of branch~198, namely branches~195 and~196, are detected in $60.0\%$ and $42.9\%$ of replicates. This is not a lack of signal on branches~197 and~198 but a consequence of non-identifiability: shifts on these two root-adjacent branches are not statistically identifiable from each other, so their detection mass is spread across the pair. This observation is consistent with our analysis in the \emph{Evolutionary shifts in optimal value and variance} subsection. Overall, these results highlight how the structure of the phylogeny and the position of shifts critically influence the detectability of evolutionary changes, with variance shifts requiring more informative configurations for reliable inference compared to shifts in optima. A complementary scatterplot of the same detection proportions against each branch's start time is provided in Supplementary Figure~\ref{fig:detection-vs-time}.

\subsection{The influence of estimation of $\alpha$}\label{sec:alpha-misest}
We conducted simulations to examine how misestimating $\alpha$ affects ShiVa's estimation accuracy. The true value of $\alpha$ was fixed at $1$. To assess the impact of a wrong working value, we ran ShiVa with a range of fixed $\hat{\alpha} \in \{0.7, 1, 1.5, 3.5, 7, 14, 35\}$. We considered two scenarios: (a) a single shift in the optimal value on branch~71, with shift magnitude $\beta \in \{1, 3, 5, 7\}$, and (b) a single shift in the variance on branch~195, with shift magnitude $\gamma \in \{1, 1.5, 2, 3, 5\}$. Each configuration was repeated over $100$ simulation replicates. We evaluate performance using MISE of $\sigma^2$ and MISE of $\mu$ (defined in the \emph{Comparison of performance} subsection), which measure how accurately the inferred per-branch diffusion variance and optimal trait value reproduce the truth. For reference, the competing methods ($\ell$1ou, phyloEM and PCMFit) do not use $\alpha$ in the same parametric form, so their performance is unaffected by $\hat{\alpha}$ and serves as a baseline.

\begin{figure}[htbp]

    \centering
    % Panel (a): only mean shift
    \begin{subfigure}[b]{0.9\textwidth}
        \centering
        \begin{tikzpicture}
            \node[anchor=south west, inner sep=0] (img) {\includegraphics[width=\textwidth,height=0.25\textheight,keepaspectratio]{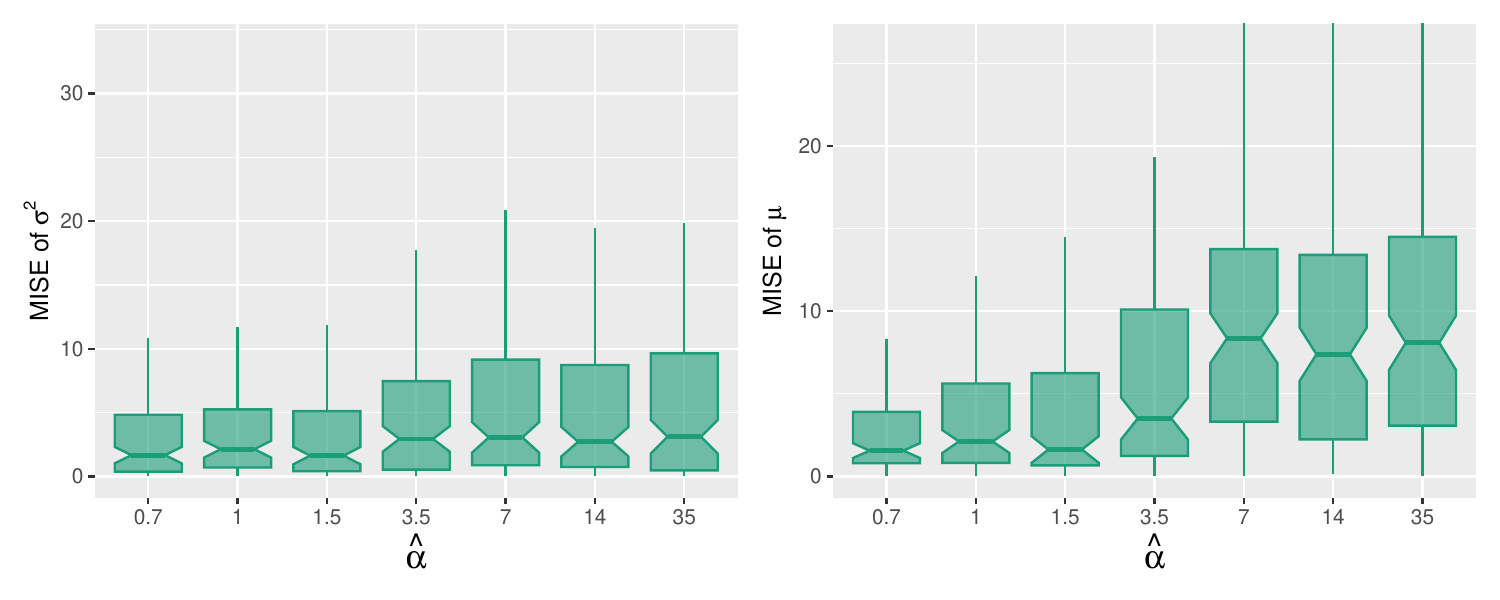}};
            \node[anchor=north west, inner sep=1pt, xshift=-10pt, font=\footnotesize] at (img.north west) {a)};
        \end{tikzpicture}
        \label{fig:2-alpha-mean}
    \end{subfigure}

    % Panel (b): only variance shift
    \begin{subfigure}[b]{0.9\textwidth}
        \centering
        \begin{tikzpicture}
            \node[anchor=south west, inner sep=0] (img) {\includegraphics[width=\textwidth,height=0.25\textheight,keepaspectratio]{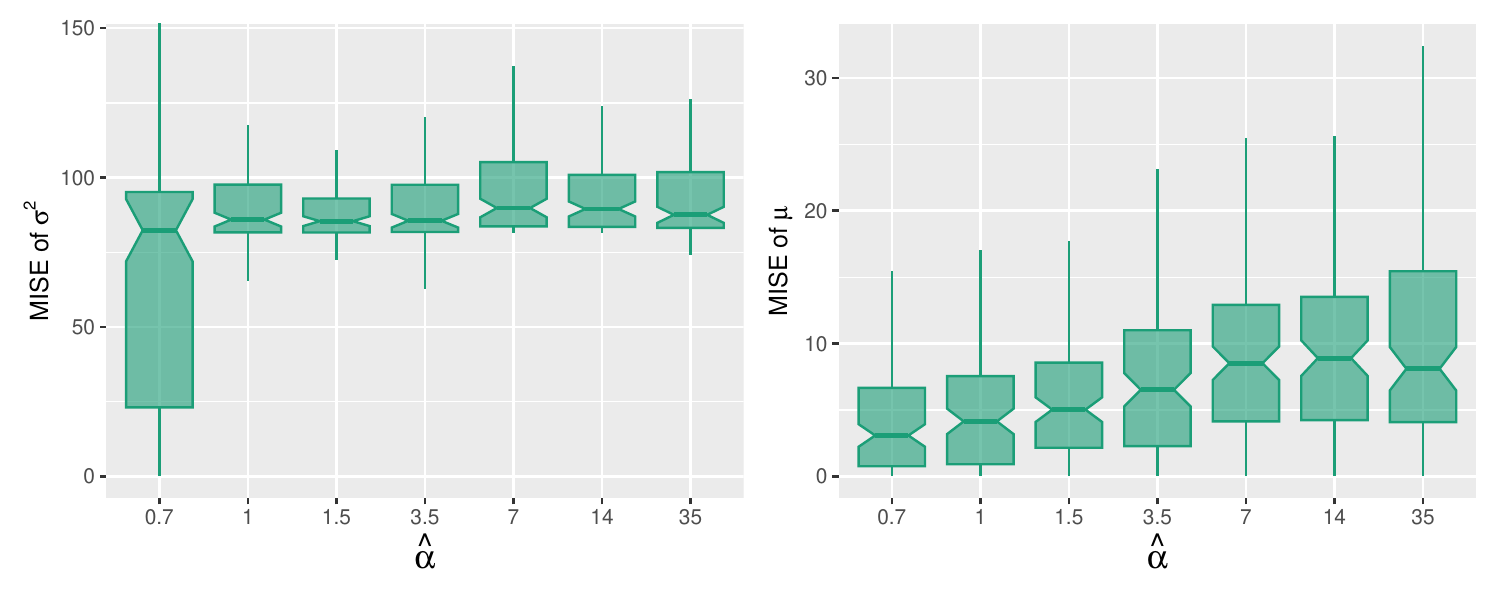}};
            \node[anchor=north west, inner sep=1pt, xshift=-10pt, font=\footnotesize] at (img.north west) {b)};
        \end{tikzpicture}
        \label{fig:2-alpha-var}
    \end{subfigure}

    \caption{MISE of ShiVa under misestimated $\hat{\alpha}$ (true $\alpha = 1$) for two scenarios: a) 1 shift in optimal value ($\beta = 3$); b) 1 shift in variance ($\gamma = 3$). Each panel shows notched boxplots over 100 replicates of MISE of $\sigma^2$ (left) and MISE of $\mu$ (right) on a log scale; the $x$-axis lists the working value~$\hat{\alpha}$.}
    \label{fig:2-alpha}
\end{figure}

Figure~\ref{fig:2-alpha} summarises the results. When $\hat{\alpha}$ deviates from the true value, ShiVa's MISE tends to increase, with the effect more pronounced for MISE of $\mu$ than for MISE of $\sigma^2$. Nevertheless, the errors remain bounded across the tested range of $\hat{\alpha}$, indicating that ShiVa is broadly robust to misspecification of $\alpha$. This supports the practice of using a simple ad-hoc estimate of $\alpha$---such as the null-model estimate obtained by \texttt{phylolm}---in applications, as moderate estimation error in $\alpha$ has limited impact on ShiVa's accuracy. The corresponding predictive log-likelihoods, which convey a similar message, are reported in Supplementary Figure~\ref{fig:supp-alpha-loglik}.

 \bigskip

\section{Case study} \setcurrentname{Case study}\label{sec:case}

To illustrate the application of our method, we analyzed two empirical datasets. The first dataset concerns floral diameter in the parasitic plant family \emph{Rafflesiaceae} (3 species, with 22 \emph{Euphorbiaceae} outgroups), using the phylogeny \texttt{flowerTree} from \citet{davis2007floral}. The transition to parasitism in \emph{Rafflesiaceae} is hypothesized to have relaxed constraints on flower size present in autotrophic relatives, providing a natural test case for detecting a shift in evolutionary variance rather than in the optimum alone. The second dataset involves buccal morphology in \emph{Centrarchidae} sunfishes, based on a time-calibrated phylogeny of 28 species from \citet{Revell_Collar_2009}. We applied ShiVa to log-transformed floral diameter in the \emph{Rafflesiaceae} dataset and log-transformed buccal length in the \emph{Centrarchidae} dataset, comparing results with $\ell$1ou+pBIC, $\ell$1ou+BIC, PhylogeneticEM, and PCMFit. Notably, ShiVa incorporates cross-validation during parameter tuning, introducing some randomness into the results. To ensure robustness, we recommend running ShiVa three times and selecting the model with the lowest BIC, as we did in our empirical analyses.

\begin{figure}
  \centering
    \includegraphics[width=\textwidth]{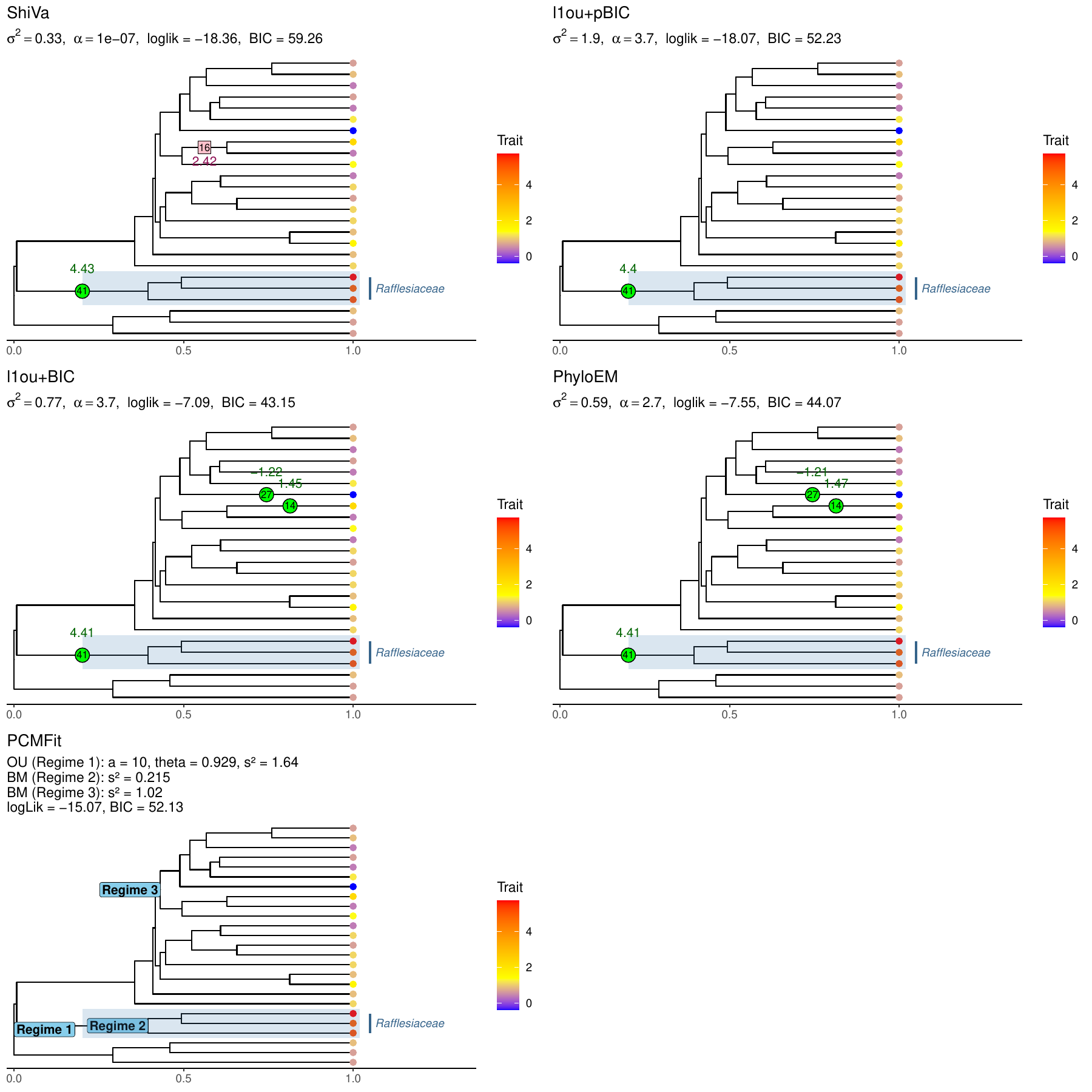}
     \caption{\label{case study-flower} The detection results on the flower dataset using different methods are shown. Green circles indicate shifts in optimal trait values, while pink squares denote shifts in evolutionary variance. The numbers inside the markers correspond to the node indices where shifts occur, and the colored numbers next to them represent the estimated magnitudes of those shifts.}
\end{figure}

Figure~\ref{case study-flower} presents the shift detection results on the flower dataset. ShiVa identified one shift in the optimal value and one shift in evolutionary variance, with estimated values of \( \sigma^2 = 0.33 \) and \( \gamma = 2.42 \), respectively. Compared to the results of $\ell$1ou+pBIC, ShiVa detects one additional variance shift. The shift size of 2.42 is substantial, especially when considered relative to the original variance, highlighting a significant change in evolutionary dynamics.

The estimated value $\hat{\alpha} \approx 10^{-7}$ shown in Figure~\ref{case study-flower} comes from the null-model \texttt{phylolm} fit on the trait without accounting for shifts; when the trait actually contains shifts, their variance is absorbed into $\sigma_0^2$ in a no-shift fit, biasing $\hat{\alpha}$ downward. The package's \texttt{refit\_alpha} option (enabled by default) addresses this by re-estimating $\alpha$ with \texttt{phylolm} \emph{conditional on the detected mean shifts} and refitting the full model at the updated value. For the flower dataset, this yields $\hat{\alpha} = 11.6$, which is in the same OU range as the estimates from the other methods (2.7--10). The detected shifts --- the mean shift on the branch leading to \emph{Rafflesiaceae} (edge~41) and the variance shift on edge~16 --- are unchanged.

Although ShiVa appears to yield a lower log-likelihood and higher BIC than $\ell$1ou+pBIC, this comparison includes the estimation of the adaptation rate parameter \( \alpha \). When comparing detection results under a fixed \( \alpha \) (the null-model estimate obtained by \texttt{phylolm}, used here to place all methods on a common scale, since empirical data has no ``true'' $\alpha$), ShiVa achieves a higher log-likelihood and lower BIC than $\ell$1ou+pBIC, suggesting that modeling variance shifts improves model fit. Notably, compared to $\ell$1ou+pBIC, ShiVa detects an additional variance shift on branch 16. To evaluate the significance of this shift, we performed a likelihood ratio test. The log-likelihood of the model with only a mean shift on branch 41 is $-21.89$, while the model that also includes a variance shift on branch 16 achieves a log-likelihood of $-18.36$. The resulting test statistic $-2(\ell_0 - \ell_1) = 7.06$ is compared to a $\chi^2_2$ distribution (the nested models differ by the branch identity and magnitude of the extra variance shift), giving a $p$-value of $0.0293$, indicating that the variance shift on branch 16 is statistically significant. PhyloEM and $\ell$1ou+BIC yield identical shift detections and attain the highest log-likelihood among all methods. 

To better understand the behavior of each method, we grouped the detected models into four representative types:  
(1) ShiVa: one shift in optimal value and one shift in variance;  
(2) $\ell$1ou+pBIC: one shift in optimal value;  
(3) $\ell$1ou+BIC / PhyloEM: three shifts in optimal value;  
(4) PCMFit: changes in all parameters on two shifted branches.
To assess how well each method performs when a given model is the true generative process, we simulate 100 datasets from each estimated model and re-apply all five methods to the simulated data. We then compute the predictive log-likelihood to compare model fit across methods. As shown in Table~\ref{flower-table}, ShiVa consistently achieves relatively high and stable predictive log-likelihood, even when the data are generated from models estimated by other methods. This suggests that ShiVa produces robust and reasonable detection results across a variety of underlying evolutionary scenarios, supporting its credibility and generalizability.

\begin{table}[ht]
\centering
\small
\caption{\label{flower-table} The median predictive log-likelihood of different methods, evaluated on datasets simulated from each of the estimated models. (flower dataset)
}
\begin{tabular}{l@{\hskip 3pt}r@{\hskip 3pt}r@{\hskip 3pt}r@{\hskip 3pt}r@{\hskip 3pt}r@{\hskip 3pt}r}
\toprule
Model Simulated & ShiVa & $\ell$1ou+pBIC & $\ell$1ou+BIC & phyloEM & PCMFit & True \\
\midrule
(1) & \textbf{-23.88} & -28.84 & -34.81 & -28.66 & -167.60 & -18.37 \\
(2) & \textbf{-21.25} & -23.02 & -29.67 & -23.02 & -153.10 & -18.07 \\
(3) & -20.03 & -21.14 & \textbf{-18.16} & -20.70 & -234.62 & -7.07 \\
(4) & \textbf{-23.29} & -26.33 & -70.25 & -26.19 & -284.80 & -13.50 \\
\bottomrule
\end{tabular}
\end{table}

\begin{figure}
  \centering
    \includegraphics[width=\textwidth]{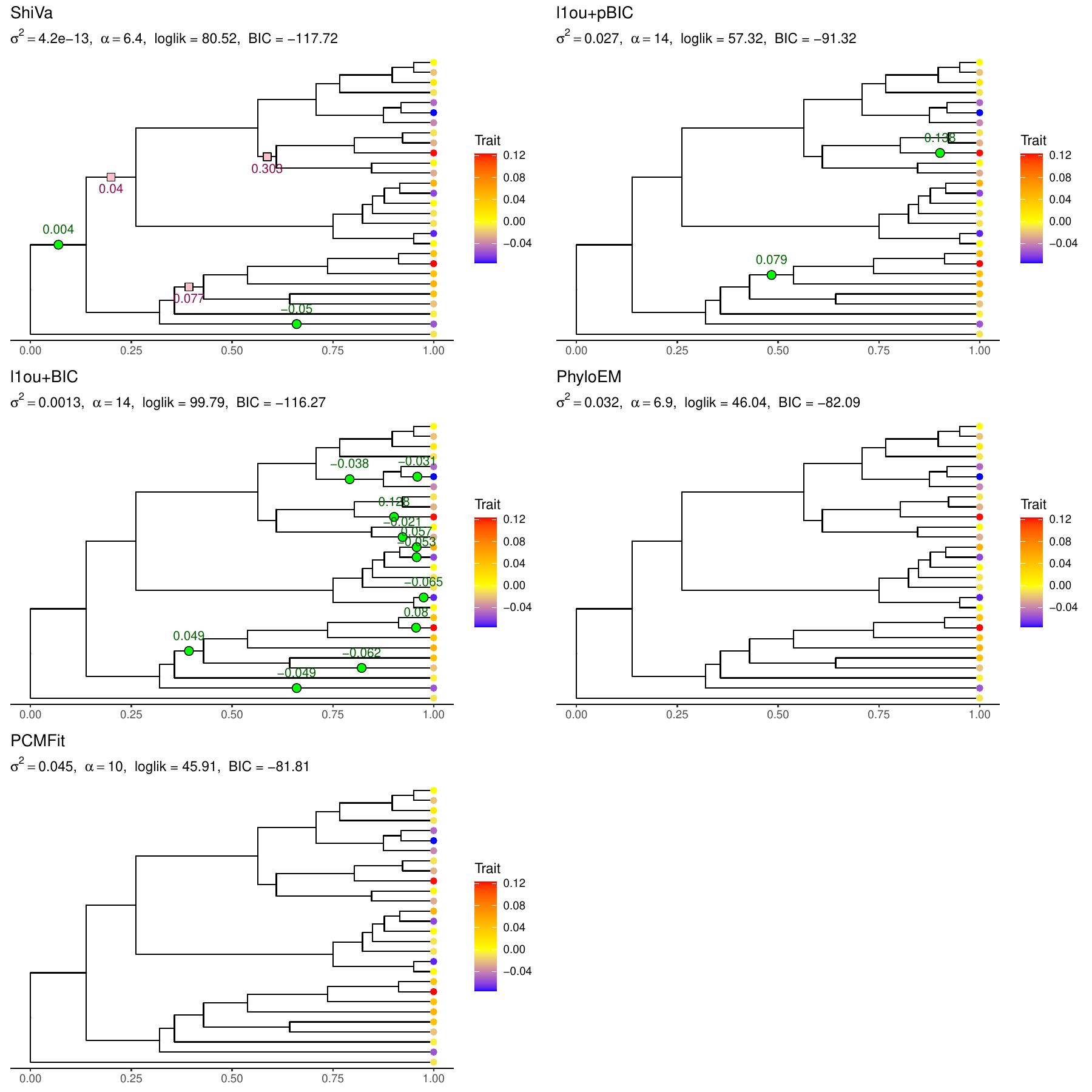}
     \caption{\label{case study}The detection results on the centrarchidae data using different methods are shown. Green circles indicate shifts in optimal values, and pink squares indicate shifts in evolutionary variance. The numbers inside the markers represent the branch indices where shifts occur, while the colored numbers next to them denote the estimated shift magnitudes.}
\end{figure}

Figure~\ref{case study} summarizes the results from the \emph{Centrarchidae} analysis. ShiVa detected two shifts in the optimal trait value and three shifts in evolutionary variance. It estimated a saturated model by assigning an extremely small value to the original $\sigma^2$. Among the three variance shifts, the one on branch 48—with a magnitude of 0.309—is the most substantial, while the other two primarily restore typical levels of diffusion variance. $\ell$1ou+pBIC identified two shifts in the optimal trait value, while $\ell$1ou+BIC inferred 11 such shifts. PhylogeneticEM and PCMFit detected no shifts under their respective model selection criteria.

We compare the methods using both log-likelihood and BIC, since log-likelihood alone rewards models with more parameters and is therefore not a fair basis for model comparison on its own. ShiVa achieved a higher log-likelihood than $\ell$1ou+pBIC, PhylogeneticEM, and PCMFit, and its BIC is lower than each of these three. $\ell$1ou+BIC attained the highest log-likelihood but did so by detecting a large number of optimal-value shifts—including several within the clade descending from branch~48, where ShiVa detects a variance shift of substantial magnitude~($0.309$); specifically, $\ell$1ou+BIC places two shifts (branches~29 and~31) inside that clade and $\ell$1ou+pBIC places one (branch~29). Because of this parameter cost, $\ell$1ou+BIC's BIC is higher (i.e., worse) than ShiVa's, so the combined log-likelihood and BIC comparison favours ShiVa rather than $\ell$1ou+BIC.

Using this dataset, we grouped the estimated models into four representative types:
(1) ShiVa: 2 shifts in optimal value and 3 shifts in variance;
(2) $\ell$1ou+pBIC: 2 shifts in optimal value;
(3) $\ell$1ou+BIC: 11 shifts in optimal value;
(4) PhyloEM / PCMFit: no shifts detected.

As in the previous analysis, we simulated 100 datasets from each estimated model and re-applied all five methods to the simulated data. We then computed the predictive log-likelihood to evaluate model fit across methods. As shown in Table~\ref{sunfish-table}, ShiVa consistently achieves relatively high and stable predictive log-likelihoods, except for model~(3). Its performance is affected in scenarios with a large number of shifts in the optimal value.

\begin{table}[ht]
\centering
\small
\caption{\label{sunfish-table} The median predictive log-likelihood of each method, evaluated on simulated datasets generated from each estimated model. (sunfish dataset)}
\begin{tabular}{l@{\hskip 3pt}r@{\hskip 3pt}r@{\hskip 3pt}r@{\hskip 3pt}r@{\hskip 3pt}r@{\hskip 3pt}r}
\toprule
Model Simulated & ShiVa & $\ell$1ou+pBIC & $\ell$1ou+BIC & phyloEM & PCMFit & True \\
\midrule
(1) & \textbf{27.74} & 17.79 & 2.21 & 19.39 & 19.93 & 72.31 \\
(2) & 43.29 & \textbf{49.17} & 43.80 & 41.23 & 40.97 & 57.32 \\
(3) & 25.96 & 49.83 & \textbf{85.67} & 39.01 & 46.26 & 99.82 \\
(4) & 41.87 & 41.32 & 32.00 & 43.17 & \textbf{44.75} & 46.05 \\
\bottomrule
\end{tabular}
\end{table}

\bigskip
\section{Conclusion} \setcurrentname{Conclusion}\label{sec5}
In this article, we have used a multi-optima multi-variance OU process to describe an evolutionary process where abrupt shifts can occur in either optimal value or variance. We then proposed a new method to simultaneously detect shifts in optimal value and shifts in variance. We implemented the method in \texttt{R}. Our package is available from the first author's GitHub page \url{https://github.com/WenshaZ/ShiVa}. Furthermore, we have conducted simulation studies to show the effectiveness of our method to detect both kinds of shifts and compared it to methods which only detect shifts in optimal value.

Our results showed that ShiVa effectively balances True Positives (TP) and False Positives (FP) across different scenarios, and recovers the per-branch evolutionary parameters with low mean integrated squared error (MISE), which measures the cumulative branch-wise error in estimation of optima and variance that accounts for misestimated but adjacent shifts. When there was only a shift in the optimal value, ShiVa performed comparably to other methods, such as $\ell$1ou and phyloEM, maintaining similar predictive accuracy while achieving a better balance between TP and FP compared to PCMFit, which tended to overfit in weak signal scenarios. When a shift in variance was present, ShiVa demonstrated a significant advantage over other methods, achieving higher predictive log-likelihood and effectively distinguishing between variance and mean shifts. This led to fewer False Positives compared to $\ell$1ou and phyloEM, which often misinterpreted variance shifts as multiple shifts in optimal value. Although PCMFit achieved higher TP, it generated significantly more FP, indicating overfitting. A plausible reason PCMFit's penalty is insufficient to prevent this overfitting is the geometry of its model-selection criterion. PCMFit uses the Akaike information criterion to score candidate regime partitions \citep{Mitov_Bartoszek_Stadler_2019}, which imposes a fixed per-parameter cost that does not scale with sample size. Because a full OU regime adds several parameters, the per-shift threshold is small, and it is easily beaten by chance when the greedy search examines every internal branch as a candidate shift position. In addition, the threshold does not depend on the magnitude of the fitted parameter jump, so a spurious regime whose fitted $\theta$ differs only slightly from its neighbours is admitted at the same cost as a well-supported one. ShiVa's LASSO+BIC combination is stricter on both fronts: the $\ell_1$ penalty is applied to the shift magnitudes themselves and continuously shrinks small fitted values to zero, and the BIC selection criterion imposes a per-parameter penalty that grows with the sample size. A rigorous comparison of the two penalty schemes lies beyond the scope of this paper and is left for future work.

Our empirical analyses on floral diameter in \emph{Rafflesiaceae} and buccal morphology in \emph{Centrarchidae} sunfishes demonstrate the practical utility of ShiVa. In both datasets, ShiVa successfully identified meaningful variance shifts that were missed or misinterpreted by methods focusing only on optimal value shifts. While ShiVa may yield slightly lower likelihoods in some cases, it achieves better model fit when accounting for variance shifts and maintains strong predictive performance across models. These results underscore ShiVa's robustness and its ability to capture complex evolutionary dynamics that may be overlooked by existing approaches.

For model simplicity, we assume that the variance parameter is constant along each branch and that all shifts in variance happen at the beginning of the branch. Simulations of cases where these assumptions are violated show that our method is robust to misspecification in these assumptions. Biologically, environmental changes need not align with speciation events, so a real variance shift could occur mid-branch; in such cases the beginning-of-branch assumption introduces some model misspecification, although the simulations suggest that the overall detection results are not strongly affected. An interesting future research direction is to extend our model to allow shifts in variance at internal positions along a branch. This would allow us to estimate the exact time of a shift in variance. Writing the likelihood for this case is straightforward, but more work may be needed to ensure stability of the estimates. Another direction for future work is to improve the computational efficiency of the method. The likelihood calculation involves a large number of matrix inverse computations,which can be computationally expensive for large trees. \citet{bastide_ho_baele_lemey_suchard_2021} provide an efficient algorithm to calculate the log likelihood and its derivatives for certain phylogenetic models. If this method could be adapted to our method, it would lead to a major improvement in the computational speed, allowing our method to scale to larger phylogenies.
\bigskip

\newpage

\section{Data availability statement}
The phylogenetic tree of Anolis lizards is provided by \citet{anolis_data} and it can be accessed via the R package \texttt{l1ou} (\url{https://github.com/khabbazian/l1ou}). The phylogenetic tree and trait data for the \emph{Rafflesiaceae} flower dataset (with \emph{Euphorbiaceae} outgroups) \citep{davis2007floral} are available through the R package \texttt{phylolm}, and the \emph{Centrarchidae} sunfish dataset \citep{Revell_Collar_2009} can be accessed via the R package \texttt{phytools}. Both packages are available on CRAN. The datasets used in this study are also archived on Dryad (\url{https://doi.org/10.5061/dryad.v41ns1scc}). Our R package \texttt{ShiVa} is available at \url{https://github.com/WenshaZ/ShiVa}, and the simulation and case study code can be found at \url{https://github.com/WenshaZ/ShiVa-Experiments}.

\bigskip 

\section{Author contributions}
All authors contributed to conceive the ideas, design the methodology and the simulations. WZ conducted the simulations, analyzed the results and wrote the manuscript under the guidance of LSTH and TK. LSTH and TK reviewed and edited the manuscript.

\bigskip
\section{Acknowledgement}
LSTH was supported by the Canada Research Chairs program, the NSERC Discovery Grant RGPIN-2018-05447, and the NSERC Discovery Launch Supplement.
TK was supported by the NSERC Discovery Grant RGPIN-2023-03332.

\bigskip\bigskip

\section*{Supplemental Materials}
\renewcommand{\thesubsection}{\Alph{subsection}}

\subsection{Identifiability of variance shifts along a branch}
\label{sec:supp-identifiability}

In the main text we noted that the position of a single variance shift along a branch is identifiable, while two variance shifts on the same branch are not. Here we provide the derivation underlying these claims.

Consider a clade in which species $i$, $j$, $k$, and $l$ descend from a common ancestor $a$, and suppose a single variance shift occurs at time $t$ on the branch from $a$ to its child $a'$:

\begin{center}
\includegraphics[width=0.4\textwidth]{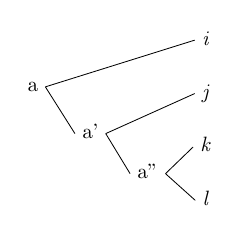}
\end{center}

Two differences in covariances among the descendants are
\begin{align*}
\mathbf{\Sigma}_{jk} - \mathbf{\Sigma}_{ij} &= \frac{1}{2\alpha} \left[\sigma_0^2 \left(e^{-2\alpha(T - t_{a'})} - e^{-2\alpha(T - t_a)}\right) + \Delta\sigma^2 \left(e^{-2\alpha(T - t_{a'})} - e^{-2\alpha(T - t)}\right)\right], \\
\mathbf{\Sigma}_{kl} - \mathbf{\Sigma}_{ij} &= \frac{1}{2\alpha} \left[\sigma_0^2 \left(e^{-2\alpha(T - t_{a''})} - e^{-2\alpha(T - t_{a})}\right) + \Delta\sigma^2 \left(e^{-2\alpha(T - t_{a''})} - e^{-2\alpha(T - t)}\right)\right].
\end{align*}
These two equations are linearly independent in $(t, \Delta\sigma^2)$ and can be solved for both unknowns, so a single variance shift is identifiable.

When two variance shifts occur on the same branch at times $t_1$ and $t_2$ with magnitudes $\Delta\sigma_1^2$ and $\Delta\sigma_2^2$, their joint contribution to the covariance between any two species descending from the branch is
\[
\frac{e^{-2\alpha (T - t_a)}}{2\alpha} \left[\left(1 - e^{-2\alpha (t_a - t_1)}\right) \Delta\sigma_1^2 + \left(1 - e^{-2\alpha (t_a - t_2)}\right) \Delta\sigma_2^2 \right].
\]
This expression is a sum of contributions, and a single shift with appropriately chosen time and magnitude can reproduce the same covariance. The two-shift configuration is therefore not identifiable.

\subsection{Detection proportion versus branch position}
\label{sec:supp-detection-vs-time}

\begin{figure}[H]
    \centering
    \includegraphics[width=\textwidth]{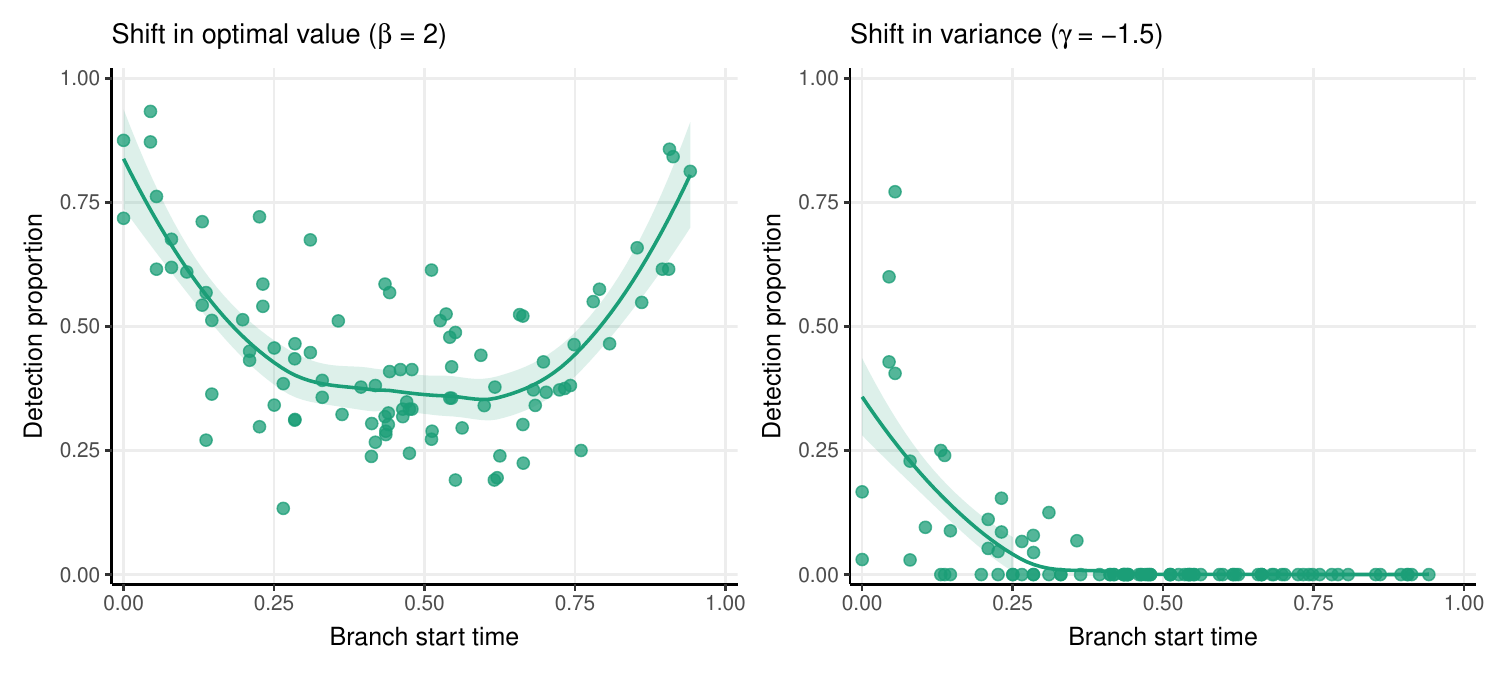}
    \caption{Per-branch detection proportion (ShiVa) against the branch's start time on the normalised lizard phylogeny. Left: shifts in optimal value ($\beta = 2$); right: shifts in variance ($\gamma = -1.5$). Curves are loess fits with $95\%$ pointwise confidence bands.}
    \label{fig:detection-vs-time}
\end{figure}

To complement the tree visualisation of Figure~\ref{fig: 2-detection-distribution}, Supplementary Figure~\ref{fig:detection-vs-time} plots, for each branch, the ShiVa detection proportion against the branch's start time (parent depth). The mean-shift panel ($\beta = 2$) shows a clear U-shape: detection is high for branches near the root (start time close to $0$) and for branches near the tips (start time close to $1$), and lowest for branches in the middle of the tree. The variance-shift panel ($\gamma = -1.5$) shows uniformly low detection once the branch start time exceeds about $0.25$, consistent with variance shifts being harder to detect at this magnitude.

\subsection{The effect of shift position within a branch}
In \emph{Trait evolution with shifts in both optimal value and variance}, we noted that our assumptions about the shifts in variance---at most one shift will occur on any branch; and the shifts occur at the beginning of the branches---are restrictions on the space of possible models, so in particular if these assumptions are not satisfied, then the model is misspecified. In this subsection, we conduct simulations in which these assumptions are violated. 

\begin{figure}[htbp]
  \centering
    \includegraphics[width=\textwidth]{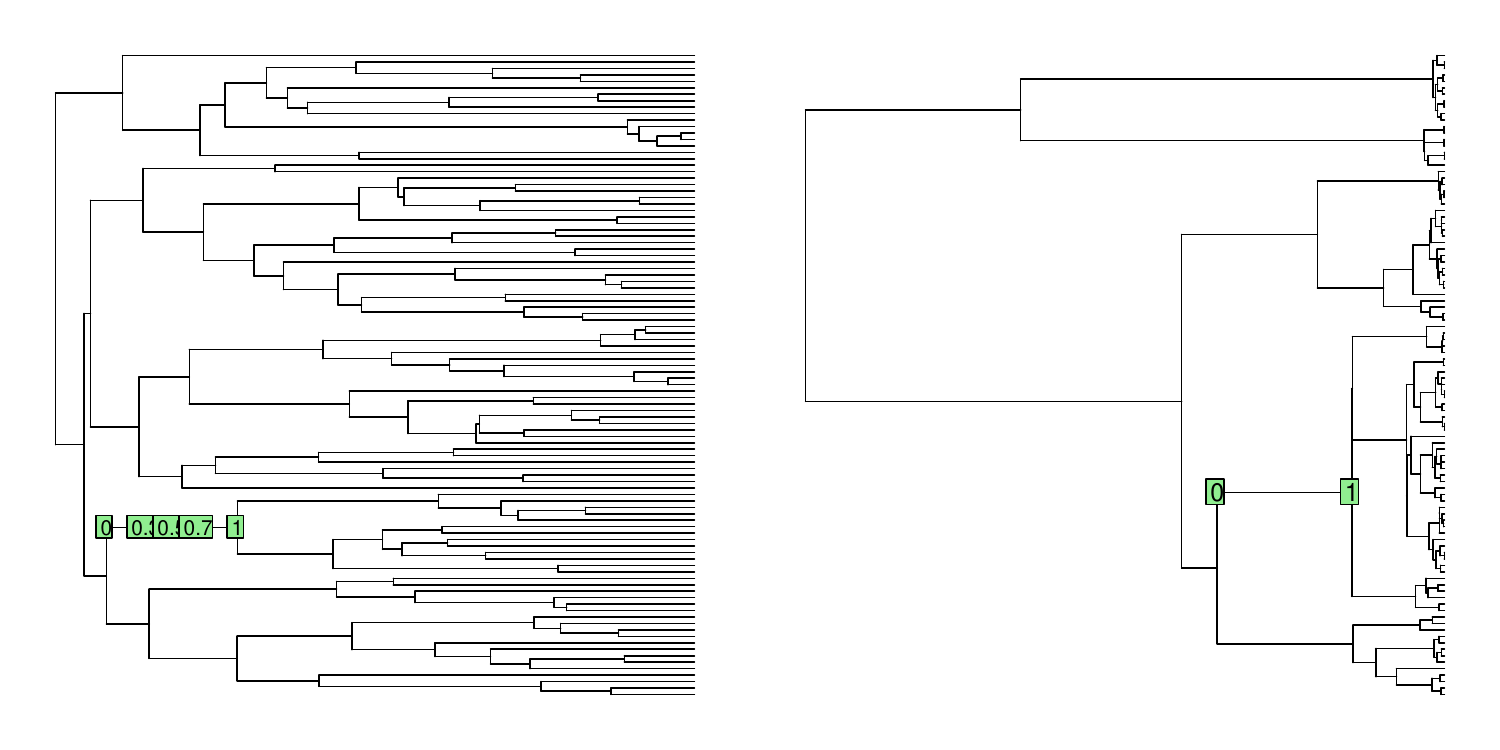}
     \caption{\label{4-loc}Left: The different locations of shifts on the same branch; Right: Two opposite shifts occur on the same branch}
\end{figure}

\begin{figure}[htbp]
  \centering
    \includegraphics[width=\textwidth]{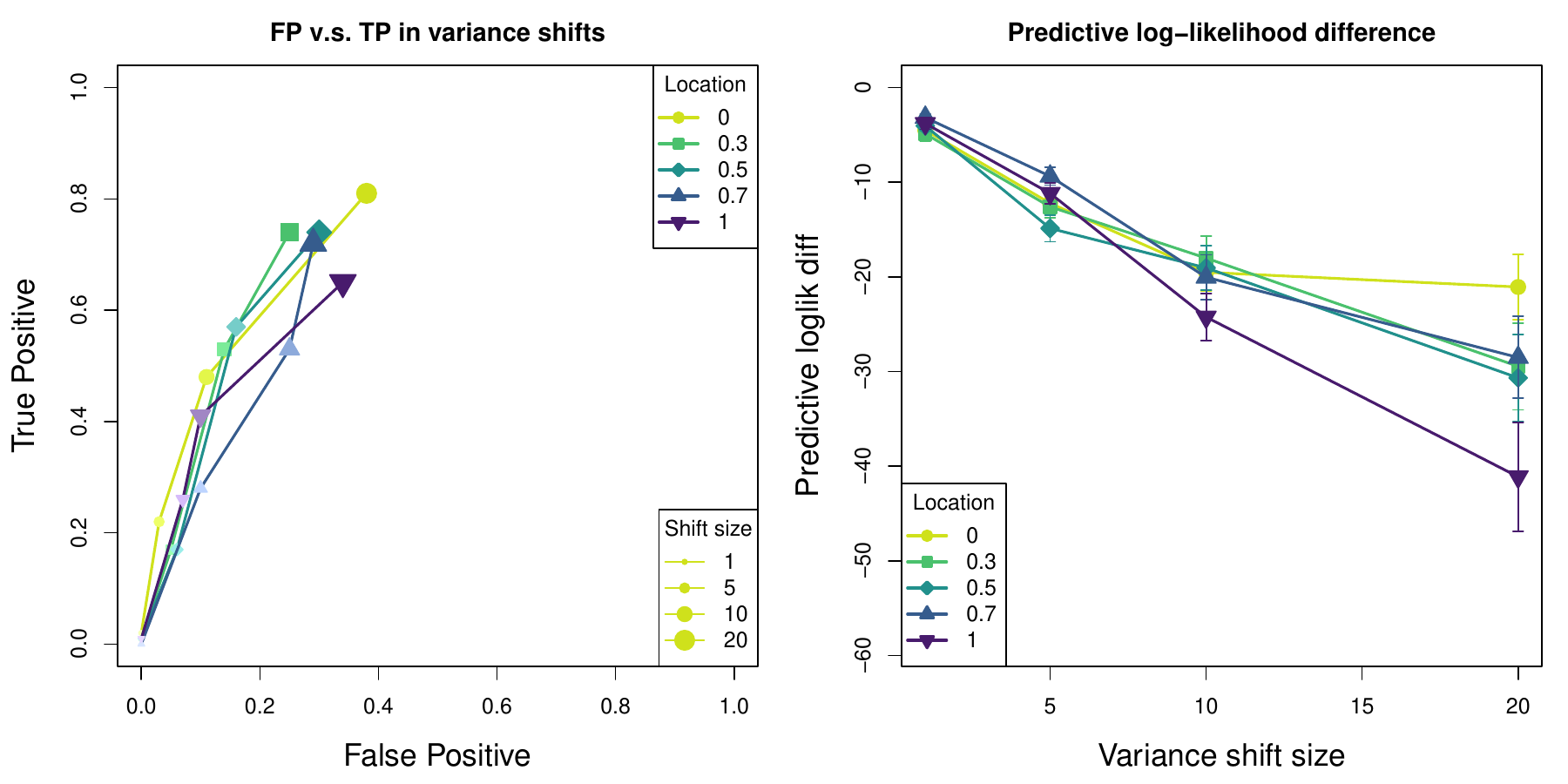}
     \caption{\label{4}The true positive v.s. false positive and the predictive log-likelihood difference of ShiVa with shifts at different locations}
\end{figure}

Firstly, we conduct a series of simulations with the shift occuring in different positions along a fixed branch. The different locations of the shift are shown in Figure~\ref{4-loc} (left). Figure~\ref{4} shows the True Positives v.s. False Positives and the difference of predictive log-likelihood between ShiVa and the true model with shifts in different locations. Location here is a parameter ranging from 0 to 1, with 0 meaning the beginning of the branch and 1 meaning the end of the branch. The ability of the method to detect the shift in variance is not greatly impacted by the position of the shift along the branch. Furthermore, when the shift size is small, the log-likelihood of the model is not much reduced in situations where the location of the shift is not at the beginning of the branch. However, when the magnitude of the shift is larger, the model misspecification does cause a substantial decrease in log-likelihood when the shift is not located at the begining of the branch. Overall, our method is robust to violations of our assumption that shifts occur at the beginning of a branch.

Secondly, we simulate scenarios with two shifts on a single branch. If shifts are both positive or both negative, the data will show a stronger signal than when just one shift occurs. Therefore, we simulate situations where two opposite shifts occur (at locations = 0 and 1) (Figure~\ref{4-loc} right). To compare, we also simulate situations where only one shift occurs at the beginning of that branch. Table~\ref{tab:multi} shows that when the two opposite shifts occur on that branch, our method cannot accurately detect the shift in variance on that branch and the false positive number of shifts in optimal value increases. In this case, the violation of the assumption causes some difficulties for our method. However, in the general case, it is not a major concern.

%\begin{figure}
%  \centering
%    \includegraphics[width=\textwidth]{4-loc-2.pdf}
%     \caption{\label{4-loc-2}Two opposite shifts occur on the same branch}
%     \textcolor{red}{combine this figure with figure 12}
% \end{figure}

\renewcommand{\arraystretch}{1.5}
\FloatBarrier
\vspace{0.5em}
\noindent\begin{table}[H]
    \caption{The performance of ShiVa when two opposite shifts occur on the same branch 
    %\textcolor{red}{font size is too small, hard to read}
    }
    \label{tab:multi}
%\resizebox{0.9\textwidth}{!}{%
%\small
\begin{tabular}{lll|clclclclcl}
\multicolumn{3}{l|}{\textbf{Shift Size}}                                                                                      & \multicolumn{2}{c}{\textbf{20}} & \multicolumn{2}{c}{\textbf{15}} & \multicolumn{2}{c}{\textbf{10}} & \multicolumn{2}{c}{\textbf{5}} & \multicolumn{2}{c}{\textbf{1}} \\ \hline
%\multicolumn{1}{l|}{\multirow{2}{*}{\textbf{True positive(optimal value)}}}   & \multicolumn{2}{l|}{\textbf{opposite shifts}} & \multicolumn{2}{c}{0}           & \multicolumn{2}{c}{0}           & \multicolumn{2}{c}{0}           & \multicolumn{2}{c}{0}          & \multicolumn{2}{c}{0}          \\
%\multicolumn{1}{l|}{}                                                         & \multicolumn{2}{l|}{\textbf{single shift}}    & \multicolumn{2}{c}{0}           & \multicolumn{2}{c}{0}           & \multicolumn{2}{c}{0}           & \multicolumn{2}{c}{0}          & \multicolumn{2}{c}{0}          \\ \hline
\multicolumn{1}{l|}{\multirow{2}{*}{\textbf{False  positive(optimal value)}}} & \multicolumn{2}{l|}{\textbf{opposite shifts}} & \multicolumn{2}{c}{0.66}        & \multicolumn{2}{c}{0.63}        & \multicolumn{2}{c}{0.49}        & \multicolumn{2}{c}{0.48}       & \multicolumn{2}{c}{0.44}       \\
\multicolumn{1}{l|}{}                                                         & \multicolumn{2}{l|}{\textbf{single shift}}    & \multicolumn{2}{c}{0.4}         & \multicolumn{2}{c}{0.5}         & \multicolumn{2}{c}{0.26}        & \multicolumn{2}{c}{0.49}       & \multicolumn{2}{c}{0.43}       \\ \hline
\multicolumn{1}{l|}{\multirow{2}{*}{\textbf{True positive(variance)}}}        & \multicolumn{2}{l|}{\textbf{opposite shifts}} & \multicolumn{2}{c}{0}           & \multicolumn{2}{c}{0}           & \multicolumn{2}{c}{0}           & \multicolumn{2}{c}{0}          & \multicolumn{2}{c}{0}          \\
\multicolumn{1}{l|}{}                                                         & \multicolumn{2}{l|}{\textbf{single shift}}    & \multicolumn{2}{c}{0.9}         & \multicolumn{2}{c}{0.86}        & \multicolumn{2}{c}{0.84}        & \multicolumn{2}{c}{0.41}       & \multicolumn{2}{c}{0.01}       \\ \hline
\multicolumn{1}{l|}{\multirow{2}{*}{\textbf{False positive(variance)}}}       & \multicolumn{2}{l|}{\textbf{opposite shifts}} & \multicolumn{2}{c}{0.01}        & \multicolumn{2}{c}{0}           & \multicolumn{2}{c}{0.02}        & \multicolumn{2}{c}{0}          & \multicolumn{2}{c}{0}          \\
\multicolumn{1}{l|}{}                                                         & \multicolumn{2}{l|}{\textbf{single shift}}    & \multicolumn{2}{c}{0.73}        & \multicolumn{2}{c}{0.7}         & \multicolumn{2}{c}{0.57}        & \multicolumn{2}{c}{0.21}       & \multicolumn{2}{c}{0.02}       \\ \hline
\end{tabular}%
%    }

\end{table}

\FloatBarrier

\subsection{Simulations with varying $\alpha$}
\label{sec:supp-vary-alpha}

To check whether the conclusions of the main simulations are specific to the baseline setting $\alpha = 1$, we repeated the main-text simulations for a range of true adaptation rates~$\alpha \in \{0.7, 1, 1.5, 3.5, 7, 14, 35\}$ with $\sigma^2_0 = 2$ held fixed. For each true~$\alpha$ we applied all five methods as in the main experiment (with ShiVa estimating~$\alpha$ from a null model using \texttt{phylolm} and refitting after shift detection). Table~\ref{tab:mise-alpha} reports the median MISE of $\sigma^2$ and MISE of $\mu$ across 100 replicates and shift magnitudes. ShiVa's MISE values stay in the same range as the other shift-detection methods across all tested~$\alpha$, with no systematic degradation as~$\alpha$ grows; in panel~b), where the data actually contain a variance shift, ShiVa achieves the lowest MISE of $\sigma^2$ in five out of seven rows. On MISE of $\mu$, phyloEM is the most accurate in every row, with $\ell$1ou+pBIC close behind; ShiVa is within a small constant factor of these two. All methods' errors decrease as~$\alpha$ grows because the strong OU pull compresses the trait around its optima, reducing the variability of the estimated parameters.

\begin{table}[htbp]
\centering
\small
\setlength{\tabcolsep}{3pt}
\setlength{\aboverulesep}{0pt}
\setlength{\belowrulesep}{0pt}
\caption{\label{tab:mise-alpha} Median MISE under varying true adaptation rate~$\alpha$ with $\sigma^2_0 = 2$ held fixed. Values are medians across 100 replicates and across the shift magnitudes used in the main experiment. Best (lowest) value per row within each half in \textbf{bold}.}
\begin{tabular}{l|rrrrr|rrrrr}
\toprule
 & \multicolumn{5}{c|}{\textbf{MISE of $\sigma^2$}} & \multicolumn{5}{c}{\textbf{MISE of $\mu$}} \\
\textbf{$\alpha$} & ShiVa & $\ell$1ou+pBIC & $\ell$1ou+BIC & phyloEM & PCMFit & ShiVa & $\ell$1ou+pBIC & $\ell$1ou+BIC & phyloEM & PCMFit \\
\midrule
 & \multicolumn{5}{l|}{\textit{(a) Only mean shifts on branch~71}} & \multicolumn{5}{c}{} \\
0.7  & 7.29  & \textbf{6.83} & 7.74 & 7.28          & 19.6          & 3.57   & 3.18   & 5.79   & \textbf{3.10}    & 383 \\
1    & 6.16  & 5.10          & 7.57 & \textbf{4.97} & 17.1          & 2.46   & 2.27   & 4.11   & \textbf{2.23}    & 314 \\
1.5  & 6.30  & \textbf{6.19} & 9.35 & 6.85          & 16.8          & 1.96   & 1.82   & 2.66   & \textbf{1.80}    & 146 \\
3.5  & 16.1  & 13.0          & 16.4 & \textbf{10.8} & 26.9          & 0.603  & 0.388  & 0.834  & \textbf{0.378}   & 20.8 \\
7    & 44.5  & 30.7          & 38.7 & 30.7          & \textbf{20.1} & 0.241  & 0.0921 & 0.186  & \textbf{0.0863}  & 0.465 \\
14   & 92.2  & 60.4          & 58.6 & 58.6          & \textbf{23.2} & 0.130  & 0.0417 & 0.0734 & \textbf{0.0400}  & 0.163 \\
35   & 131   & 37.8          & \textbf{32.8} & 38.0 & 90.5          & 0.0788 & 0.0174 & 0.0275 & \textbf{0.0171}  & 0.0565 \\
 & \multicolumn{5}{l|}{\textit{(b) Only variance shifts on branch~195}} & \multicolumn{5}{c}{} \\
0.7  & \textbf{36.9} & 40.0  & 43.8 & 40.3 & 55.7          & 3.20   & 2.76   & 8.33   & \textbf{1.97}     & 21.3 \\
1    & \textbf{39.5} & 44.7  & 48.1 & 43.7 & 73.3          & 2.68   & 2.44   & 6.62   & \textbf{1.56}     & 25.7 \\
1.5  & \textbf{37.0} & 42.6  & 50.5 & 40.9 & 66.5          & 1.48   & 1.55   & 3.98   & \textbf{0.497}    & 23.3 \\
3.5  & \textbf{41.1} & 45.6  & 48.8 & 45.6 & 66.3          & 0.477  & 0.350  & 1.33   & \textbf{0.143}    & 3.14 \\
7    & 81.8          & 91.8  & 104  & 90.4 & \textbf{54.6} & 0.219  & 0.124  & 0.440  & \textbf{0.0613}   & 0.975 \\
14   & 124           & 137   & 142  & 140  & \textbf{60.7} & 0.0588 & 0.0392 & 0.177  & \textbf{0.0219}   & 0.297 \\
35   & \textbf{75.3} & 82.5  & 82.0 & 81.3 & 162           & 0.0238 & 0.0143 & 0.0478 & \textbf{0.00853}  & 0.0921 \\
\bottomrule
\end{tabular}
\end{table}

For completeness, we also show the predictive log-likelihood version of the misestimation comparison that complements Figure~\ref{fig:2-alpha}:

\begin{figure}[htbp]
    \centering
    \begin{subfigure}[b]{0.9\textwidth}
        \centering
        \begin{tikzpicture}
            \node[anchor=south west, inner sep=0] (img) {\includegraphics[width=\textwidth,height=0.21\textheight,keepaspectratio]{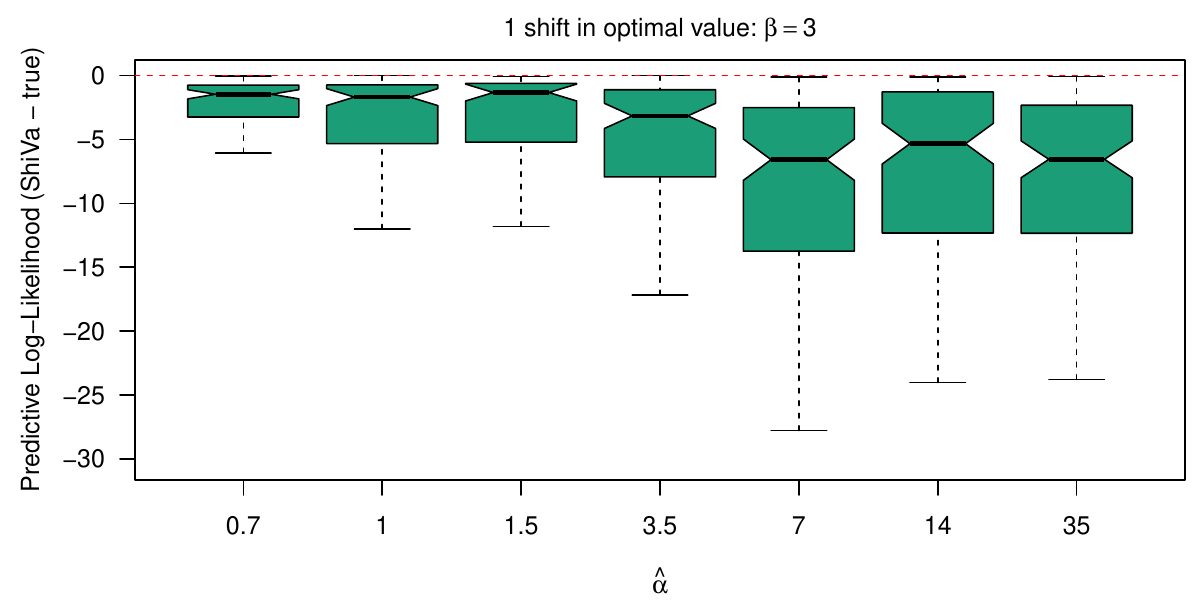}};
            \node[anchor=north west, inner sep=1pt, xshift=-10pt, font=\footnotesize] at (img.north west) {a)};
        \end{tikzpicture}
    \end{subfigure}

    \begin{subfigure}[b]{0.9\textwidth}
        \centering
        \begin{tikzpicture}
            \node[anchor=south west, inner sep=0] (img) {\includegraphics[width=\textwidth,height=0.21\textheight,keepaspectratio]{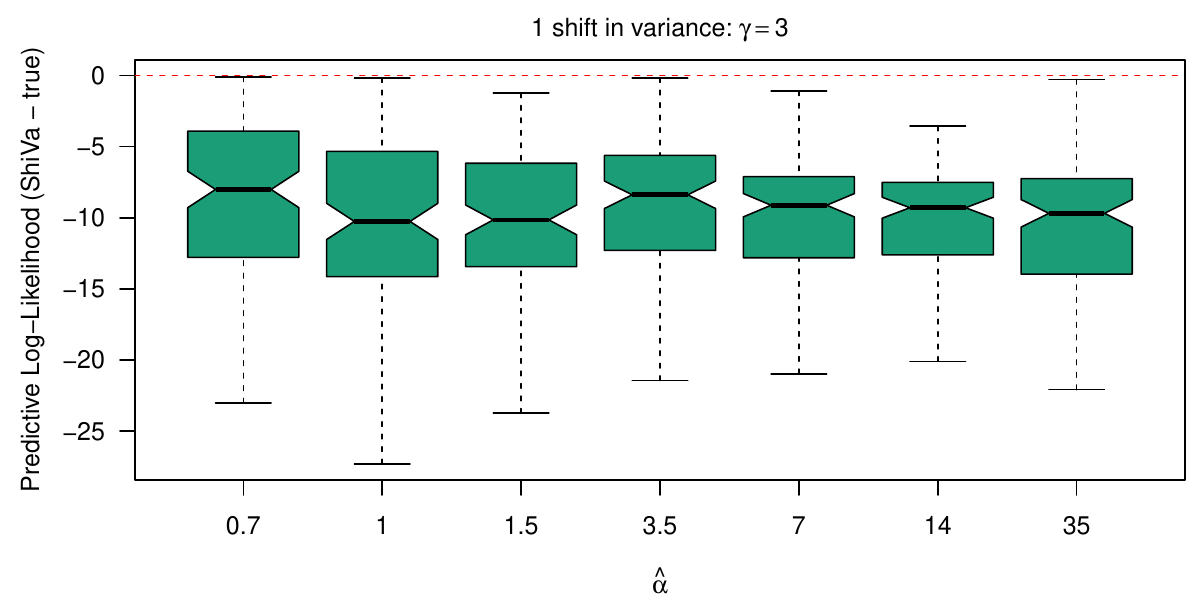}};
            \node[anchor=north west, inner sep=1pt, xshift=-10pt, font=\footnotesize] at (img.north west) {b)};
        \end{tikzpicture}
    \end{subfigure}

    \caption{Predictive log-likelihood of ShiVa under misestimated $\hat{\alpha}$ (true $\alpha = 1$) for two scenarios: a) 1 shift in optimal value ($\beta = 3$); b) 1 shift in variance ($\gamma = 3$). The $x$-axis gives $\log(\hat{\alpha})$.}
    \label{fig:supp-alpha-loglik}
\end{figure}

\FloatBarrier

\subsection{Supplementary figures and tables}

\begin{figure}[!tp]
    \centering
    \begin{subfigure}[b]{0.38\textwidth}\centering
        \begin{tikzpicture}
            \node[anchor=south west, inner sep=0] (img) {\includegraphics[width=\textwidth]{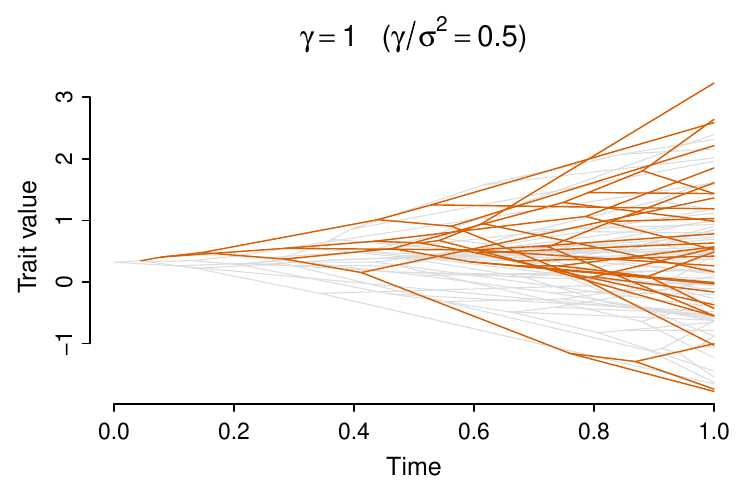}};
            \node[anchor=north west, inner sep=1pt, xshift=-10pt, font=\footnotesize] at (img.north west) {a)};
        \end{tikzpicture}
    \end{subfigure}%
    \hfill
    \begin{subfigure}[b]{0.38\textwidth}\centering
        \begin{tikzpicture}
            \node[anchor=south west, inner sep=0] (img) {\includegraphics[width=\textwidth]{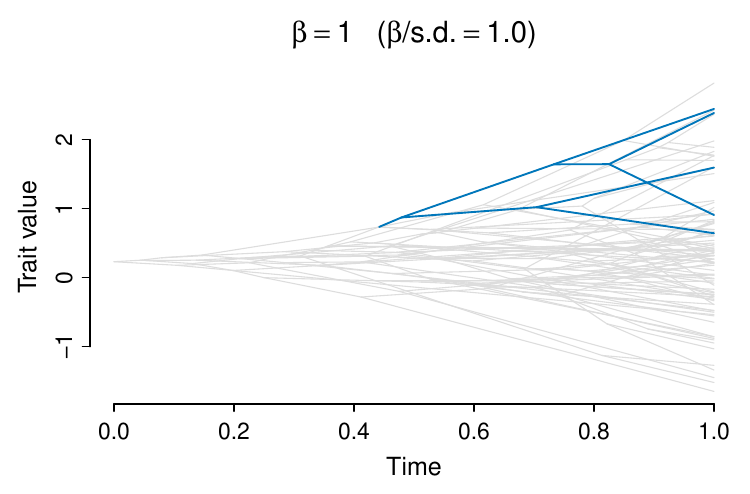}};
            \node[anchor=north west, inner sep=1pt, xshift=-10pt, font=\footnotesize] at (img.north west) {b)};
        \end{tikzpicture}
    \end{subfigure}

    \begin{subfigure}[b]{0.38\textwidth}\centering
        \begin{tikzpicture}
            \node[anchor=south west, inner sep=0] (img) {\includegraphics[width=\textwidth]{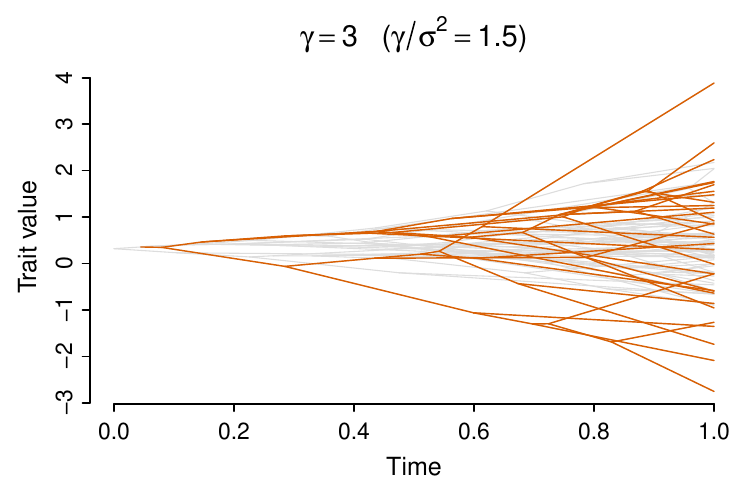}};
            \node[anchor=north west, inner sep=1pt, xshift=-10pt, font=\footnotesize] at (img.north west) {c)};
        \end{tikzpicture}
    \end{subfigure}%
    \hfill
    \begin{subfigure}[b]{0.38\textwidth}\centering
        \begin{tikzpicture}
            \node[anchor=south west, inner sep=0] (img) {\includegraphics[width=\textwidth]{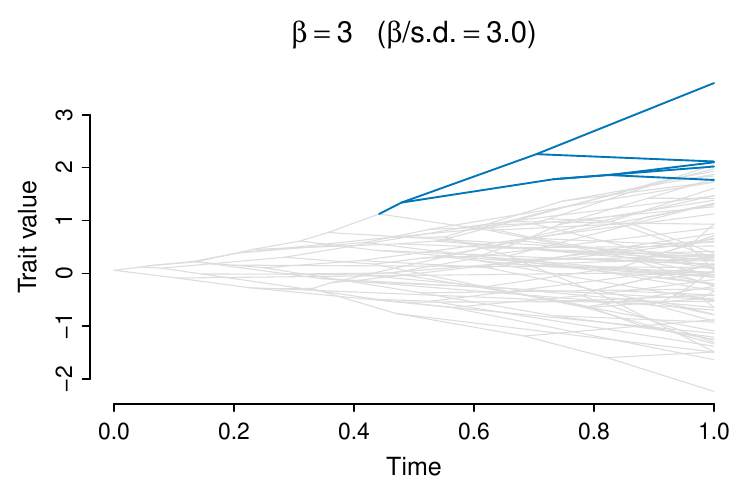}};
            \node[anchor=north west, inner sep=1pt, xshift=-10pt, font=\footnotesize] at (img.north west) {d)};
        \end{tikzpicture}
    \end{subfigure}

    \begin{subfigure}[b]{0.38\textwidth}\centering
        \begin{tikzpicture}
            \node[anchor=south west, inner sep=0] (img) {\includegraphics[width=\textwidth]{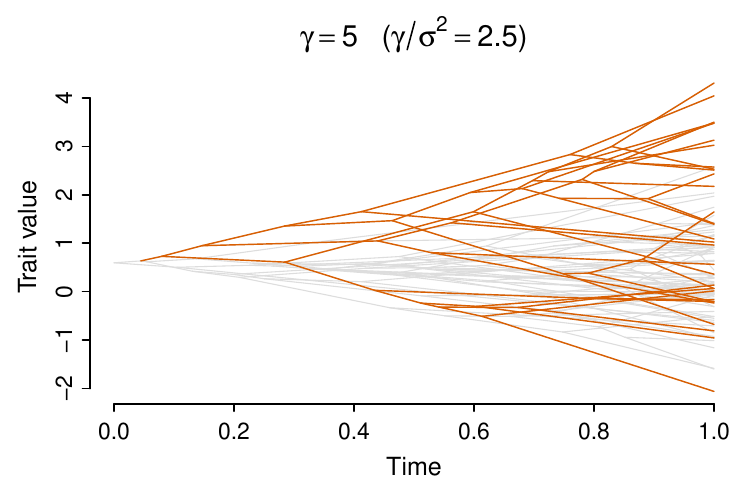}};
            \node[anchor=north west, inner sep=1pt, xshift=-10pt, font=\footnotesize] at (img.north west) {e)};
        \end{tikzpicture}
    \end{subfigure}%
    \hfill
    \begin{subfigure}[b]{0.38\textwidth}\centering
        \begin{tikzpicture}
            \node[anchor=south west, inner sep=0] (img) {\includegraphics[width=\textwidth]{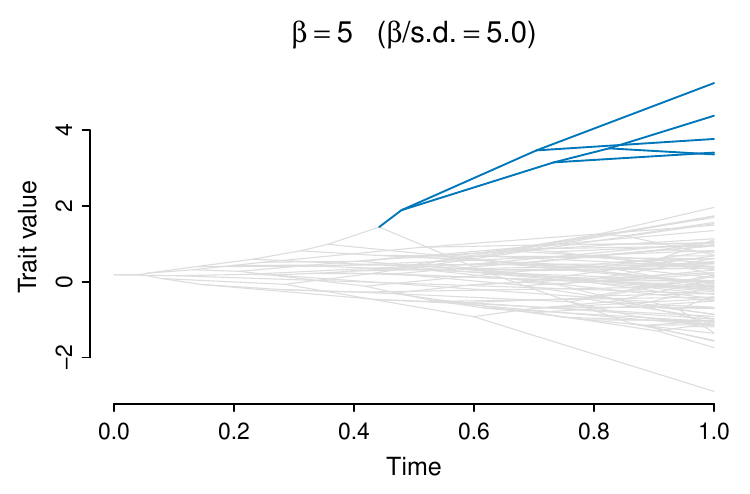}};
            \node[anchor=north west, inner sep=1pt, xshift=-10pt, font=\footnotesize] at (img.north west) {f)};
        \end{tikzpicture}
    \end{subfigure}

    \begin{subfigure}[b]{0.38\textwidth}\centering
        \begin{tikzpicture}
            \node[anchor=south west, inner sep=0] (img) {\includegraphics[width=\textwidth]{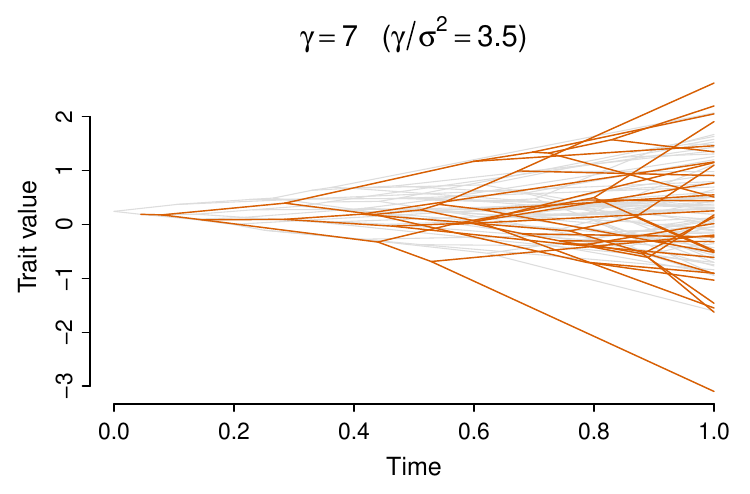}};
            \node[anchor=north west, inner sep=1pt, xshift=-10pt, font=\footnotesize] at (img.north west) {g)};
        \end{tikzpicture}
    \end{subfigure}%
    \hfill
    \begin{subfigure}[b]{0.38\textwidth}\centering
        \begin{tikzpicture}
            \node[anchor=south west, inner sep=0] (img) {\includegraphics[width=\textwidth]{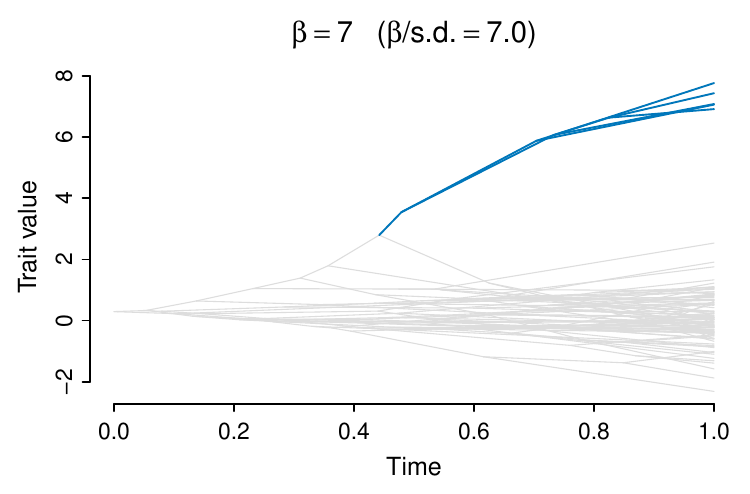}};
            \node[anchor=north west, inner sep=1pt, xshift=-10pt, font=\footnotesize] at (img.north west) {h)};
        \end{tikzpicture}
    \end{subfigure}
    \caption{Traitgrams of simulated data under variance shifts on branch~195 (left column, orange) and mean shifts on branch~71 (right column, blue), paired by shift magnitude, showing: a) variance shift $\gamma=1$; b) mean shift $\beta=1$; c) variance shift $\gamma=3$; d) mean shift $\beta=3$; e) variance shift $\gamma=5$; f) mean shift $\beta=5$; g) variance shift $\gamma=7$; h) mean shift $\beta=7$. Parameters: $\alpha = 1$, $\sigma^2 = 2$, stationary standard deviation~$= 1$. Gray lines show unshifted lineages; coloured lines show the shifted clade.}
    \label{fig:traitgram-supp}
\end{figure}

\begin{figure}[!tp]

    \centering
    \begin{subfigure}[b]{0.9\textwidth}
        \centering
        \begin{tikzpicture}
            \node[anchor=south west, inner sep=0] (img) {\includegraphics[width=\textwidth,height=0.21\textheight,keepaspectratio]{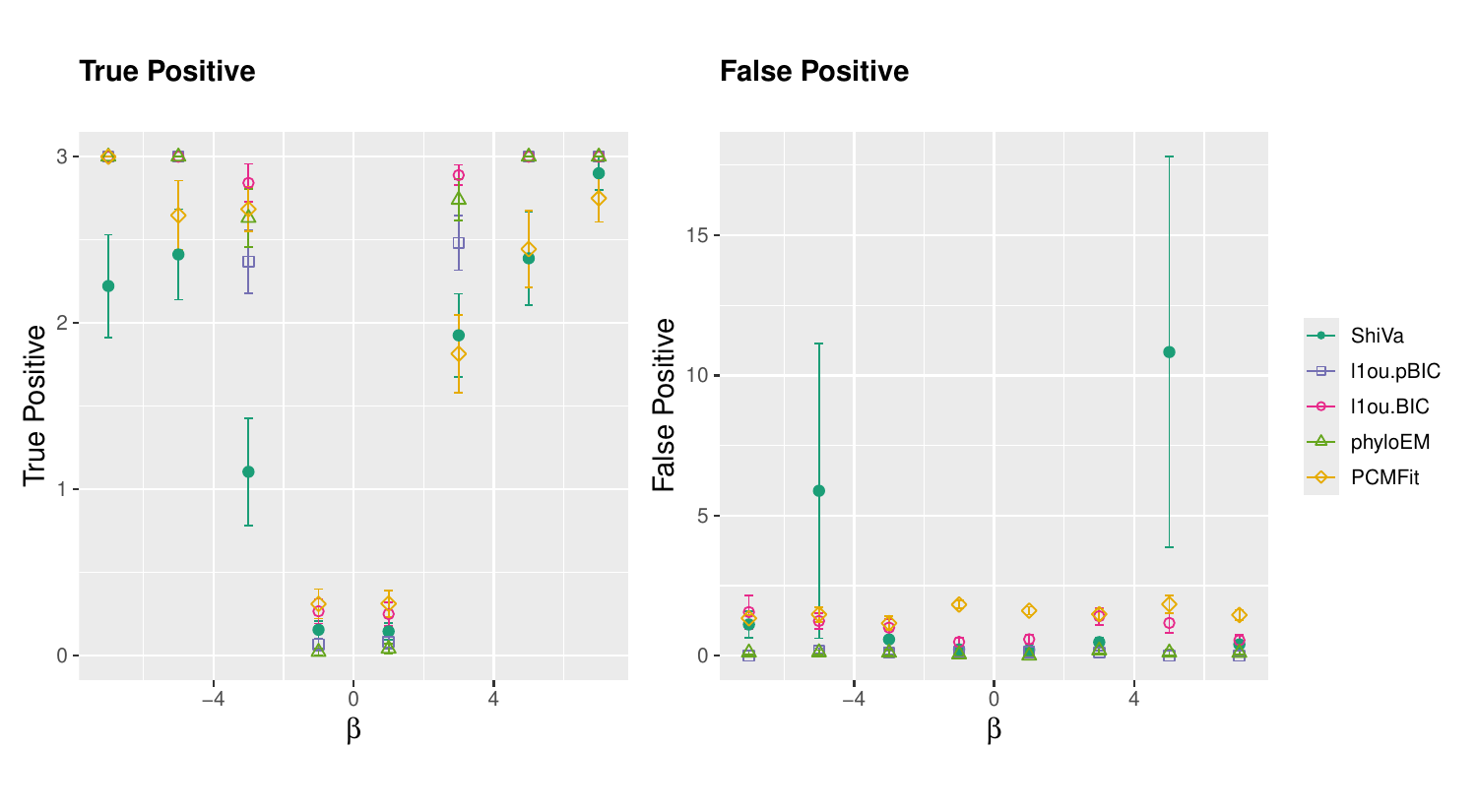}};
            \node[anchor=north west, inner sep=1pt, xshift=-10pt, font=\footnotesize] at (img.north west) {a)};
        \end{tikzpicture}
    \end{subfigure}

    \begin{subfigure}[b]{0.9\textwidth}
        \centering
        \begin{tikzpicture}
            \node[anchor=south west, inner sep=0] (img) {\includegraphics[width=\textwidth,height=0.21\textheight,keepaspectratio]{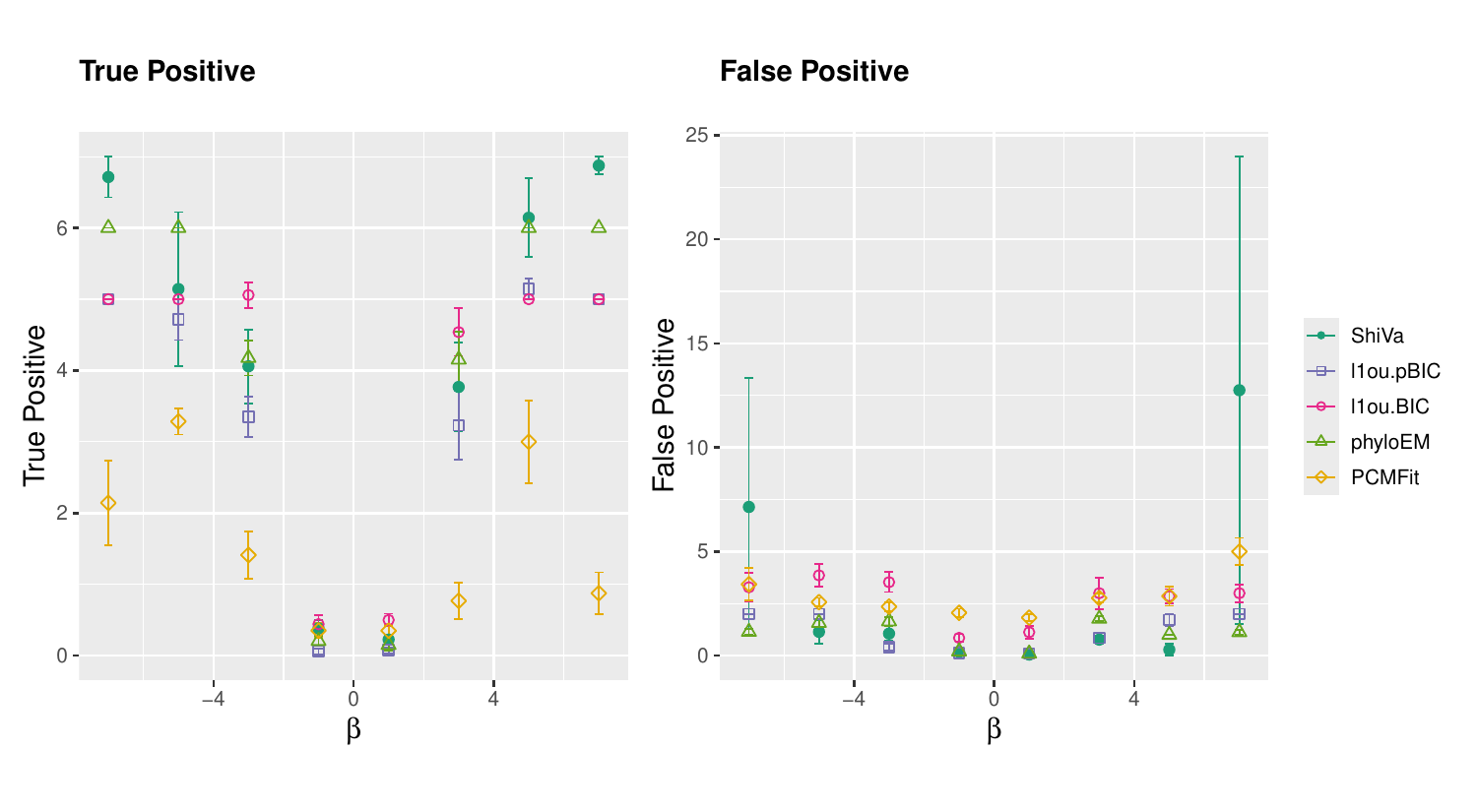}};
            \node[anchor=north west, inner sep=1pt, xshift=-10pt, font=\footnotesize] at (img.north west) {b)};
        \end{tikzpicture}
    \end{subfigure}

    \begin{subfigure}[b]{0.9\textwidth}
        \centering
        \begin{tikzpicture}
            \node[anchor=south west, inner sep=0] (img) {\includegraphics[width=\textwidth,height=0.21\textheight,keepaspectratio]{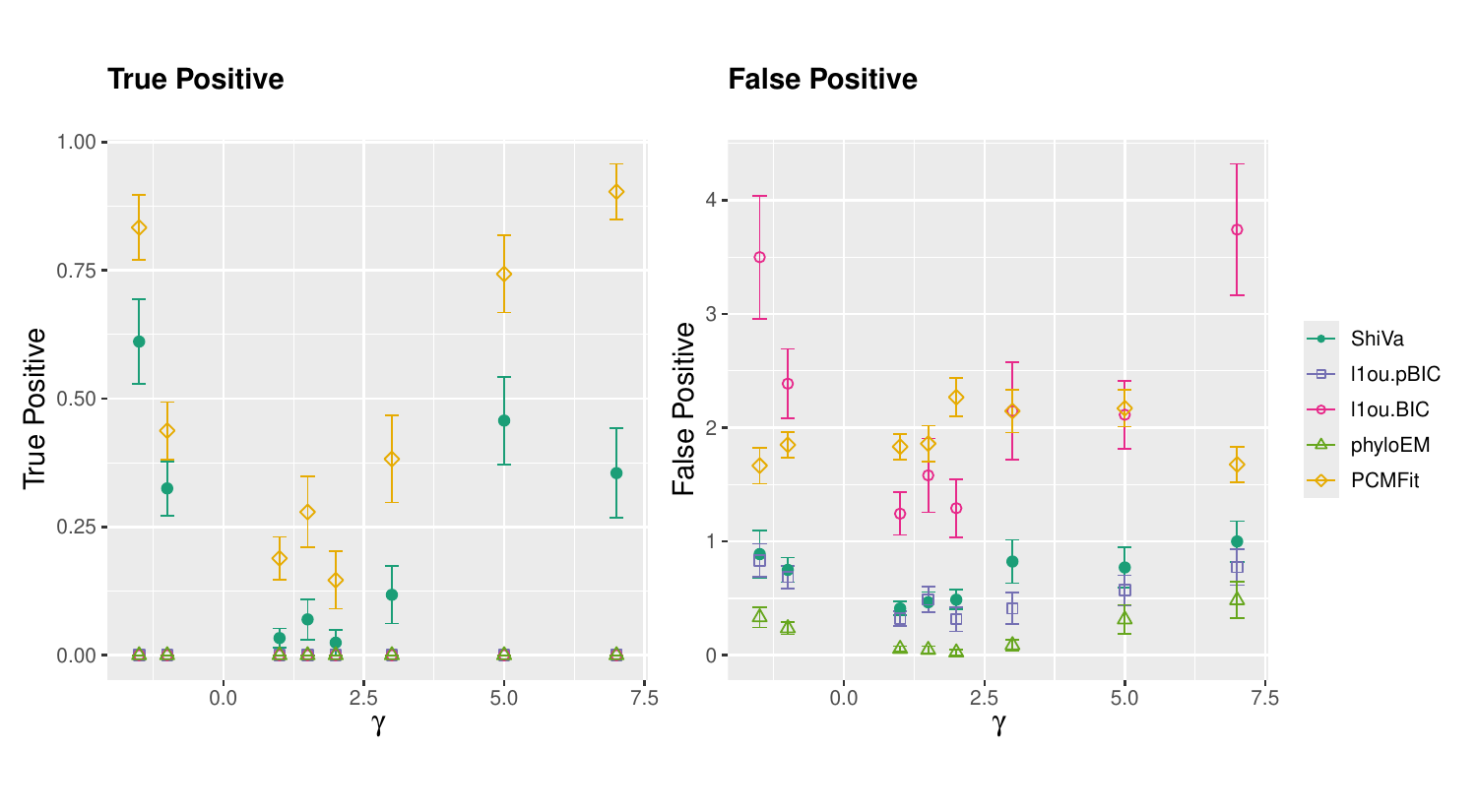}};
            \node[anchor=north west, inner sep=1pt, xshift=-10pt, font=\footnotesize] at (img.north west) {c)};
        \end{tikzpicture}
    \end{subfigure}

    \begin{subfigure}[b]{0.9\textwidth}
        \centering
        \begin{tikzpicture}
            \node[anchor=south west, inner sep=0] (img) {\includegraphics[width=\textwidth,height=0.21\textheight,keepaspectratio]{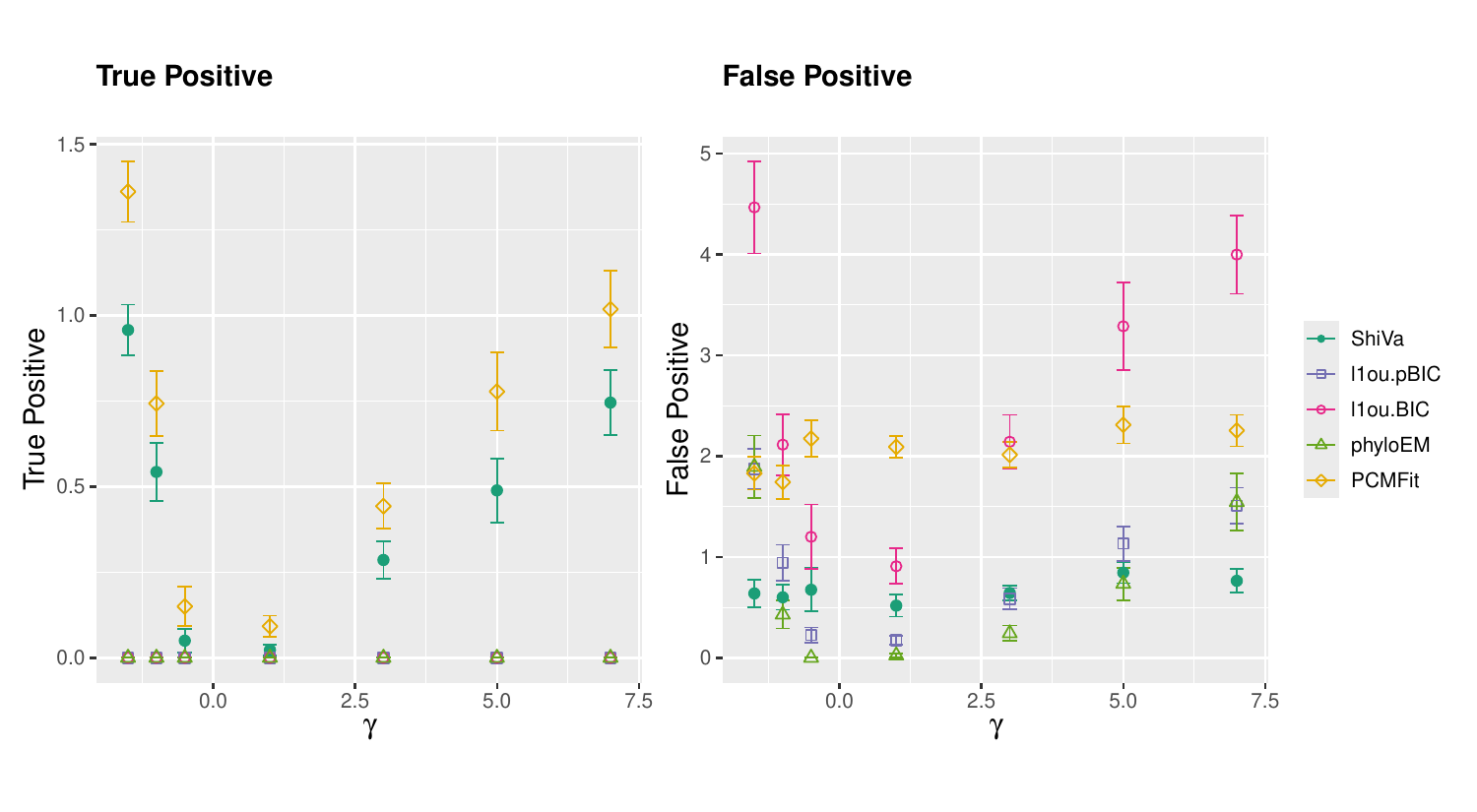}};
            \node[anchor=north west, inner sep=1pt, xshift=-10pt, font=\footnotesize] at (img.north west) {d)};
        \end{tikzpicture}
    \end{subfigure}
    \caption{Comparison of True Positive and False Positive of different methods under additional shift configurations: a)~3 shifts in optimal value (branches 43, 71, 192), b)~7 shifts in optimal value, c)~1 variance shift at a different position (branch 196), d)~2 variance shifts (branches 195 and 130).}
    \label{fig:supp-TP-FP}
\end{figure}

\begin{table}[ht]
\centering
\small
\setlength{\tabcolsep}{3pt}
\setlength{\aboverulesep}{0pt}
\setlength{\belowrulesep}{0pt}
\caption{\label{tab:mise-all} Median MISE across the replicate datasets for the five methods, corresponding to the boxplots of Figure~\ref{fig:mise-main}. The left half reports MISE of $\sigma^2$ and the right half MISE of $\mu$. Rows are grouped by simulation scenario (a)--(d), matching the panels of Figure~\ref{fig:2-TP-FP}; within each block rows give the shift magnitude. Best (lowest) value per row within each half in \textbf{bold}.}
\begin{tabular}{l|rrrrr|rrrrr}
\toprule
 & \multicolumn{5}{c|}{\textbf{MISE of $\sigma^2$}} & \multicolumn{5}{c}{\textbf{MISE of $\mu$}} \\
\textbf{Shift size} & ShiVa & $\ell$1ou+pBIC & $\ell$1ou+BIC & phyloEM & PCMFit & ShiVa & $\ell$1ou+pBIC & $\ell$1ou+BIC & phyloEM & PCMFit \\
\midrule
 & \multicolumn{5}{l|}{\textit{(a) Only mean shifts, varying $\beta$}} & \multicolumn{5}{c}{} \\
-7 & \textbf{5.79} & 6.52 & 11.7 & 6.52 & 38.5 & 2.19 & 1.53 & 3.46 & \textbf{1.46} & 3,600 \\
-5 & 5.08 & \textbf{4.68} & 6.23 & 5.67 & 10.1 & 2.15 & 1.89 & 2.97 & \textbf{1.81} & 1,710 \\
-3 & 6.91 & 7.17 & 8.99 & \textbf{6.69} & 10.9 & 2.45 & 2.42 & 4.94 & \textbf{2.21} &  427 \\
-1 & \textbf{3.67} &  4.4 & 4.44 & 4.04 & 17.8 & 4.83 & 4.26 &  5.3 & \textbf{3.23} & 83.6 \\
1 & 5.91 &  6.3 & 6.69 & \textbf{   5} & 22.5 & 3.25 & 3.08 & 4.05 & \textbf{3.03} &  103 \\
3 & 7.89 & 5.48 & \textbf{4.92} & 5.82 & 16.6 & 1.83 & \textbf{1.61} & 3.18 & 1.61 &  375 \\
5 & 8.51 & \textbf{4.66} & 8.42 & 4.66 & 23.8 & 2.26 & 1.29 & 3.88 & \textbf{ 1.2} &  743 \\
7 & 5.04 & \textbf{4.24} & 8.53 & 7.76 & 52.5 & 2.62 & \textbf{1.54} & 4.58 &  1.8 &  297 \\
 & \multicolumn{5}{l|}{\textit{(b) Only variance shifts, varying $\gamma$}} & \multicolumn{5}{c}{} \\
-1.5 & \textbf{16.8} &   25 & 31.7 & 27.4 & 33.6 & 1.73 & 1.63 & 6.58 & \textbf{1.21} & 11.1 \\
-1 & 12.8 & 12.5 & 13.9 & \textbf{12.4} & 29.8 & 1.14 & 0.958 & 3.42 & \textbf{0.842} & 23.4 \\
1 & 20.3 & \textbf{19.2} & 20.9 & 20.5 & 46.8 & 3.99 & 1.75 & 7.23 & \textbf{0.77} & 36.2 \\
1.5 & 29.7 & \textbf{  29} & 34.4 &   33 & 75.3 & 3.18 & 2.79 & 6.17 & \textbf{2.05} & 30.2 \\
2 & \textbf{43.9} & 45.9 & 47.1 &   44 & 72.3 & 2.29 & 1.85 &    6 & \textbf{1.11} & 50.9 \\
3 & \textbf{91.6} & 95.4 &  106 & 95.4 &  147 & 3.67 & 2.41 & 10.5 & \textbf{0.843} & 21.4 \\
5 & \textbf{ 230} &  253 &  264 &  247 &  266 & 3.17 & 5.55 &   22 & \textbf{2.21} & 21.2 \\
7 & \textbf{ 305} &  478 &  471 &  479 &  369 & \textbf{3.45} & 11.2 & 25.5 & 5.01 & 13.7 \\
 & \multicolumn{5}{l|}{\textit{(c) Both shifts; $\gamma=5$ fixed, varying $\beta$}} & \multicolumn{5}{c}{} \\
-10 &  266 &  267 &  258 & \textbf{ 254} & 3,290 & 9.14 & 5.74 & 18.2 & \textbf{ 5.5} &  930 \\
-7 & \textbf{ 249} &  254 &  294 &  250 &  543 & 6.55 & \textbf{6.17} & 25.7 & 7.18 & 2,730 \\
-5 & \textbf{ 244} &  255 &  256 &  257 &  455 & \textbf{5.71} & 7.76 & 22.5 & 6.73 &  912 \\
-3 & \textbf{ 244} &  255 &  267 &  251 &  298 & \textbf{4.74} & 11.5 & 20.9 & 14.7 &  371 \\
-1 & \textbf{ 230} &  253 &  254 &  258 &  304 & \textbf{5.19} & 8.09 & 18.2 & 6.13 & 32.1 \\
1 & \textbf{ 233} &  257 &  257 &  261 &  304 & 7.63 & 10.1 & 20.3 & \textbf{7.59} & 32.8 \\
3 & \textbf{ 232} &  253 &  268 &  254 &  322 & \textbf{3.63} & 15.7 & 21.7 & 16.1 &  122 \\
5 & \textbf{ 252} &  254 &  270 &  256 &  287 & \textbf{5.48} & 8.65 & 21.7 & 9.03 &  367 \\
7 & \textbf{ 247} &  257 &  261 &  259 &  293 & \textbf{6.77} & 8.97 & 24.2 & 9.43 & 80.9 \\
10 & \textbf{ 240} &  262 &  278 &  263 &  403 & 8.31 & \textbf{ 5.7} & 20.4 &  7.7 & 73.9 \\
 & \multicolumn{5}{l|}{\textit{(d) Both shifts; $\beta=5$ fixed, varying $\gamma$}} & \multicolumn{5}{c}{} \\
-1.5 & 24.5 & 24.3 & 27.3 & \textbf{24.1} & 28.6 & 2.26 & 1.76 & 7.56 & \textbf{1.38} &  306 \\
-1 & 14.3 & \textbf{13.5} & 19.8 &   14 &   25 & 1.87 & \textbf{1.47} &    5 &  1.5 &  459 \\
1 & 20.9 & \textbf{  15} & 20.5 & 15.5 & 31.7 & 2.74 & 2.02 & 6.36 & \textbf{1.76} &  900 \\
1.5 & 32.3 & 29.1 & 31.4 & \textbf{28.4} & 50.8 & 2.46 & \textbf{1.76} & 7.52 & 1.83 &  912 \\
2 & 45.9 & \textbf{41.8} &   46 & 43.5 & 77.2 & 3.19 & \textbf{2.19} &  9.2 & 2.24 &  609 \\
3 & 98.6 & 92.4 & 96.3 & \textbf{92.1} &  134 & 3.62 & 3.47 & 12.6 & \textbf{3.22} &  408 \\
5 & \textbf{ 252} &  254 &  270 &  256 &  287 & \textbf{5.48} & 8.65 & 21.7 & 9.03 &  367 \\
7 & \textbf{ 297} &  483 &  506 &  476 &  310 & \textbf{3.84} & 10.2 & 30.6 & 10.1 &  120 \\
\bottomrule
\end{tabular}
\end{table}

\begin{figure}[!tp]

    \centering
    \begin{subfigure}[b]{0.9\textwidth}
        \centering
        \begin{tikzpicture}
            \node[anchor=south west, inner sep=0] (img) {\includegraphics[width=\textwidth,height=0.21\textheight,keepaspectratio]{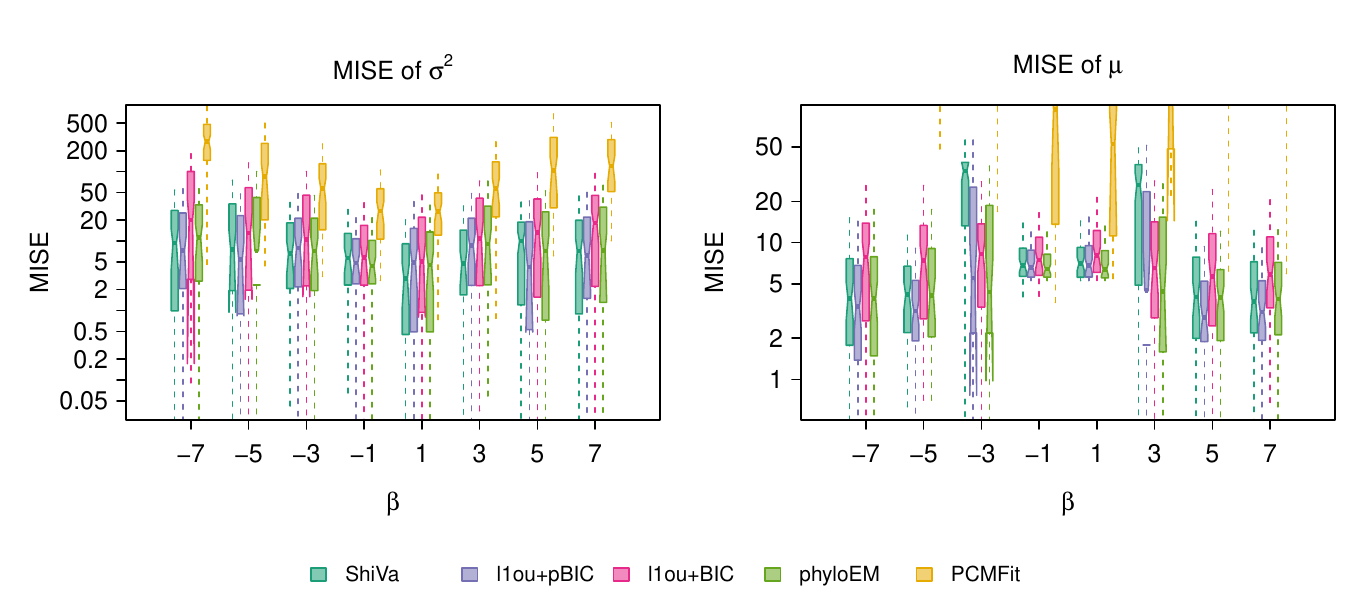}};
            \node[anchor=north west, inner sep=1pt, xshift=-10pt, font=\footnotesize] at (img.north west) {a)};
        \end{tikzpicture}
    \end{subfigure}

    \begin{subfigure}[b]{0.9\textwidth}
        \centering
        \begin{tikzpicture}
            \node[anchor=south west, inner sep=0] (img) {\includegraphics[width=\textwidth,height=0.21\textheight,keepaspectratio]{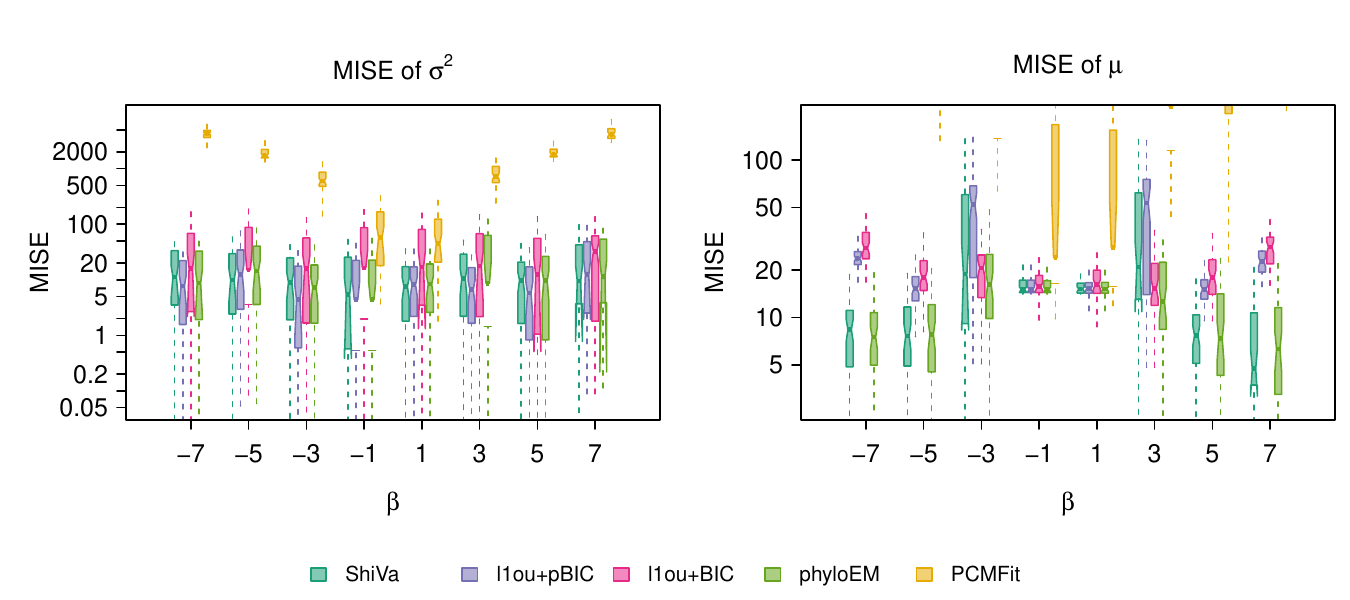}};
            \node[anchor=north west, inner sep=1pt, xshift=-10pt, font=\footnotesize] at (img.north west) {b)};
        \end{tikzpicture}
    \end{subfigure}

    \begin{subfigure}[b]{0.9\textwidth}
        \centering
        \begin{tikzpicture}
            \node[anchor=south west, inner sep=0] (img) {\includegraphics[width=\textwidth,height=0.21\textheight,keepaspectratio]{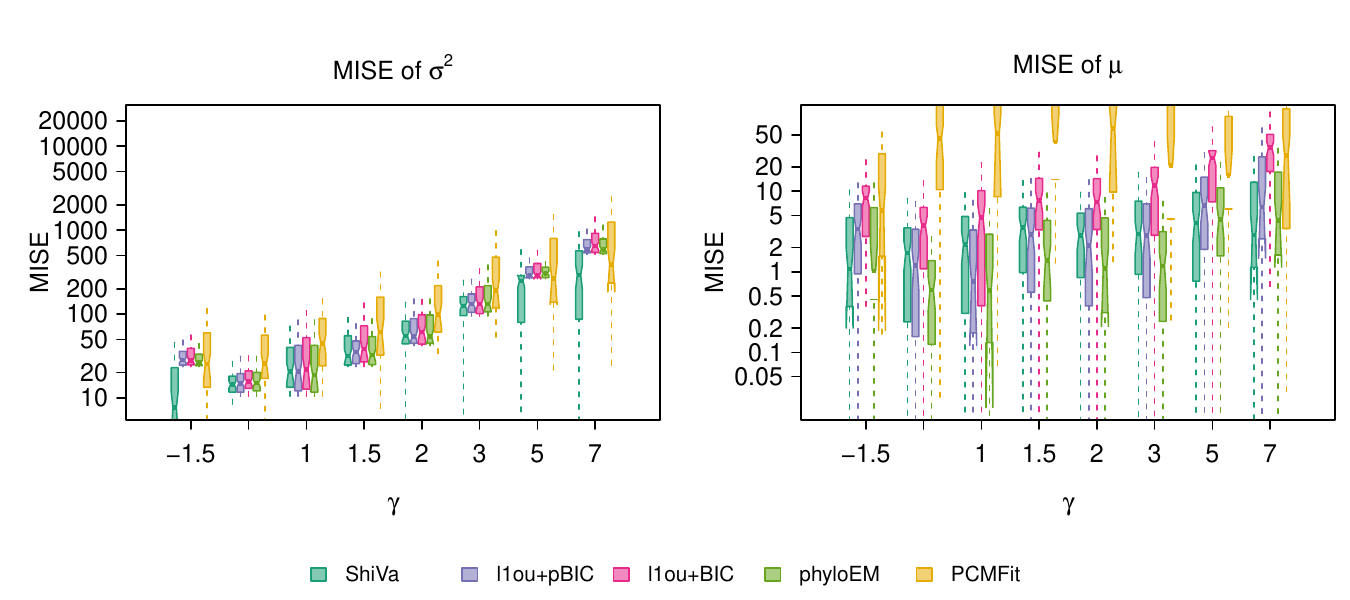}};
            \node[anchor=north west, inner sep=1pt, xshift=-10pt, font=\footnotesize] at (img.north west) {c)};
        \end{tikzpicture}
    \end{subfigure}

    \begin{subfigure}[b]{0.9\textwidth}
        \centering
        \begin{tikzpicture}
            \node[anchor=south west, inner sep=0] (img) {\includegraphics[width=\textwidth,height=0.21\textheight,keepaspectratio]{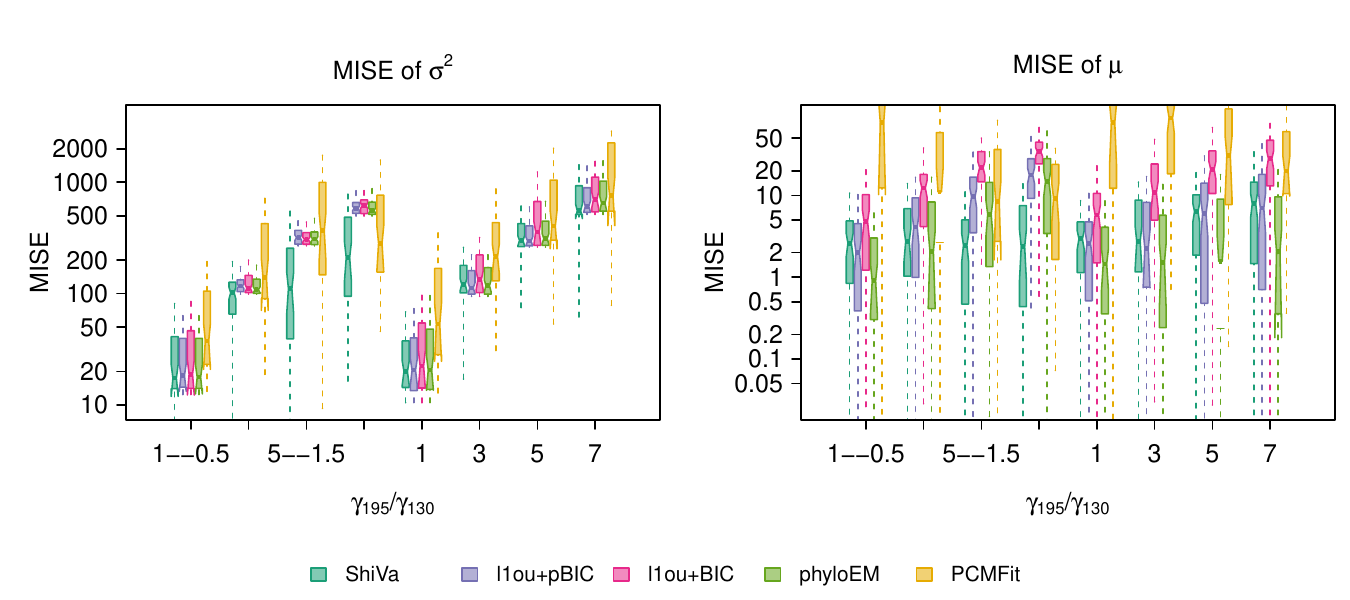}};
            \node[anchor=north west, inner sep=1pt, xshift=-10pt, font=\footnotesize] at (img.north west) {d)};
        \end{tikzpicture}
    \end{subfigure}
    \caption{Comparison of MISE of $\sigma^2$ and MISE of $\mu$ of different methods (Supplementary simulations) across four configurations: a) 3 shifts in optimal value; b) 7 shifts in optimal value; c) 1 shift in variance (different position); d) 2 shifts in variance. The $y$-axis is on a log scale, with limits chosen from the four shift-detection methods so that their boxplots remain readable; PCMFit values that exceed this range are clipped from view.}
\end{figure}

\begin{table}[htbp]
\caption{Computational Time Comparison for Different Methods (Supplementary simulations)}
\adjustbox{max width=\textwidth}{
\begin{tabular}{ccccccccc}
\toprule
\multicolumn{4}{c}{} & \multicolumn{5}{c}{\textbf{Average Computation Time (seconds)}} \\ 
\cmidrule(lr){5-9} 
\makecell{\textbf{Shift Position} \\ \textbf{(mean)}} & \textbf{Beta} & \makecell{\textbf{Shift Position} \\ \textbf{(variance)}} & \textbf{Gamma} & \makecell{ \textbf{ShiVa}} & \makecell{\textbf{l1ou-pBIC}} & \makecell{\textbf{l1ou-BIC}} & \makecell{ \textbf{phyloEM}} & \makecell{ \textbf{PCMFit}} \\
\midrule
43,71,192   & 1    & 0   & 0    & 86.80  & 7.66   & 8.43   & 69.64   & 11561.58  \\
43,71,192   & 3    & 0   & 0    & 74.95  & 8.37   & 8.36   & 68.09   & 27477.04  \\
43,71,192   & 5    & 0   & 0    & 88.87  & 8.76   & 8.86   & 68.05   & 36900.89  \\
43,71,192   & -1   & 0   & 0    & 71.88  & 7.99   & 8.40   & 69.21   & 13952.25  \\
43,71,192   & -3   & 0   & 0    & 67.89  & 8.61   & 8.56   & 70.58   & 35867.14  \\
43,71,192   & -5   & 0   & 0    & 76.54  & 9.13   & 9.16   & 72.69   & 40202.81  \\
6,71,197,3,88,191,98   & 1    & 0   & 0    & 117.94  & 7.80   & 8.05   & 70.53   & 14863.06  \\
6,71,197,3,88,191,98   & 3    & 0   & 0    & 80.94  & 6.06   & 6.08   & 70.63   & 31135.83  \\
6,71,197,3,88,191,98   & 5    & 0   & 0    & 81.68  & 8.55   & 8.86   & 67.94   & 59322.66  \\
6,71,197,3,88,191,98   & -1   & 0   & 0    & 83.20  & 7.79   & 8.04   & 69.02   & 15652.61  \\
6,71,197,3,88,191,98   & -3   & 0   & 0    & 98.53  & 7.42   & 6.84   & 69.47   & 29763.69  \\
6,71,197,3,88,191,98   & -5   & 0   & 0    & 69.55  & 7.99   & 7.85   & 70.15   & 60076.49  \\
0   & 0    & 196  & 1    & 68.06  & 8.08   & 8.35   & 69.46   & 14515.10  \\
0   & 0    & 196  & 1.5  & 72.20  & 8.03   & 8.31   & 69.51   & 15649.65  \\
0   & 0    & 196  & 2    & 83.65  & 8.39   & 8.54   & 70.18   & 17266.36  \\
0   & 0    & 196  & 3    & 90.27  & 7.61   & 8.14   & 69.24   & 18907.37  \\
0   & 0    & 196  & 5    & 89.08  & 8.27   & 8.42   & 69.61   & 18364.71  \\
0   & 0    & 196  & 7    & 101.44  & 8.03   & 8.51   & 67.09   & 18858.03  \\
0   & 0    & 196  & -1   & 99.09  & 8.26   & 8.66   & 67.89   & 16773.86  \\
0   & 0    & 196  & -1.5 & 113.54  & 8.21   & 8.72   & 67.30   & 19419.61  \\
0   & 0    & 195,130  & 1,-0.5  & 80.75  & 8.37   & 8.68   & 70.94   & 16275.18  \\
0   & 0    & 195,130  & 3,-1    & 93.39  & 8.11   & 8.39   & 70.21   & 18297.14  \\
0   & 0    & 195,130  & 5,-1.5  & 159.13  & 12.06   & 12.58   & 116.55   & 36602.57  \\
0   & 0    & 195,130  & 7,-1.5  & 114.15  & 8.60   & 8.86   & 69.08   & 22939.48  \\
0   & 0    & 195,130  & 1      & 85.99  & 8.15   & 8.31   & 70.17   & 15881.60  \\
0   & 0    & 195,130  & 3      & 91.66  & 8.10   & 8.38   & 70.03   & 17969.21  \\
0   & 0    & 195,130  & 5      & 104.25  & 8.69   & 9.17   & 71.10   & 25327.97  \\
0   & 0    & 195,130  & 7      & 95.08  & 8.63   & 8.94   & 69.17   & 23545.64  \\
\bottomrule
\end{tabular}
}
\end{table}

%%%%%%%%%%%%%%%%%%%% REFERENCES %%%%%%%%%%%%%%%%%%
\bibliographystyle{apalike}
\bibliography{reference_abbr}

\end{document}